\documentclass[twocolumn, twocolappendix]{aastex631}
\usepackage{lineno}

\newcommand{\bat}{Swift/BAT}
\newcommand{\NAGNbass}{858} 

\newcommand{\todo}{\ifmmode \text{\color{red}\Huge{\(\bullet\)}} \else {\color{red}{\Huge$\bullet$}}\fi}
\newcommand{\tido}{\ifmmode {{\color{red}\bullet}} \else {\color{red}$\bullet$}\fi}

\newcommand{\E        }[1]{\ifmmode 10^{#1} \else $10^{#1}$\fi}
\newcommand{\tE        }[1]{\ifmmode \times10^{#1} \else $\times10^{#1}$\fi}
\newcommand{\til}{\ifmmode \sim \else $\sim$\fi}
\renewcommand{\~} {\ifmmode \sim \else $\sim$\fi}

\newcommand{\logNH }{\ifmmode \log (N_{\rm H}/{\rm cm}^{-2}) \else $\log (N_{\rm H}/{\rm cm}^{-2})$\fi}

\newcommand{\Mbh   }{\ifmmode M_{\rm BH} \else $M_{\rm BH}$\fi}

\newcommand{\Catrip}{\ifmmode \left[{\rm Ca}\,\textsc{ii}\right\,\lambda8498, 8542, 8662 \else Ca\,\textsc{ii} $\,\lambda8498, 8542, 8662$\fi}

\newcommand{\pc}	{\ifmmode {\rm pc} \else pc\fi}
\newcommand{\ld}	{\ifmmode {\rm l.d.} \else l.d.\fi}
\newcommand{\kms}	{\ifmmode {\rm km\,s}^{-1} \else km\,s$^{-1}$\fi}
\newcommand{\cc}	{\ifmmode {\rm cm}^{-3}    \else cm$^{-3}$\fi}
\newcommand{\cmii}	{\ifmmode {\rm cm}^{-2}    \else cm$^{-2}$\fi}
\newcommand{\ergs}	{\ifmmode {\rm erg\,s}^{-1} \else erg s$^{-1}$\fi}
\newcommand{\ergcms}	{\ifmmode {\rm erg\,cm}^{-2}\,{\rm s}^{-1} \else erg\,cm$^{-2}$\,s$^{-1}$\fi}
\newcommand{\ergcmsA}	{\ifmmode {\rm erg\,cm}^{-2}\,{\rm s}^{-1}\,{\rm\AA}^{-1}
\else erg\,cm$^{-2}$\,s$^{-1}$\,\AA$^{-1}$\fi}
\newcommand{  \ergcmsHz  }{\ifmmode{\rm erg\,cm}^{-2}\,{\rm s}^{-1}\,{\rm Hz}^{-1}
                       \else ergs\,cm$^{-2}$\,s$^{-1}$\,Hz$^{-1}$\fi}
\newcommand{\kev}	{\ifmmode {\rm keV} \else keV\fi}

\newcommand{\mic}	{\ifmmode {\rm \mu m} \else $\mu$m\fi}
\newcommand{\vFWHM}	{\ifmmode v_{\mbox{\tiny FWHM}} \else $v_{\mbox{\tiny FWHM}}$\fi}
\newcommand{\vBLR}	{\ifmmode v_{\mbox{\tiny BLR}} \else $v_{\mbox{\tiny BLR}}$\fi}
\newcommand{\sigBLR}	{\ifmmode \sigma_{\mbox{\tiny BLR}} \else $\sigma_{\mbox{\tiny BLR}}$\fi}
\newcommand{\vNLR}	{\ifmmode v_{\mbox{\tiny NLR}} \else $v_{\mbox{\tiny NLR}}$\fi}
\newcommand{\tauBLR}	{\ifmmode \tau_{\mbox{\tiny BLR}} \else $\tau_{\mbox{\tiny BLR}}$\fi}

\newcommand{\Hubble}	{\ifmmode {\rm km\,s}^{-1}\,{\rm Mpc}^{-1} \else km\,s$^{-1}$\,Mpc$^{-1}$\fi}
\newcommand{\NDunit}	{\ifmmode {\rm Mpc}^{-3} \else Mpc$^{-3}$\fi}
\newcommand{\LFunit}	{\ifmmode {\rm Mpc}^{-3}\,{\rm mag}^{-1} \else Mpc$^{-3}$\,mag$^{-1}$\fi}
\newcommand{\MFunit}	{\ifmmode {\rm Mpc}^{-3}\,{\rm dex}^{-1} \else Mpc$^{-3}$\,dex$^{-1}$\fi}

\newcommand{\Msun}{\ifmmode M_{\odot} \else $M_{\odot}$\fi}
\newcommand{\Lsun}{\ifmmode L_{\odot} \else $L_{\odot}$\fi}
\newcommand{\Zsun}{\ifmmode Z_{\odot} \else $Z_{\odot}$\fi}
\newcommand{\mpyr}{\ifmmode \Msun\,{\rm yr}^{-1} \else $\Msun\,{\rm yr}^{-1}$\fi}

\newcommand{\qnote}{\ifmmode q_{0} \else $q_{0}$\fi}
\newcommand{\Hnote}{\ifmmode H_{0} \else $H_{0}$\fi}
\newcommand{\hnote}{\ifmmode h_{0} \else $h_{0}$\fi}
\newcommand{\anote}{\ifmmode a_{0} \else $a_{0}$\fi}

\newcommand{  \Halpha   }{\ifmmode {\rm H}\alpha \else H$\alpha$\fi}
\newcommand{  \ha   	}{\ifmmode {\rm H}\alpha \else H$\alpha$\fi}
\newcommand{  \Hbeta    }{\ifmmode {\rm H}\beta \else H$\beta$\fi}
\newcommand{  \hb    	}{\ifmmode {\rm H}\beta \else H$\beta$\fi}
\newcommand{  \Hgamma   }{\ifmmode {\rm H}\gamma \else H$\gamma$\fi}
\newcommand{  \Hdelta   }{\ifmmode {\rm H}\delta \else H$\delta$\fi}
\newcommand{  \Lya      }{\ifmmode {\rm Ly}\alpha \else Ly$\alpha$\fi}
\newcommand{  \Lyb      }{\ifmmode {\rm Ly}\beta \else Ly$\beta$\fi}
\newcommand{  \Pa       }{\ifmmode {\rm P}\alpha \else P$\alpha$\fi}
\newcommand{  \Pb       }{\ifmmode {\rm P}\beta \else P$\beta$\fi}
\newcommand{  \Paa       }{\ifmmode {\rm P}\alpha \else P$\alpha$\fi}
\newcommand{  \Pab       }{\ifmmode {\rm Pa}\beta \else Pa$\beta$\fi}
\newcommand{  \Bra      }{\ifmmode {\rm Br}\alpha \else Br$\alpha$\fi}
\newcommand{  \Brg      }{\ifmmode {\rm Br}\gamma \else Br$\gamma$\fi}
\newcommand{  \hii      }{\ifmmode {\rm H}\,\textsc{ii} \else H\,\textsc{ii}\fi}
\newcommand{  \hei      }{\ifmmode {\rm He}\,\textsc{i} \else He\,\textsc{i}\fi}
\newcommand{  \heii     }{\ifmmode {\rm He}\,\textsc{ii} \else He\,\textsc{ii}\fi}
\newcommand{  \HeIIuv   }{\ifmmode {\rm He}\,\textsc{ii}\,\lambda1640 \else He\,\textsc{ii}\,$\lambda1640$\fi}
\newcommand{  \HeIIop   }{\ifmmode {\rm He}\,\textsc{ii}\,\lambda4686 \else He\,\textsc{ii}\,$\lambda4686$\fi}
\newcommand{  \cii      }{\ifmmode {\rm C}\,\textsc{ii}  \else C\,\textsc{ii}\fi}
\newcommand{  \ciii     }{\ifmmode {\rm C}\,\textsc{iii}\right] \else C\,\textsc{iii}]\fi}
\newcommand{  \CIII     }{\ifmmode {\rm C}\,\textsc{iii}\right]\,\lambda1909 \else C\,\textsc{iii}]\,$\lambda1909$\fi}
\newcommand{  \civ      }{\ifmmode {\rm C}\,\textsc{iv}  \else C\,\textsc{iv}\fi}
\newcommand{  \CIV      }{\ifmmode {\rm C}\,\textsc{iv}\,\lambda1549 \else C\,\textsc{iv}\,$\lambda1549$\fi}
\newcommand{  \nii      }{\ifmmode [{\rm N}\,\textsc{ii}]  \else [N\,\textsc{ii}]\fi}
\newcommand{  \niii     }{\ifmmode {\rm N}\,\textsc{iii} \else N\,\textsc{iii}\fi}
\newcommand{  \niv      }{\ifmmode {\rm N}\,\textsc{iv}  \else N\,\textsc{iv}\fi}
\newcommand{  \NIVuv    }{\ifmmode {\rm N}\,\textsc{iv}\,\lambda1486 \else N\,\textsc{iv}\,$\lambda1486$\fi}
\newcommand{  \nv       }{\ifmmode {\rm N}\,\textsc{v}   \else N\,\textsc{v}\fi}
\newcommand{\oi}{\ifmmode \left[{\rm O}\,\textsc{i}\right] \else [O\,{\sc i}]\fi}
\newcommand{\OI}{\ifmmode \left[{\rm O}\,\textsc{i}\right]\,\lambda6300 \else [O\,{\sc i}]$\,\lambda6300$\fi}

\newcommand{\oii}{\ifmmode \left[{\rm O}\,\textsc{ii}\right] \else [O\,{\sc ii}]\fi}
\newcommand{\OII}{\ifmmode \left[{\rm O}\,\textsc{ii}\right]\,\lambda3727 \else [O\,{\sc ii}]\,$\lambda3727$\fi}
\newcommand{\oiii}{\ifmmode \left[{\rm O}\,\textsc{iii}\right] \else [O\,{\sc iii}]\fi}
\newcommand{\OIII}{\ifmmode \left[{\rm O}\,\textsc{iii}\right]\,\lambda5007 \else [O\,{\sc iii}]\,$\lambda5007$\fi}

\newcommand{\NII}{\ifmmode \left[{\rm N}\,\textsc{ii}\right]\,\lambda6583 \else [N\,{\sc ii}]$\,\lambda6583$\fi}

\newcommand{\NeIII}{\ifmmode \left[{\rm Ne}\,\textsc{iii}\right]\,\lambda3968 \else [Ne\,{\sc iii}]$\,\lambda3968$\fi}

\newcommand{\NeV}{\ifmmode \left[{\rm Ne}\,\textsc{v}\right]\,\lambda3426 \else [Ne\,{\sc v}]$\,\lambda3426$\fi}

\newcommand{\HeII}{\ifmmode {\rm He}\,\textsc{ii}\,\lambda4686 \else He\,{\sc ii}$\,\lambda4686$\fi}

\newcommand{\sii}{\ifmmode \left[{\rm S}\,\textsc{ii}\right] \else [S\,{\sc ii}]\fi}

\newcommand{\SII}{\ifmmode \left[{\rm S}\,\textsc{ii}\right]\,\lambda6717,6731 \else [S\,{\sc ii}]$\,\lambda6717,6731$\fi}

\newcommand{  \OIIIuv   }{\ifmmode {\rm O}\,\textsc{iii}\,\lambda1663 \else O\,\textsc{iii}\,$\lambda1663$\fi}
\newcommand{  \oiv      }{\ifmmode {\rm O}\,\textsc{iv}  \else O\,\textsc{iv}\fi}
\newcommand{  \OIVuv    }{\ifmmode {\rm O}\,\textsc{iv}\,\lambda1402  \else O\,\textsc{iv}\,$\lambda1402$\fi}
\newcommand{  \OIVIR    }{\ifmmode {\rm O}\,\textsc{iv}\,25.9\,\mu {\rm m} \else O\,\textsc{iv}\,$25.9\,\mu$m\fi}
\newcommand{  \ovi      }{\ifmmode {\rm O}\,\textsc{vi}   \else O\,\textsc{vi}\fi}
\newcommand{  \Ovi      }{\ifmmode {\rm O}\,\textsc{vi}\,\lambda1035 \else O\,\textsc{vi}\,$\lambda1035$\fi}
\newcommand{  \nei      }{\ifmmode {\rm Ne}\,\textsc{i}   \else Ne\,\textsc{i}\fi}
\newcommand{  \neii     }{\ifmmode {\rm Ne}\,\textsc{ii}  \else Ne\,\textsc{ii}\fi}
\newcommand{  \NeiiIR   }{\ifmmode {\rm Ne}\,\textsc{ii}\,12.8\,\mu {\rm m} \else Ne\,\textsc{ii}\,$12.8\,\mu$m\fi}
\newcommand{  \neiii    }{\ifmmode {\rm Ne}\,\textsc{iii} \else Ne\,\textsc{iii}\fi}
\newcommand{  \neiv     }{\ifmmode {\rm Ne}\,\textsc{iv}  \else Ne\,\textsc{iv}\fi}
\newcommand{  \nev      }{\ifmmode {\rm Ne}\,\textsc{v}   \else Ne\,\textsc{v}\fi}
\newcommand{  \NevIR    }{\ifmmode {\rm Ne}\,\textsc{v}\,24.3\,\mu {\rm m} \else Ne\,\textsc{v}\,$24.3\,\mu$m\fi}
\newcommand{  \nevi     }{\ifmmode {\rm Ne}\,\textsc{vi}  \else Ne\,\textsc{vi}\fi}
\newcommand{  \mgi      }{\ifmmode {\rm Mg}\,\textsc{i}   \else Mg\,\textsc{i}\fi}
\newcommand{  \mgii     }{\ifmmode {\rm Mg}\,\textsc{ii}  \else Mg\,\textsc{ii}\fi}
\newcommand{  \MgII     }{\ifmmode {\rm Mg}\,\textsc{ii}\,\lambda2798 \else Mg\,\textsc{ii}\,$\lambda2798$\fi}

\newcommand{  \siii     }{\ifmmode {\rm S}\,\textsc{iii} \else S\,\textsc{iii}\fi}
\newcommand{  \siv      }{\ifmmode {\rm S}\,\textsc{iv}  \else S\,\textsc{iv}\fi}
\newcommand{  \sili     }{\ifmmode {\rm Si}\,\textsc{i}   \else Si\,\textsc{i}\fi}
\newcommand{  \silii    }{\ifmmode {\rm Si}\,\textsc{ii}  \else Si\,\textsc{ii}\fi}
\newcommand{  \Siliv    }{\ifmmode {\rm Si}\,\textsc{iv}  \else Si\,\textsc{iv}\fi}
\newcommand{  \SilIVuv  }{\ifmmode {\rm Si}\,\textsc{iv}\,\lambda1400  \else Si\,\textsc{iv}\,$\lambda1400$\fi}

\newcommand{  \caii     }{\ifmmode {\rm Ca}\,\textsc{ii}   \else Ca\,\textsc{ii}\fi}
 \newcommand{\Mgb}{\ifmmode \left{\rm Mg}\,\textsc{i}\right\,\lambda5175 \else Mg\,{\sc i}\,$\lambda5175$\fi}
\newcommand{\Cahk}{\ifmmode \left[{\rm Ca H+K}\,\textsc{ii}\right\,\lambda3935,3968 \else Ca H+K$\,\lambda3935,3968$\fi}
\newcommand{  \feii     }{\ifmmode {\rm Fe}\,\textsc{ii}  \else Fe\,\textsc{ii}\fi}
\newcommand{  \feiii    }{\ifmmode {\rm Fe}\,\textsc{iii} \else Fe\,\textsc{iii}\fi}

\newcommand{ \Lhb   }{\ifmmode L\left(\hb\right) \else $L\left(\hb\right)$\fi}
\newcommand{ \fwhb  }{\ifmmode {\rm FWHM}\left(\hb\right) \else FWHM(\hb)\fi}
\newcommand{ \Lha   }{\ifmmode L\left(\ha\right) \else $L\left(\ha\right)$\fi}
\newcommand{ \fwha  }{\ifmmode {\rm FWHM}\left(\ha\right) \else FWHM(\ha)\fi}
\newcommand{ \Lmg   }{\ifmmode L\left(\mgii\right) \else $L\left(\mgii\right)$\fi}
\newcommand{ \fwmg  }{\ifmmode {\rm FWHM}\left(\mgii\right) \else FWHM(\mgii)\fi}
\newcommand{ \Lciv  }{\ifmmode L\left(\civ\right) \else $L\left(\civ\right)$\fi}
\newcommand{ \fwciv }{\ifmmode {\rm FWHM}\left(\civ\right) \else FWHM(\civ)\fi}
\newcommand{ \fwhm  }{\ifmmode {\rm FWHM} \else FWHM\fi} 
\newcommand{ \voff  }{\ifmmode v_{\rm off} \else $v_{\rm off}$\fi} 

\newcommand{ \mumg  }{\ifmmode \mu\left(\mgii\right) \else $\mu\left(\mgii\right)$\fi}
\newcommand{ \fmg   }{\ifmmode f\left(\mgii\right) \else $f\left(\mgii\right)$\fi}
\newcommand{ \muciv }{\ifmmode \mu\left(\civ\right) \else $\mu\left(\civ\right)$\fi}
\newcommand{ \fciv  }{\ifmmode f\left(\civ\right) \else $f\left(\civ\right)$\fi}

\newcommand{  \auvo     }{\ifmmode \alpha_{\nu,{\rm UVO}} \else $\alpha_{\nu,{\rm UVO}}$\fi}
\newcommand{  \Ledd     }{\ifmmode L_{\rm Edd} \else $L_{\rm Edd}$\fi}
\newcommand{  \lamLlam  }{\ifmmode \lambda L_{\lambda} \else $\lambda L_{\lambda}$\fi}
\newcommand{  \lLl      }{\ifmmode \lambda L_{\lambda} \else $\lambda L_{\lambda}$\fi}
\newcommand{  \nuLnu    }{\ifmmode \nu L_{\nu} \else $\nu L_{\nu}$\fi}
\newcommand{  \nLn      }{\ifmmode \nu L_{\nu} \else $\nu L_{\nu}$\fi}
\newcommand{  \Luv      }{\ifmmode L_{1450} \else $L_{1450}$\fi}
\newcommand{  \Lop      }{\ifmmode L_{5100} \else $L_{5100}$\fi}
\newcommand{  \lLop     }{\ifmmode \log\left(\Lop/\ergs\right) \else $\log\left(\Lop/\ergs\right)$\fi}
\newcommand{  \Lthree   }{\ifmmode L_{3000} \else $L_{3000}$\fi}
\newcommand{  \lLthree  }{\ifmmode \log\left(\Lthree/\ergs\right) \else $\log\left(\Lthree/\ergs\right)$\fi}

\newcommand{\Fthree}{\ifmmode F_{3000} \else $F_{3000}$\fi}
\newcommand{\fuv}{\ifmmode f_{\lambda}\left(1450{\rm \AA}\right) \else $f_{\lambda}\left(1450 {\rm \AA}\right)$\fi}
\newcommand{\fthree}{\ifmmode f_{\lambda}\left(3000{\rm \AA}\right) \else $f_{\lambda}\left(3000{\rm \AA}\right)$\fi}
\newcommand{\fH}{\ifmmode f_{\lambda}\left(1.65\micron\right) \else
$f_{\lambda}\left(1.65\micron\right)$\fi}

\newcommand{\fbol}{\ifmmode f_{\rm bol} \else $f_{\rm bol}$\fi}
\newcommand{\fbolwv}{\ifmmode f_{\rm bol}\left(\lambda\right) \else $f_{\rm bol}\left(\lambda\right)$\fi}
\newcommand{\fbolopt}{\ifmmode f_{\rm bol}\left(5100{\rm \AA}\right) \else $f_{\rm bol}\left(5100{\rm \AA}\right)$\fi}
\newcommand{\fbolthree}{\ifmmode f_{\rm bol}\left(3000{\rm \AA}\right) \else $f_{\rm bol}\left(3000{\rm \AA}\right)$\fi}
\newcommand{\fboluv}{\ifmmode f_{\rm bol}\left(1450{\rm \AA}\right) \else $f_{\rm bol}\left(1450{\rm \AA}\right)$\fi}

\newcommand{  \mbh      }{\ifmmode M_{\rm BH} \else $M_{\rm BH}$\fi}
\newcommand{  \lmbh     }{\ifmmode \log\left(\mbh/\Msun\right) \else $\log\left(\mbh/\Msun\right)$\fi} 
\newcommand{  \lledd    }{\ifmmode L_{\rm bol}/L_{\rm Edd} \else $L_{\rm bol}/L_{\rm Edd}$\fi}
\newcommand{  \Lbol     }{\ifmmode L_{\rm bol} \else $L_{\rm bol}$\fi}
\newcommand{  \lbol     }{\ifmmode L_{\rm bol} \else $L_{\rm bol}$\fi}
\newcommand{  \lLbol    }{\ifmmode \log\left(\Lbol/\ergs\right) \else $\log\left(\Lbol/\ergs\right)$\fi} 
\newcommand{  \Lagn     }{\ifmmode L_{\rm AGN} \else $L_{\rm AGN}$\fi}
\newcommand{  \lagn     }{\ifmmode L_{\rm AGN} \else $L_{\rm AGN}$\fi}

\newcommand{  \tgrow     }{\ifmmode t_{\rm growth} \else $t_{\rm growth}$\fi}
\newcommand{  \tUni      }{\ifmmode t_{\rm Universe} \else $t_{\rm Universe}$\fi}

\newcommand{  \Mindot	}{\ifmmode \dot{M}_{\rm infall} \else $\dot{M}_{\rm infall}$\fi}
\newcommand{  \Mbhdot	}{\ifmmode \dot{M}_{\rm BH} \else $\dot{M}_{\rm BH}$\fi}
\newcommand{  \Maddot	}{\ifmmode \dot{M}_{\rm AD} \else $\dot{M}_{\rm AD}$\fi}

\newcommand{  \as	}{\ifmmode a_{\rm *} 		\else $a_{\rm *}$\fi}
\newcommand{  \avec	}{\ifmmode \vec{a}_{\rm *} 	\else $\vec{a}_{\rm *}$\fi}
\newcommand{  \re	}{\ifmmode \eta      	\else $\eta$\fi}

\newcommand{  \mseed    }{\ifmmode M_{\rm seed} \else $M_{\rm seed}$\fi}
\newcommand{  \mbul     }{\ifmmode M_{\rm Bulge} \else $M_{\rm Bulge}$\fi} 
\newcommand{  \mstar    }{\ifmmode M_{*} \else $M_{*}$\fi} 
\newcommand{  \mgal     }{\ifmmode M_{*} \else $M_{*}$\fi} 
\newcommand{  \mhost    }{\ifmmode M_{\rm Host} \else $M_{\rm Host}$\fi}
\newcommand{  \mm       }{\ifmmode M_{*}/M_{\rm BH} \else $M_{*}/M_{\rm BH}$\fi}
\newcommand{  \mmsmall  }{\ifmmode M_{\rm BH}/M_{*} \else $M_{\rm BH}/M_{*}$\fi}
\newcommand{  \mmlarge  }{\ifmmode M_{*}/M_{\rm BH} \else $M_{*}/M_{\rm BH}$\fi}
\newcommand{  \mmwp     }{\ifmmode \left(M_{*}/M_{\rm BH}\right) \else $\left(M_{*}/M_{\rm BH}\right)$\fi}
\newcommand{  \ml       }{\ifmmode M_{*}/L_{*} \else $M_{*}/L_{*}$\fi}
\newcommand{  \mlwp     }{\ifmmode \left(M_{*}/L\right) \else $\left(M_{*}/L\right)$\fi}
\newcommand{  \mlk      }{\ifmmode \left(M_{*}/L_{K}\right) \else $\left(M_{*}/L_{K}\right)$\fi}
\newcommand{  \sigs     }{\ifmmode \sigma_{*} \else $\sigma_{*}$\fi}
\newcommand{  \Reff     }{\ifmmode R_{\rm e} \else $R_{\rm e}$\fi}

\newcommand{\bj}{\ifmmode b_{\rm J} \else $b_{\rm J}$\fi}

\newcommand{\iab}{\ifmmode i_{\rm AB} \else $i_{\rm AB}$\fi}

\newcommand{\jab}{\ifmmode J_{\rm AB} \else $J_{\rm AB}$\fi}
\newcommand{\hab}{\ifmmode H_{\rm AB} \else $H_{\rm AB}$\fi}
\newcommand{\kab}{\ifmmode K_{\rm AB} \else $K_{\rm AB}$\fi}

\newcommand{\jveg}{\ifmmode J_{\rm Vega} \else $J_{\rm Vega}$\fi}
\newcommand{\hveg}{\ifmmode H_{\rm Vega} \else $H_{\rm Vega}$\fi}
\newcommand{\kveg}{\ifmmode K_{\rm Vega} \else $K_{\rm Vega}$\fi}

\def\arcsec{\hbox{$^{\prime\prime}$}}

\newcommand{  \Chisq    }{\ifmmode \chi^{2} \else $\chi^{2}$}
\newcommand{  \nelec    }{\ifmmode n_{e} \else $n_{e}$\fi}     
\newcommand{\nh}{\ensuremath{N_\mathrm{H}}}
\newcommand{  \Ncol     }{\ifmmode N_{col} \else $N_{col}$\fi} 
\newcommand{  \NH       }{\ifmmode N_{\rm H} \else $N_{\rm H}$\fi}     

\def\sun{\hbox{$\odot$}}

\def\arcsec{\hbox{$^{\prime\prime}$}}

\def\ion#1#2{#1$\;${\small\rm\@Roman{#2}}\relax}

\newcommand{\OIIIa}{\ifmmode \left[{\rm O}\,\textsc{iii}\right]\,\lambda4959 \else [O\,{\sc iii}]\,$\lambda4959$\fi}
\newcommand{\NIIa}{\ifmmode \left[{\rm N}\,\textsc{ii}\right]\,\lambda6548 \else [N\,{\sc ii}]\,$\lambda6548$\fi}
\newcommand{\SIIa}{\ifmmode \left[{\rm S}\,\textsc{ii}\right]\,\lambda6716 \else [S\,{\sc ii}]\,$\lambda6716$\fi}
\newcommand{\SIIb}{\ifmmode \left[{\rm S}\,\textsc{ii}\right]\,\lambda6732 \else [S\,{\sc ii}]\,$\lambda6731$\fi}
\newcommand{\NeVa}{\ifmmode \left[{\rm Ne}\,\textsc{v}\right]\,\lambda3346 \else [Ne\,{\sc v}]\,$\lambda3346$\fi}
\newcommand{\NeVb}{\ifmmode \left[{\rm Ne}\,\textsc{v}\right]\,\lambda3426 \else [Ne\,{\sc v}]\,$\lambda3426$\fi}
\newcommand{\NeIIIa}{\ifmmode \left[{\rm Ne}\,\textsc{iii}\right]\,\lambda3869 \else [Ne\,{\sc iii}]\,$\lambda3869$\fi}
\newcommand{\NeIIIb}{\ifmmode \left[{\rm Ne}\,\textsc{iii}\right]\,\lambda3968 \else [Ne\,{\sc iii}]\,$\lambda3968$\fi}

\newcommand{\mgb}{\ifmmode \left{\rm Mg}\,\textsc{i}\right \else Mg\,{\sc i}\fi}

\def\arcsec{{\mbox{$^{\prime \prime}$}}}

\def\erg{{\rm\thinspace erg}}

\def\g{{\rm\thinspace g}}

\def\K{{\rm\thinspace K}}

\def\Lsun{\hbox{$\rm\thinspace L_{\odot}$}}
\def\m{{\rm\thinspace m}}

\def\pc{{\rm\thinspace pc}}

\def\s{{\rm\thinspace s}}

\newcommand{\HeIIir}{\ifmmode {\rm He}\,\textsc{ii}\,\lambda8237 \else He\,{\sc ii}$\,\lambda8237$\fi}
\newcommand{\HeIir}{\ifmmode {\rm He}\,\textsc{i}\,\lambda10830 \else He\,{\sc i}$\,\lambda10830$\fi}

\newcommand{\SIII}{\ifmmode \left[{\rm S}\,\textsc{iii}\right]\,\lambda9531 \else [S\,\textsc{ii}]\,$\lambda9531$\fi}

\newcommand {\Lsoftint} {\ifmmode L^{\rm int}_{\mathrm{2-10\ keV}} \else $L^{\rm int}_{\mathrm{2-10\ keV}}$\fi}

\newcommand {\ergpersec} {\ifmmode {\rm erg~s}^{-1} \else erg~s$^{-1}$ \fi}

\newcommand {\nhunit} {cm$^{-2}$\xspace}

\def\micron{{\mbox{$\mu{\rm m}$}}}
\def\arcsec{{\mbox{$^{\prime \prime}$}}}

\def\arcsec{{\mbox{$^{\prime \prime}$}}}

\def\erg{{\rm\thinspace erg}}

\def\g{{\rm\thinspace g}}

\def\K{{\rm\thinspace K}}

\def\Lsun{\hbox{$\rm\thinspace L_{\odot}$}}
\def\m{{\rm\thinspace m}}

\def\pc{{\rm\thinspace pc}}

\def\s{{\rm\thinspace s}}

\def\ergps{\hbox{$\erg\s^{-1}\,$}}

\def\apjl{\hbox{ApJL}}
\def\apjs{\hbox{ApJS}}

\def\aap{\hbox{AAP}}
\def\micron{{\mbox{$\mu{\rm m}$}}}
\def\arcsec{{\mbox{$^{\prime \prime}$}}}

\newcommand{\nuvr}{\ifmmode {\rm NUV}-r \else NUV-$r$\fi}
\newcommand{\mh}{\ifmmode M_{\rm H_2} \else $M_{\rm H_2}$\fi}
\newcommand{\mhi}{\ifmmode M_{\rm HI} \else $M_{\rm HI}$\fi}

\newcommand{\must}{\ifmmode \mu_{\ast} \else $\mu_{\ast}$\fi}
\newcommand{\hmol}{\ifmmode H_2 \else $H_2$\fi}
\newcommand{\rmol}{\ifmmode R_{\rm mol} \else $R_{\rm mol}$\fi}
\newcommand{\tdep}{\ifmmode t_{\rm dep}({\rm H_2}) \else $t_{\rm dep}({\rm H_2})$\fi}
\newcommand{\tdepHI}{\ifmmode t_{\rm dep}({\rm HI}) \else $t_{\rm dep}({\rm HI})$\fi}
\newcommand{\fgas}{\ifmmode f_{\rm H_2} \else $f_{\rm H_2}$\fi}
\newcommand{\fhi}{\ifmmode f_{\rm HI} \else $f_{\rm HI}$\fi}
\newcommand{\xco}{\ifmmode \alpha_{\rm CO} \else $\alpha_{\rm CO}$\fi}

\newcommand{\SiX}{\ifmmode \left[{\rm Si}\,\textsc{x}\right]\,\lambda14300 \else [Si\,{\sc x}]\,$\lambda14300$\fi}
\newcommand{\SiVI}{\ifmmode \left[{\rm Si}\,\textsc{vi}\right]\,\lambda19640 \else [Si\,{\sc vi}]\,$\lambda19640$\fi}
\newcommand{\SXI}{\ifmmode \left[{\rm S}\,\textsc{xi}\right]\,\lambda19196 \else [S\,{\sc xi}]\,$\lambda19196$\fi}
\newcommand{\SVIII}{\ifmmode \left[{\rm S}\,\textsc{viii}\right]\,\lambda9915 \else [S\,{\sc viii}]\,$\lambda9915$\fi}
\newcommand{\SIX}{\ifmmode \left[{\rm S}\,\textsc{ix}\right]\,\lambda12520 \else [S\,{\sc ix}]\,$\lambda12520$\fi}
\newcommand{\FeXIII}{\ifmmode \left[{\rm Fe}\,\textsc{xiii}\right]\,\lambda10747 \else [Fe\,{\sc xiii}]\,$\lambda10747$\fi}
\newcommand{\SiXI}{\ifmmode \left[{\rm Si}\,\textsc{xi}\right]\,\lambda19320 \else [Si\,{\sc xi}]\,$\lambda19320$\fi}

\def\arcsec{{\mbox{$^{\prime \prime}$}}}

\def\erg{{\rm\thinspace erg}}

\def\g{{\rm\thinspace g}}

\def\K{{\rm\thinspace K}}

\def\Lsun{\hbox{$\rm\thinspace L_{\odot}$}}
\def\m{{\rm\thinspace m}}

\def\pc{{\rm\thinspace pc}}

\def\s{{\rm\thinspace s}}

\def\ergps{\hbox{$\erg\s^{-1}\,$}}

\def\apjl{\hbox{ApJL}}
\def\apjs{\hbox{ApJS}}

\def\aap{\hbox{AAP}}
\def\micron{{\mbox{$\mu{\rm m}$}}}
\def\arcsec{{\mbox{$^{\prime \prime}$}}}

\shorttitle{Quantifying X-ray Selection Effects in Deep Surveys with BASS}
\shortauthors{Tokayer et al.}

\graphicspath{{./}{figures/}}

\usepackage{natbib,xspace}
\begin{document}


\title{BASS XLV: Quantifying AGN Selection Effects in the Chandra COSMOS-Legacy Survey with BASS}

\correspondingauthor{Yarone M. Tokayer}
\email{yarone.tokayer@yale.edu}

\author[0000-0002-0430-5798]{Yarone M. Tokayer}
\affiliation{Department of Physics, Yale University, P.O. Box 208121, New Haven, CT 06520, USA}

\author[0000-0002-7998-9581]{Michael J. Koss}
\affil{Eureka Scientific, 2452 Delmer Street Suite 100, Oakland, CA 94602-3017, USA}

\author[0000-0002-0745-9792]{C. Megan Urry}
\affiliation{Yale Center for Astronomy \& Astrophysics and Department of Physics, Yale University, P.O. Box 208120, New Haven, CT 06520-8120, USA}

\author[0000-0002-0745-9792]{Priyamvada Natarajan}
\affiliation{Yale Center for Astronomy \& Astrophysics and Department of Physics, Yale University, P.O. Box 208120, New Haven, CT 06520-8120, USA}

\author[0000-0002-7962-5446]{Richard Mushotzky}
\affiliation{Department of Astronomy, University of Maryland, College Park, MD 20742, USA}
\affiliation{Joint Space-Science Institute, University of Maryland, College Park, MD 20742, USA}

\author[0000-0003-0476-6647]{Mislav Balokovi\'{c}}
\affiliation{Yale Center for Astronomy \& Astrophysics and Department of Physics, Yale University, P.O. Box 208120, New Haven, CT 06520-8120, USA}

\author[0000-0002-8686-8737]{Franz E. Bauer}
\affiliation{Instituto de Astrof\'{\i}sica  and Centro de Astroingenier{\'{\i}}a, Facultad de F\'{i}sica, Pontificia Universidad Cat\'{o}lica de Chile, Casilla 306, Santiago 22, Chile}
\affiliation{Millennium Institute of Astrophysics (MAS), Nuncio Monse$\tilde{n}$or S{\'{o}}tero Sanz 100, Providencia, Santiago, Chile}
\affiliation{Space Science Institute, 4750 Walnut Street, Suite 205, Boulder, Colorado 80301, USA}

\author[0000-0001-9379-4716]{Peter Boorman}
\affiliation{Cahill Center for Astronomy and Astrophysics, California Institute of Technology, Pasadena, CA 91125, USA}

\author[0000-0003-2196-3298]{Alessandro Peca}
\affil{Department of Physics, University of Miami, Coral Gables, FL 33124, USA}
\affil{Eureka Scientific, 2452 Delmer Street Suite 100, Oakland, CA 94602-3017, USA}
\affiliation{Department of Physics, Yale University, P.O. Box 208121, New Haven, CT 06520, USA}

\author[0000-0001-5231-2645]{Claudio Ricci}
\affil{Instituto de Estudios Astrof\'isicos, Facultad de Ingenier\'ia y Ciencias, Universidad Diego Portales, Av. Ej\'ercito Libertador 441, Santiago, Chile}
\affil{Kavli Institute for Astronomy and Astrophysics, Peking University, Beijing 100871, People's Republic of China}

\author[0000-0001-5742-5980]{Federica Ricci}
\affiliation{Dipartimento di Fisica e Astronomia, Università di Bologna, via Gobetti 93/2, 40129 Bologna, Italy}

\author[0000-0003-2686-9241]{Daniel Stern}
\affil{Jet Propulsion Laboratory, California Institute of Technology, 4800 Oak Grove Drive, MS 169-224, Pasadena, CA 91109, USA}

\author[0000-0001-7568-6412]{Ezequiel Treister}
\affiliation{Instituto de Astrof\'isica, Facultad de F\'isica, Pontificia Universidad Cat\'olica de Chile, Casilla 306, Santiago 22, Chile}

\author[0000-0002-3683-7297]{Benny Trakhtenbrot}
\affiliation{School of Physics and Astronomy, Tel Aviv University, Tel Aviv 69978, Israel}

\begin{abstract}

    Deep extragalactic X-ray surveys, such as the Chandra COSMOS-Legacy field (CCLS), are prone to be  biased against  active galactic nuclei (AGN) with high column densities due to their lower count rates at a given luminosity.
    To quantify this selection effect, we forward model nearby ($z$$\sim$0.05) AGN from the BAT AGN Spectroscopic Survey (BASS) with well-characterized ($\gtrsim$1000 cts) broadband X-ray spectra (0.5--195 keV) to simulate the CCLS absorption distribution. 
    We utilize the BASS low-redshift analogs with similar luminosities to the CCLS (\Lsoftint$\sim10^{42-45}$\,\ergps), which  are much less affected by obscuration and low-count statistics, as the seed for our simulations, and follow the spectral fitting of the CCLS. 
    Our simulations reveal that Chandra would fail to detect the majority (53.3\%; 563/1056) of obscured ($\nh\geq10^{22}$\,\nhunit) simulated BASS AGN given the observed redshift and luminosity distribution of the CCLS.
    Even for detected sources with sufficient counts ($\geq30$) for spectral modeling,  the level of obscuration is significantly overestimated. 
    This bias is most extreme for objects whose best fit indicates a high-column density AGN ($\nh\geq10^{24}\ \mathrm{cm}^{-2}$), since the majority (66.7\%; 18/27) of  these are actually unobscured sources (\nh$<10^{22}$\,\nhunit).
    This implies that previous studies may have significantly overestimated the increase in the obscured fraction with redshift and the fraction of luminous obscured AGN.
    Our findings highlight the importance of directly considering obscuration biases and forward modeling in X-ray surveys, as well as the need for higher-sensitivity X-ray missions such as the Advanced X-ray Imaging Satellite (AXIS), and the importance of multi-wavelength indicators to estimate obscuration in distant supermassive black holes.
    
\end{abstract}


\section{Introduction} \label{sec:intro}

Active galactic nuclei (AGN)---luminous galactic centers powered by accreting supermassive black holes (SMBHs)---are powerful probes of the nature of black hole growth, galactic evolution, and cosmology.
During the course of their lifetimes, the Chandra X-ray Observatory and XMM-Newton have observed over 50 extra-galactic survey fields at depths of up to $\sim$7 Ms that cumulatively add up to years of observing time \citep{brandtDeepExtragalacticXRay2005, brandtCosmicXraySurveys2015}.
This has led to the discovery of thousands of distant black holes as AGN, and enabled resolving most of the cosmic X-ray background below 10 keV \citep{gilliSynthesisCosmicXray2007, treisterSpaceDensityComptonThick2009, brandtSurveysCosmicXRay2022}.
However, surveys with ``soft'' X-ray telescopes ($E<10$ keV) remain highly biased against heavily obscured AGN \citep[e.g.,][]{burlonThreeyearSwiftBATSurvey2011}, whose spectra peak at 20--30 keV \citep{brightmanXMMNewtonSpectralSurvey2011, ricciCOMPTONTHICKACCRETIONLOCAL2015, kossNEWPOPULATIONCOMPTONTHICK2016}.

AGN are believed to contain a torus of obscuring gas and dust, which can lead to significant attenuation of much of their electromagnetic spectra \citep{hickoxObscuredActiveGalactic2018}, depending on the viewing angle \citep[e.g.,][]{elvisStructureQuasars2000, urryAGNUnificationUpdate2004}, intrinsic luminosity \citep{durasUniversalBolometricCorrections2020, caglarLLAMAMBHsRelation2020}, Eddington ratio \citep{ricciCloseEnvironmentsAccreting2017,ricciBASSXXXVIIRole2022, ricciBASSXLIIRelation2023, caglarBASSXXXVBHs2023}, the nature of the host galaxy \citep{kossHOSTGALAXYPROPERTIES2011, kossBATAGNSpectroscopic2021}, and possibly the cosmological epoch \citep{francaHELLAS2XMMSurveyVII2005, akylasXMMNewtonChandraMeasurements2006, hasingerAbsorptionPropertiesEvolution2008, pecaCosmicEvolutionAGN2023, mattheeLittleRedDots2024}.
Detecting and characterizing the AGN population is critical to obtaining a fuller census of SMBHs and active galaxies across cosmic time.
They are also of particular interest since these optically-thick environments can be the result of galaxy mergers (e.g., \citealp{kossMergingClusteringSwift2010, kossUnderstandingDualActive2012a, ricciGrowingSupermassiveBlack2017, kossPopulationLuminousAccreting2018a, ricciHardXrayView2021}), which are believed to play a critical role in triggering AGN activity and serve as an important growth channel for both SMBHs and the galaxies in which they reside \citep[e.g.,][]{hopkinsUnifiedMergerdrivenModel2006}.

AGN obscuration is quantified by the column density of obscuring material along the line of sight, \nh, which includes obscuration near the SMBH in the torus and farther out in the host galaxy in which it resides.
We refer to sources with $\nh\geq 10^{22}\ \mathrm{cm}^{-2}$ as ``obscured'' and those with $\nh\geq 10^{24}\ \mathrm{cm}^{-2}$ as ``Compton-thick'' (CT).
In many well-studied objects, the presence of this obscuring material significantly reduces the AGN's soft X-ray, optical, and UV signatures.
There are two spectral bands, the ultra-hard X-ray ($E>10$ keV) and mid-IR (5--50 $\mu$m), that are less sensitive to the effects of obscuration and thus can survey a wide range of AGN where this obscuring material is optically thin up to the CT regime \citep[e.g.,][]{kossSTUDYINGFAINTULTRAHARD2013, ricciCOMPTONTHICKACCRETIONLOCAL2015}.

Recently, results from ALMA have suggested that nuclear millimeter emission may also be used effectively to detect highly obscured nearby AGN \citep{kawamuroBASSXXXIIStudying2022,ricciTightCorrelationMillimeter2023}.
While it is also insensitive to obscuration, strong lower frequency radio emission is found only in a fraction of AGN \citep{wilsonDifferenceRadioloudRadioquiet1995}.
Weaker radio emission in the GHz band was found in the majority of BASS AGN at 1--2\arcsec\ scales \citep{smithBATAGNSpectroscopic2020}, but has a much larger scatter with the X-ray emission ($\sim$1 dex) than in the millimeter band. 

While mid-IR selection is reliable in identifying very luminous AGN where nuclear emission is dominant, moderate luminosity AGN, such as those commonly found in the local Universe, are not easily identified because of host galaxy contamination \citep[e.g.,][]{ichikawaCOMPLETEINFRAREDVIEW2017}.
Importantly, host contamination becomes more problematic for $z\gtrsim0.5$, since the typical star forming galaxies detected at those cosmic epochs appear as luminous infrared galaxies (``LIRGs'') \citep[e.g.,][]{madauCosmicStarFormationHistory2014}.
In contrast, X-ray surveys suffer little contamination from non-nuclear emission at typical survey depths and can therefore successfully find  both low- and high-luminosity AGN \citep{kossBASSXXIData2022}.
Ultra-hard X-ray surveys performed by the \bat\ instrument \citep[14--195 keV;][]{barthelmyBurstAlertTelescope2005} and more recently by NuSTAR \citep{harrisonNUCLEARSPECTROSCOPICESCOPE2013}, have played a critical role in identifying low-luminosity, heavily obscured AGN  in the local universe. \citep[e.g.,][]{kawamuroSTUDYSWIFTBAT2016,lansburyNuSTARSerendipitousSurvey2017,annuarNuSTARObservationsFour2020}.

Multi-wavelength extragalactic surveys generally follow a ``wedding cake'' strategy \citep[e.g.,][]{khedekarPrecisionCosmologyCombination2010, groginCANDELSCOSMICASSEMBLY2011}, consisting of observations that are either shallow and wide, or deep and narrow.
With finite telescope resources, the choice is to either focus on a small area with a long exposure time or cover a wider field at a shallower depth.
Combining the two is maximally efficient: deep fields probe higher redshifts, whereas shallower all-sky surveys can probe larger volumes and are thus sensitive to the rarer, high-luminosity sources.
For example, rare high-luminosity AGN are typically found in wide-area surveys, such as Stripe 82X, the X-ray augmentation of the SDSS Stripe 82 region with Chandra and XMM-Newton \citep{lamassaFindingRareAGN2013, lamassaFindingRareAGN2013a,pecaCosmicEvolutionAGN2023, lamassaStripe82XData2024}.
The eROSITA Final Equatorial Depth Survey \citep{brunnerEROSITAFinalEquatorial2022, liuEROSITAFinalEquatorialDepth2022} is a very wide ($\sim140$ square degrees), but relatively shallow ($\sim1.2$ ks after accounting for vignetting effects), survey that has detected over 22,000 AGN at $z\lesssim3$ and relatively high luminosities \citep{nandraEROSITAFinalEquatorial2024}.
Meanwhile, some of the most sensitive pencil-beam observations are those performed with Chandra: the Chandra COSMOS-Legacy Survey \citep[CCLS; ][]{civanoCHANDRACOSMOSLEGACY2016}, and the Chandra Deep Fields North \citep[CDF-N;][]{hornschemeierXRaySourcesHubble2000} and South \citep[CDF-S;][]{giacconiFirstResultsXRay2001}.
However, in all these surveys, the majority of the detected objects are close to the detection limit and have relatively few counts.

One of the advantages of observing high-redshift AGN is that the K-correction shifts the X-ray spectrum to probe higher energies, and hence observations by, e.g., Chandra are somewhat less affected by obscuration \citep[e.g.,][]{brandtDeepExtragalacticXRay2005, hickoxObscuredActiveGalactic2018}.
However, simulations \citep{kossBroadbandObservationsComptonthick2015} have shown that a nearby CT AGN, NGC 3393, would have accurate measurements of its intrinsic luminosity only for $z<0.2$, while an unobscured AGN with similar luminosity could be measured up to $z\approx5$ with Chandra.
As we show below, traditional obscuration indicators are unreliable for low-count X-ray spectra due to the degeneracies with sources at lower column densities \citep{kossNEWPOPULATIONCOMPTONTHICK2016}. 
Furthermore, to match the intrinsic energy range of telescopes such as \bat\ (14--195 keV) or NuSTAR (3--79 keV), sources would need to be at $z\geq5$, while the CCLS and CDF-S, which represent the deepest X-ray AGN surveys to date over their respective areas, have median redshifts of  only $z\sim1.28$ \citep[see online catalog from ][]{marchesiChandraCOSMOSLegacy2016} and $z\sim1.03$ \citep[see Table 1 of ][]{tozziXraySpectralProperties2006}, respectively.
See Appendix~\ref{app:eff_area} for a discussion the effective areas of \bat\ and other missions as a function of redshift. 

Thus, two critical questions arise regarding deep AGN surveys below 10 keV, which we aim to address in this paper: (1) What fraction of distant AGN does Chandra miss due to obscuration? (2) How effectively do we recover the intrinsic spectral properties of those AGN (obscured and unobscured) that are identified?
To explore these questions, we used the \bat\ AGN Spectroscopic Survey \citep[BASS;][]{kossBATAGNSpectroscopic2017}.

The \bat\ instrument has surveyed the entire sky in ultra-hard X-rays \citep[14--195 keV;][]{barthelmyBurstAlertTelescope2005} detecting mainly nearby AGN with X-ray luminosities between $10^{41}$ and $10^{45}$ erg s$^{-1}$ \citep{baumgartner70MONTHSWIFT2013, oh105MonthSwiftBATAllsky2018}.
BASS is a follow-up survey that has measured the emission lines \citep{kossBASSXXVIDR22022,ohBASSXXIVBASS2022,mejia-restrepoBASSXXVDR22022}, X-ray properties \citep{ricciCOMPTONTHICKACCRETIONLOCAL2015,ricciBATAGNSpectroscopic2017} as well as the IR and radio properties \citep{smithBATAGNSpectroscopic2020, pfeifleBASSXXIIINew2022, denbrokBASSXXVIIINearinfrared2022, kawamuroBASSXXXIIStudying2022, bierschenkBASSXLICorrelation2024} of these AGN, which have luminosities similar to those in the Chandra and XMM deep surveys \citep{kossBASSXXIData2022}.
\cite{kossNEWPOPULATIONCOMPTONTHICK2016} demonstrate that BASS is the least biased X-ray survey in terms of obscuration, with simulations showing that the \bat\ count rate is not affected by obscuring column densities up to 10$^{23}$ cm$^{-2}$.
Therefore, BASS AGN represent ideal templates which can be matched in luminosity to simulate AGN at higher redshifts.
These well-understood simulated spectra can then be used to quantify the selection effects in deep surveys \citep{kossBASSXXIData2022}.

This paper focuses on the CCLS field, which has the largest contiguous survey area among deep-field programs ($>$100 ks effective exposure) with X-ray coverage.
The COSMOS field \citep[Cosmic Evolution Survey;][]{scovilleCosmicEvolutionSurvey2007} covers a 2 square-degree equatorial patch of the sky and consists of broadband observations from radio to X-rays.
The Chandra-COSMOS survey \citep{elvisCHANDRACOSMOSSURVEY2009,civanoCHANDRACOSMOSLEGACY2016} covers 1.5 square degrees at an effective exposure of $\sim160$ ks.
The BASS AGN luminosity function is particularly well-matched to that of CCLS.
The deep multi-wavelength and X-ray coverage of the COSMOS field is currently being augmented with JWST observations through the COSMOS-Web program \citep{caseyCOSMOSWebOverviewJWST2023}, making it of particular interest to understand how well the most sensitive IR and X-ray instruments can detect distant obscured AGN.

\cite{marchesiChandraCOSMOSLegacySurvey2016} (henceforth, M16) present the X-ray spectral analysis for 1855 CCLS sources with at least 30 counts, where they fit phenomenological models of increasing complexity, choosing those that optimize the fit statistic.
\cite{lanzuisiChandraCOSMOSLegacy2018} further model 67 of these sources that were determined to have at least a 5\% probability of being CT with the \texttt{MYTorus} model of \cite{murphyXraySpectralModel2009} combined with Markov chain Monte Carlo parameter estimation techniques.
By summing the posterior distribution of the line of sight column density of individual sources that were above 10$^{24}$\,cm$^{-2}$, and accounting for identification and classification bias, the authors find 41.9 ``effective'' CT AGN in the sample.

This paper focuses on quantifying selection effects due to obscuration in CCLS.
To fully address the question of how well we can recover intrinsic properties from Chandra observations, every study from every Chandra survey field would ideally be addressed.
Indeed, the techniques used can be extended to other deep X-ray surveys (e.g., XMM-XXL, \citealt{menzelSpectroscopicSurveyXrayselected2016}; the NuSTAR extragalactic survey, \citealt{alexanderNuSTAREXTRAGALACTICSURVEY2013}; CDF-S), as well as current and future missions (e.g., eROSITA, \citealt{predehlEROSITAXrayTelescope2021}; NewAthena, \citealt{nandraHotEnergeticUniverse2013}, \citealt{cruiseNewAthenaMissionConcept2024}; AXIS, \citealt{reynoldsOverviewAdvancedXray2023};  HEX-P, \citealt{madsenHighEnergyXray2024}).
We begin to address these questions in this paper by focusing on the CCLS, which is a reasonable starting point given its intermediate depth relative to other surveys.

Section~\ref{sec:sample_sim} describes the template spectra and the simulated data set, and Section~\ref{sec:data_anal_results} presents our data analysis and results.
We discuss the results in Section~\ref{sec:disc} and conclude in Section~\ref{sec:conc}.
Throughout the paper, we adopt a cosmology of $H_0=71\ \mathrm{km}\ \mathrm{s}^{-1}\ \mathrm{Mpc}^{-1}$, $\Omega_\mathrm{m}=0.3$, and $\Omega_\Lambda=0.7$, the same as that used in the CCLS X-ray study of M16.
To make notation less cumbersome, we use $\log\nh$ as shorthand for $\log\left(\nh/\mathrm{cm}^{-2}\right)$, the log of the obscuring column density along the line of sight in units of $\mathrm{cm}^{-2}$, and $\log L_{\text{2-10,int}}$ as shorthand for $\log\left(L_{\text{2-10,int}}/\mathrm{erg}\ \mathrm{s}^{-1}\right)$, the log of the intrinsic 2--10 keV luminosity in units of $\mathrm{erg}\ \mathrm{s}^{-1}$.


\section{Sample and simulation overview} \label{sec:sample_sim}

\subsection{\bat\ instrument and BASS}
\label{subsec:bat_survey}

The BAT 70-month AGN sample reaches depths of at least $1.34\times10^{-11}\ \mathrm{erg}\ \mathrm{s}^{-1}\ \mathrm{cm}^{-2}$ over 90\% of the sky \citep{baumgartner70MONTHSWIFT2013} and comprises \NAGNbass\ nearby ($z\lesssim0.3$ for unbeamed non-blazar) AGN \citep{kossBASSXXIIBASS2022}.
\citet[][henceforth, R17]{ricciBATAGNSpectroscopic2017} present and characterize their broadband X-ray spectra using complementary soft X-ray observations by XMM-Newton, Swift/XRT, ASCA, Chandra, and Suzaku.
Some of the properties they constrain are the intrinsic X-ray luminosity (in the 2--10, 20--50, and 14--150 keV bands), the intrinsic \nh, the slope of the X-ray power law continuum ($\Gamma$), and the temperature of the thermal plasma for obscured sources.
The phenomenological models used are broadly classified into four groups: unobscured (352), obscured (386), blazars (97), and other non-AGN models (2).
The 75 sources that were found to be CT using a phenomenological model were refit in \texttt{XSPEC} with the \texttt{BNtorus} model \citep{brightmanXMMNewtonSpectralSurvey2011} to better constrain their \nh\ values.

\subsection{Source selection and simulations}
\label{subsec:source_selection}
\begin{figure}
    \includegraphics[width=0.48\textwidth]{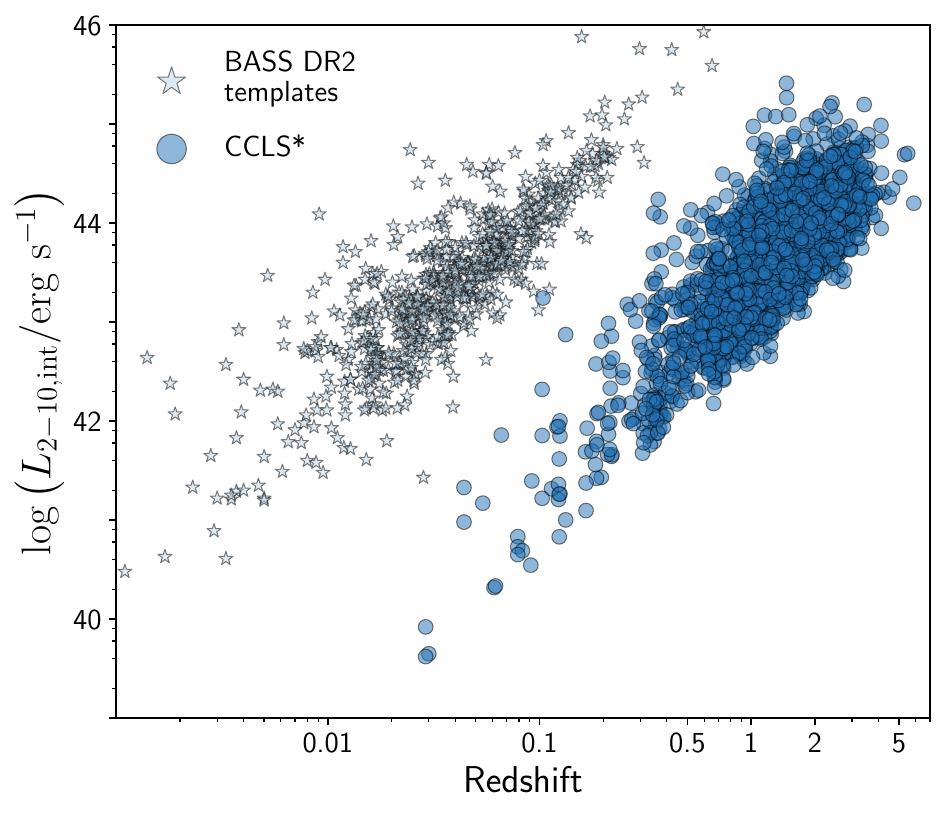}
    \caption{\textbf{Intrinsic 2-10 keV X-ray luminosity versus redshift for BASS and CCLS sources.} Distribution of the rest-frame hard X-ray luminosity versus redshift of non-blazar BASS AGN (stars) and CCLS* (dots).  Note that \bat-selected AGN cover a range of X-ray luminosities similar to that of CCLS*.
    \label{fig:survey_compare}}
\end{figure}
\begin{figure*}
    \gridline{
        \fig{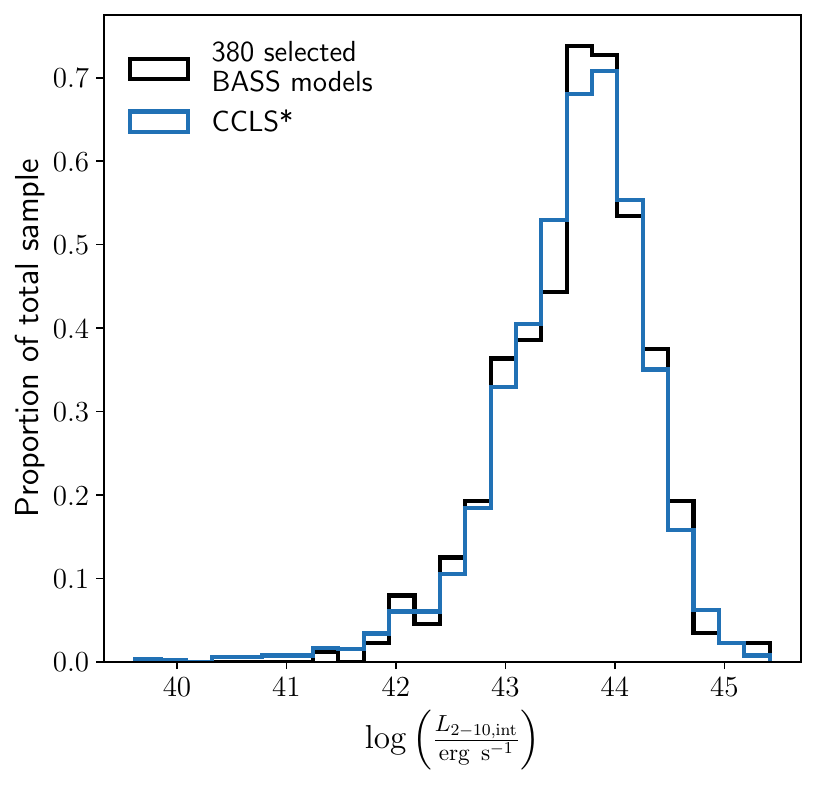}{0.32\textwidth}{(a)}
        \fig{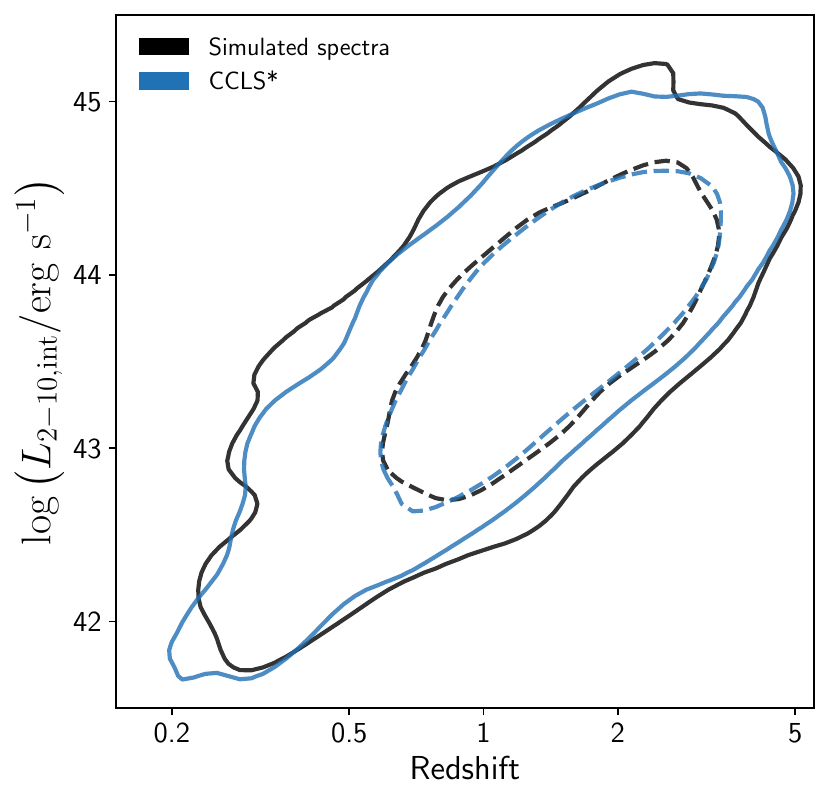}{0.32\textwidth}{(b)}
        \fig{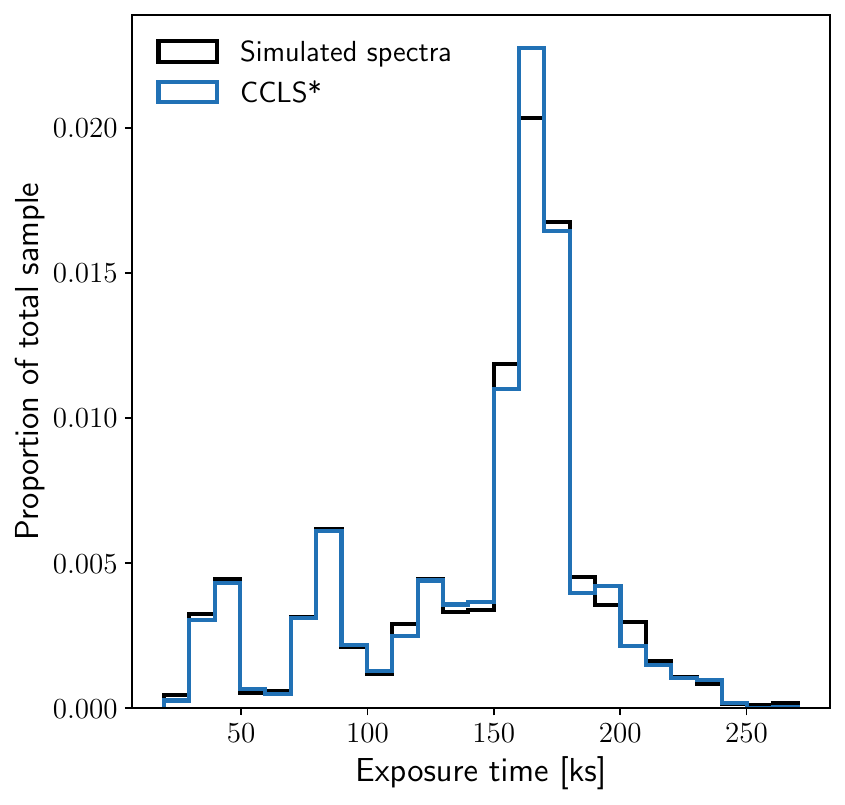}{0.32\textwidth}{(c)}
    }
    \caption{\textbf{(a) Luminosity matching.} 
    The intrinsic X-ray luminosity distributions of the CCLS* catalog (blue) and the selected template models (black). \textbf{(b) Redshift matching.} Comparison of the CCLS* catalog (blue) and the simulated data set (black) in luminosity-redshift space.  95\% of the sources lie within the solid curves and 68\% of the sources lie within the dashed curves. \textbf{(c) Exposure time matching.} Distribution of the exposure times for the 2280 simulations (black) compared with that of the 2291 spectra in CCLS* (blue).\label{fig:lum_z_dist}}
\end{figure*}

\cite{kossBATAGNSpectroscopic2017} and \cite{kossBASSXXIData2022} note that the unbeamed BAT AGN span the moderate-to-high end of the X-ray luminosity function, rendering them appropriate analogs to the AGN detected at high redshifts in deep surveys.
Indeed, the median intrinsic X-ray luminosity of the non-blazar BASS sample is $L_{\mathrm{2-10,int}} = 10^{43.41}\ \mathrm{erg}\ \mathrm{s}^{-1}$, compared to $10^{43.69}\ \mathrm{erg}\ \mathrm{s}^{-1}$ for the 2291 CCLS sources with intrinsic 2--10 keV luminosities well-constrained  from the redshifts of the optical and IR counterparts \citep{marchesiChandraCOSMOSLegacy2016}.
For this study, we focus on this subset of the CCLS sample (CCLS*) to characterize its observational biases regarding obscuration.
Fig.~\ref{fig:survey_compare} shows how the redshift-luminosity distribution of BASS non-blazar AGN compares with that of the CCLS*.
Note that the luminosity distributions are very similar despite the BASS AGN being in the local universe because the BASS survey covers the full sky.
However, since it is a flux-limited survey, there are few BASS AGN with X-ray luminosities below $10^{42}\ \mathrm{erg}\ \mathrm{s}^{-1}$.

To generate our data set, we first took models from R17 for sources that were either unobscured or obscured, excluding beamed AGN and non-AGN sources. This resulted in 738 AGN models.
We further selected models for sources with spectroscopic redshift measurements from optical counterparts and excluded five models of anomalous sources (i.e., sources with significant pileup components, contamination from the Perseus cluster, and/or with inconsistencies in the X-ray model).
This resulted in 698 AGN X-ray models from the BASS catalog, which we use as templates to simulate the CCLS* field.
These models were constructed using high-quality broadband spectra from nearby ultra-hard X-ray-selected AGN sampled from the entire sky; the median number of soft X-ray counts is 1525.5.
See Appendix~\ref{app:bass_models} for a brief overview of the R17 models.

To select a subset of BASS AGN models with a luminosity and redshift distribution well-matched to CCLS*, we adopted the intrinsic luminosity distribution of AGN in the CCLS* from \cite{marchesiChandraCOSMOSLegacy2016}, where they convert from the observed luminosities assuming an X-ray spectral index of 1.8.
Note that we did not attempt to match the CCLS* \nh\ distribution.
See Appendix~\ref{app:nh_comparison} for further discussion.
We then modeled the CCLS* intrinsic X-ray luminosity function as a probability distribution function (pdf) using kernel density estimation.
We drew random values from the estimated luminosity pdf, and selected template models with the closest $L_{\text{2-10,int}}$ (in log space), without allowing template models to be selected twice.
We found that selecting 380 template models allowed us to maximize the number of different AGN models used while reasonably matching the CCLS* luminosity distribution.
See Fig.~\ref{fig:lum_z_dist}(a).

In order to have a similar number of sources in our simulated dataset as in CCLS*, each of the 380 selected template models was used to simulate AGN spectra at six redshifts, resulting in a data set of 2280 simulated spectra.
The redshift values for each simulation were chosen based on $L_{\text{2-10,int}}$ as follows: A pdf of redshift was calculated for each of twelve $\log L_{\text{2-10,int}}$ bins using kernel density estimation from the CCLS* sources.
Then, for each BASS model, six random redshift values were drawn from the estimated pdf of its luminosity bin.
The resulting luminosity-redshift distribution is shown together with that of CCLS* in Fig.~\ref{fig:lum_z_dist}(b).

Finally, we assigned an exposure time to each simulation by drawing random samples from a pdf defined by the exposure time distribution of CCLS*.
The median exposure time of the simulated data set is 161.4 ks, while that of the CCLS* sample is 162.2 ks.
See Fig.~\ref{fig:lum_z_dist}(c).
The full list of selected BASS models and their assigned redshifts and exposure times is presented in Table~\ref{tab:templates_table}.

To simulate Chandra spectra in the CCLS field, each selected BASS model was imported into \texttt{XSPEC} \citep[v12.11;][]{arnaudXSPECFirstTen1996} and modified as follows:\begin{enumerate}
    \item The column density parameter of any component representing Galactic absorption (\texttt{phabs.nH} or \texttt{TBabs.nH}, depending on the R17 model) was changed to $N_\mathrm{H,gal} = 2.5\times10^{20}\ \mathrm{cm}^{-2}$ to place the AGN behind the same Galactic line-of-sight as the CCLS field \citep{kalberlaLeidenArgentineBonn2005}.
    \item The redshift parameter (\texttt{z} or \texttt{Redshift}) of each relevant model component was changed to the assigned simulated redshift value. 
    \item The \texttt{norm} parameter of each relevant model component was rescaled so that the flux changed according to the new luminosity distance, while the intrinsic properties of the simulated source remained unchanged. To do this, we calculated $L'_\mathrm{2-10,int}$, the intrinsic 2--10 keV luminosity using the new redshift value with the original \texttt{norm} values, and then multiplied the \texttt{norm} parameters by $L_\mathrm{2-10,int}/L'_\mathrm{2-10,int}$.
\end{enumerate}

The \texttt{fakeit} command was then used to simulate the spectra through the Chandra response, introducing only Poisson noise in the procedure.
We used Chandra-ACIS on-axis response files from cycle 14 (near when most COSMOS data were collected), and exposure times were assigned as described above.
Note that the effective exposure times account for off-axis positions of the Chandra observations \citep[see][]{civanoCHANDRACOSMOSLEGACY2016}.
See Appendix~\ref{app:background} for a discussion of potential selection effects generated by the \texttt{fakeit} procedure.

\subsection{Illustrative examples} \label{subsec:example_sims}
\begin{figure}
    \centering
    \includegraphics[width=0.48\textwidth]{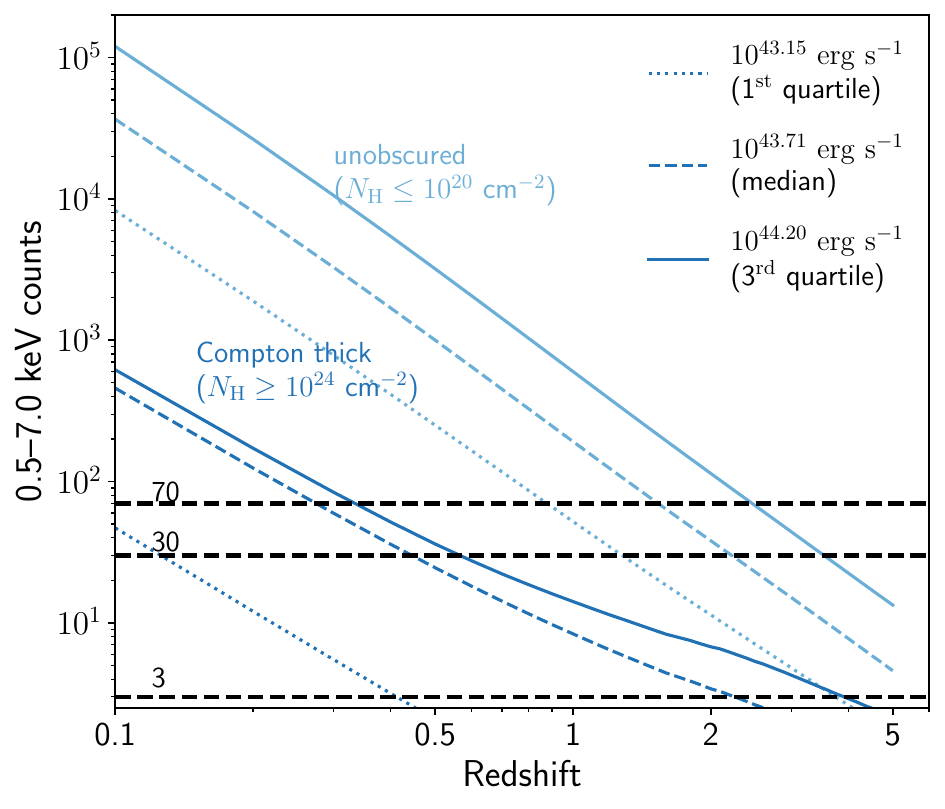}
    \caption{\textbf{Demonstration of the importance of obscuration on the detection and count fraction as a function of redshift using nearby BAT AGN.}  We compare three pairs of BASS models, each with an unobscured model (light blue) and a CT model (dark blue).  The three pairs are at $\log L_\mathrm{2-10,int}=$ 43.15 (dotted), 43.71 (dashed), and 44.20 (solid), which correspond to the first, second, and third quartiles of the CCLS* luminosity distribution. Horizontal cutoffs are shown at three counts (the faintest possible detection), 30 counts (spectral modeling with at least one free parameter), and 70 counts (spectral modeling with at least two free parameters).  Although the median luminosity unobscured source still has 30 counts up to $z\sim2$, its CT counterpart cannot be reliably detected beyond that redshift. Among CT simulations, uneven vertical spacing is due to variation in the exact \nh\ values, which are $10^{24.79}\ \mathrm{cm}^{-2}$ ($1^\mathrm{st}$ quartile luminosity), $10^{24.07}\ \mathrm{cm}^{-2}$ (median luminosity), and $10^{24.11}\ \mathrm{cm}^{-2}$ ($3^\mathrm{rd}$ quartile luminosity). The bump in the signal of the high-luminosity CT source at $z\sim1.5$ is due to photons from the reflected component beginning to enter the Chandra band pass. \label{fig:apps}}
\end{figure}

To demonstrate the effect of obscuration on the quality of Chandra observations, we selected three pairs of BASS models, with $\log L_\mathrm{2-10,int}=$ 43.15, 43.71, and 44.20.
These represent the approximate first, second, and third quartiles of the intrinsic X-ray luminosity distribution of CCLS* (and of our 380 selected template spectra).
Each pair consists of a completely unobscured model ($\nh\leq 10^{20}\ \mathrm{cm}^{-2}$) and a CT model ($\nh\geq 10^{24}\ \mathrm{cm}^{-2}$).
We recorded the net 0.5--7.0 keV Chandra counts as a function of redshift for each model using an exposure time of 162 ks, the median exposure time for the CCLS sample.
To reduce the effects of Poisson noise, we took the mean of $10^6$ simulations at each redshift.
The results are shown in Fig.~\ref{fig:apps}.
Note that while the median luminosity unobscured model has more than 30 counts at $z\sim2$, its CT counterpart cannot be reliably detected, and even the highest luminosity CT source has fewer than 30 counts for $z\gtrsim0.5$.
Also note that high-luminosity CT AGN exhibit a relative increase in data quality for $z\gtrsim1.5$,because photons with intrinsic energies less affected by obscuration begin to be detected by Chandra due to redshifiting of the spectra.
The BASS IDs used here are 643, 325, and 456 for the CT models and 774, 328, and 771 for the unobscured models.\footnote{See \url{bass-survey.com/dr2.html} for complete catalog information on these sources, including source and counterpart identifiers.}


\section{Data Analysis and Results} \label{sec:data_anal_results}
\subsection{Detection and counts} \label{subsec:detection_cts}
\begin{figure*}
    \gridline{
        \fig{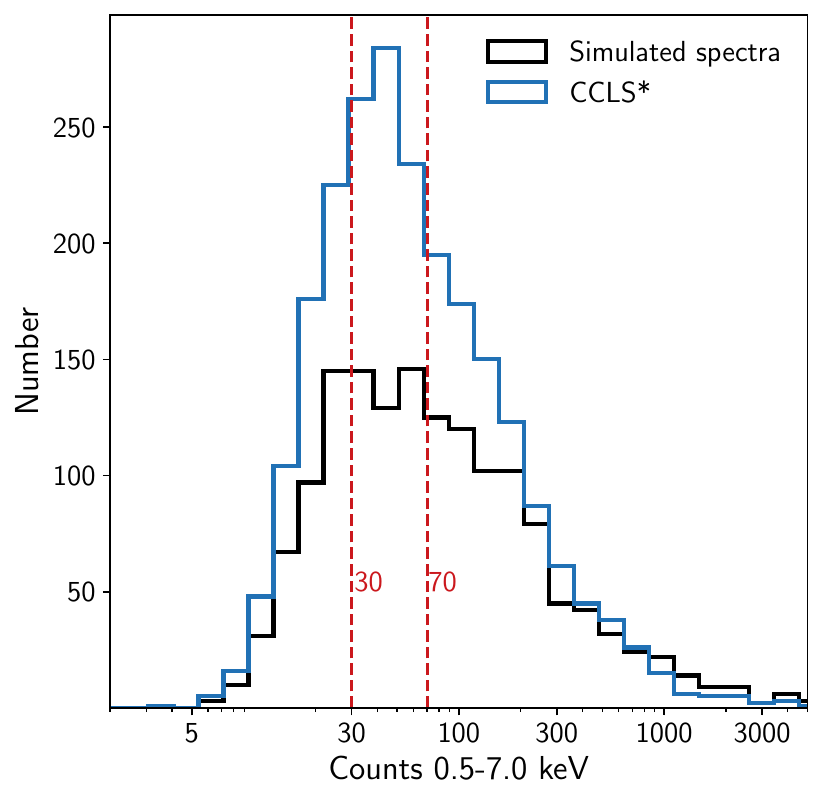}{0.325\textwidth}{(a)}
        \fig{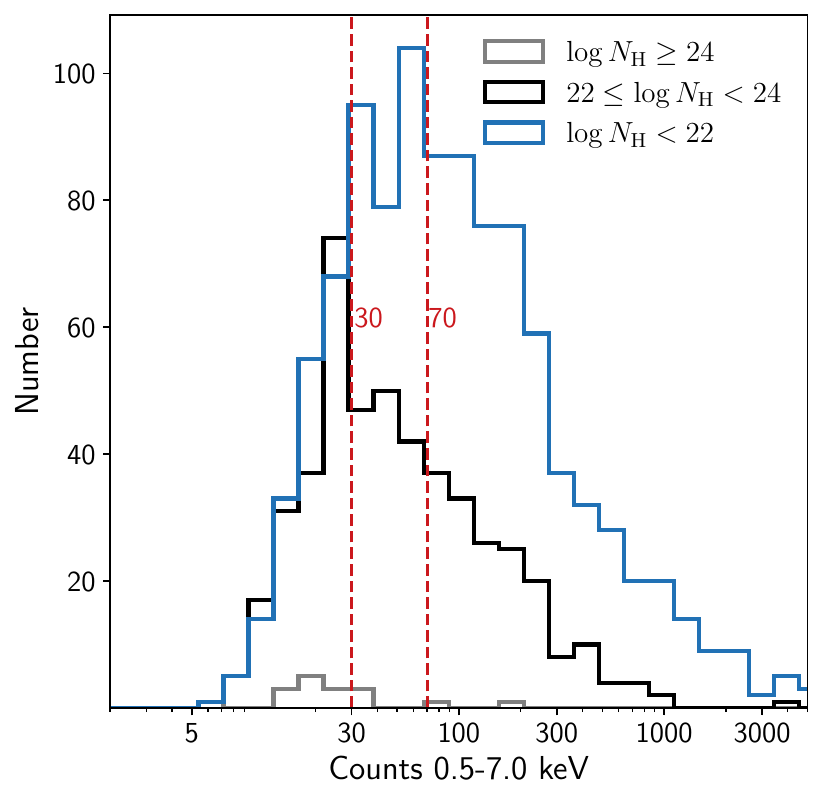}{0.325\textwidth}{(b)}
        \fig{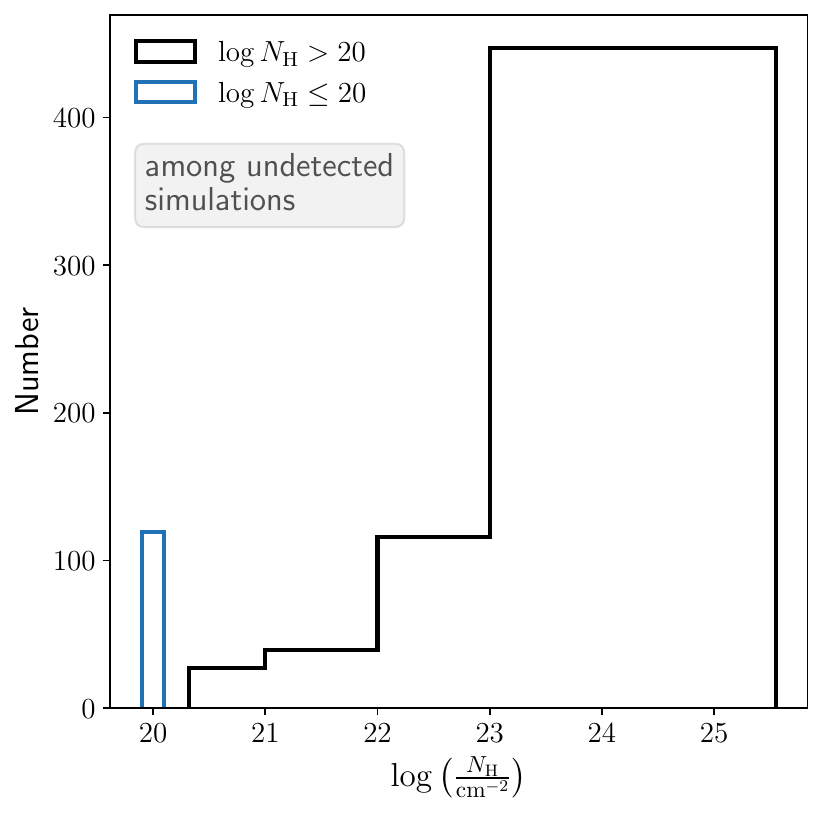}{0.31\textwidth}{(c)}
        }
    \caption{\textbf{Net counts of simulated spectra.} \textbf{(a) Comparison with CCLS*.}  The Chandra counts in the 0.5--7.0 keV band of the 1532 simulated spectra considered detected (black; see text), compared with that of the 2291 CCLS* spectra (blue). \textbf{(b) Obscured and unobscured detected spectra.} Obscured simulations (black) lean more toward the low-count regime than do their unobscured counterparts (blue), and 10/17 CT simulations (gray) have fewer than 30 counts. \textbf{(c) \nh\ distribution of sources undetected in simulations.} Undetected sources with zero counts (748) are largely heavily obscured and CT models. The spike at $\log\nh=20$ (shown separately in blue) is due to the application of the selection function below 30 counts and the fact that completely unobscured sources make up a significant portion of our simulations. 
 \label{fig:counts_hist}}
\end{figure*}
\begin{figure*}
    \gridline{\fbox{\fig{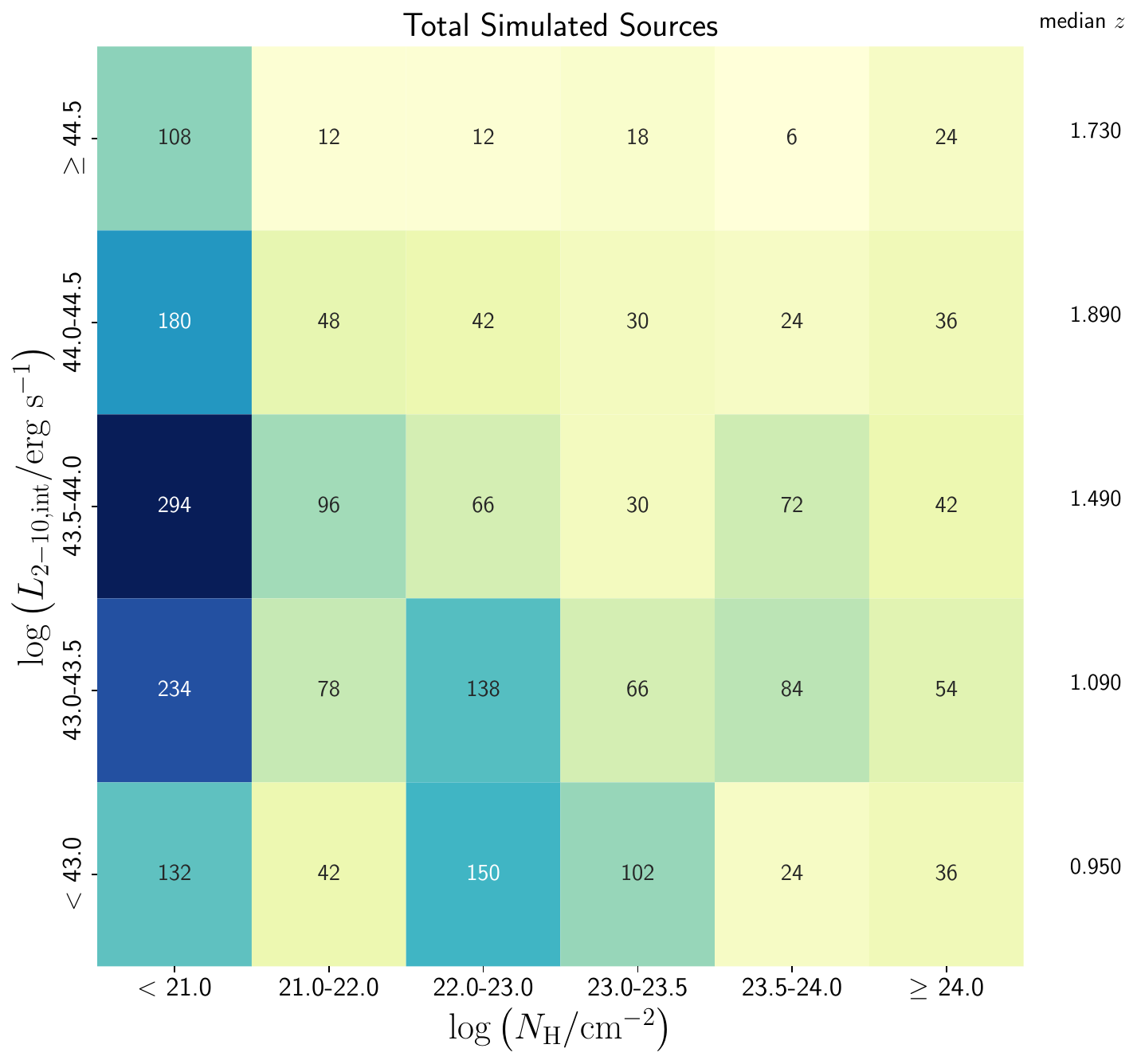}{0.49\textwidth}{(a)}}
        \fig{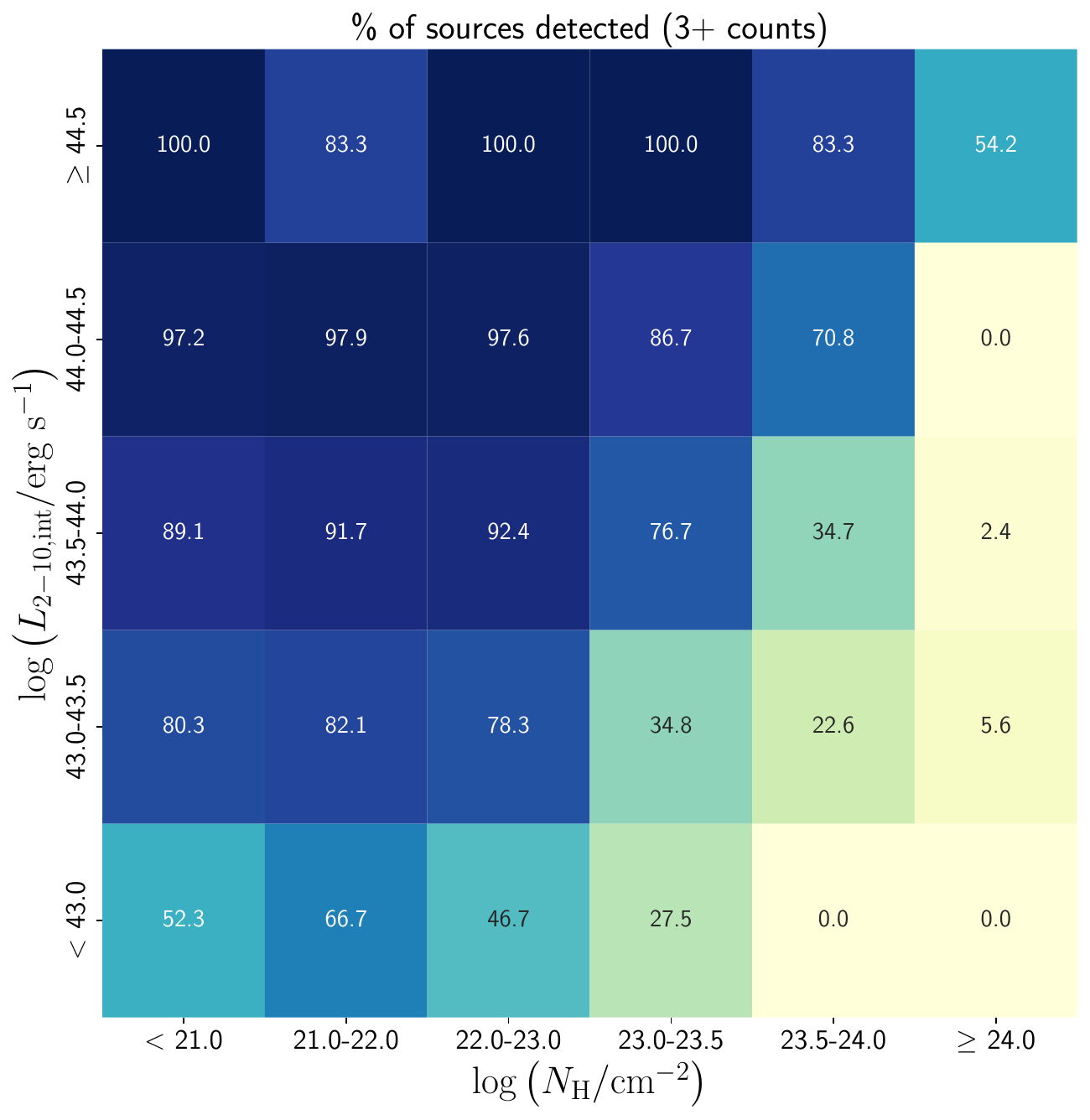}{0.49\textwidth}{(b)}}
    \gridline{\fig{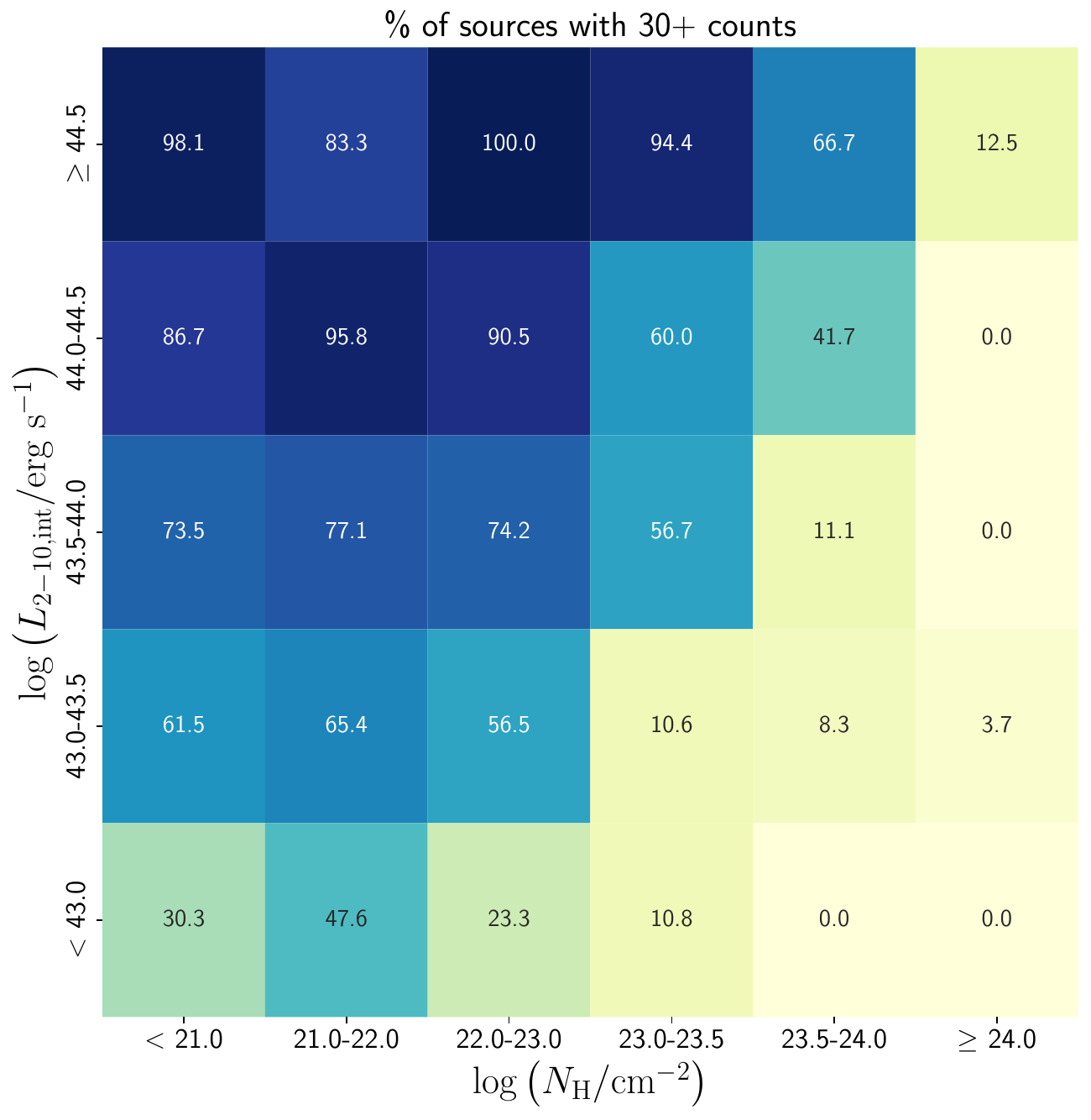}{0.49\textwidth}{(c)}
        \fig{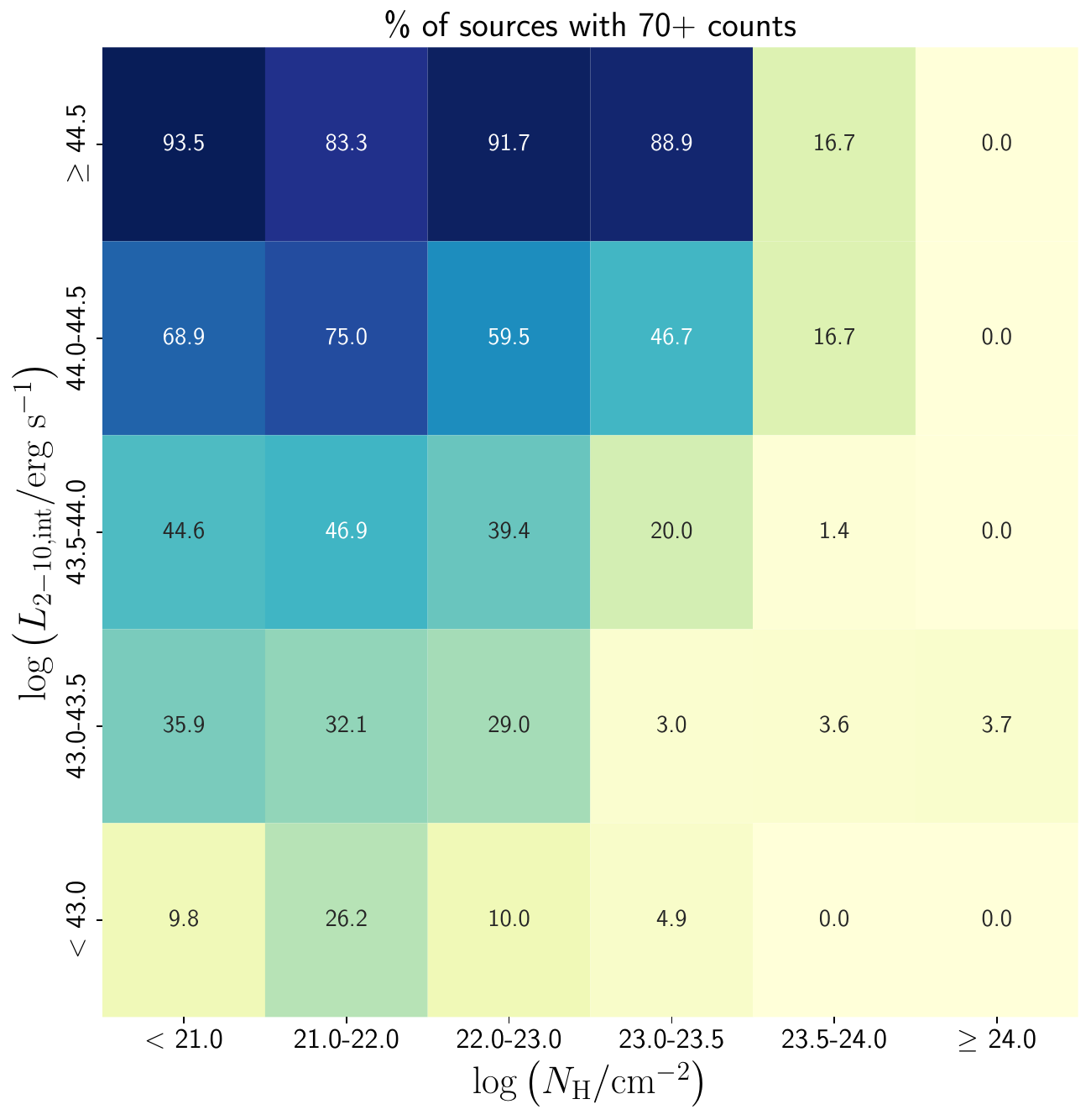}{0.49\textwidth}{(d)}}
    \caption{\textbf{Chandra detection fractions of the simulated dataset} organized in bins of intrinsic X-ray luminosity (vertical) and obscuring column density (horizontal).  Dark blue indicates high fractions and pale yellow indicates low fractions. \textbf{(a)} Total number of simulated sources and the median redshift of each luminosity bin. The expected positive correlation between redshift and luminosity is illustrated in Fig.~\ref{fig:lum_z_dist}(b). \textbf{(b)--(d) Percentage of simulations with at least 3, 30, and 70 counts.} 
    The percent values are displayed in each square. Simulated spectra with $\geq30$ counts sources were fit with at least one free parameter, and those with $\geq70$ counts were fit with at least two free parameters. \label{fig:heatmaps}}
\end{figure*}

To measure the counts of each simulated spectrum, the simulated spectra were imported into \texttt{XSPEC}, and only channels in the 0.5--7.0 keV range were noticed, following the approach of M16.
Accurately reproducing the Chandra background given the array of off-axis angles and overlapping fields was difficult and the effect is small (there are $<2$ expected background counts for an on-axis source in the CCLS data; \citealt{civanoCHANDRACOSMOSLEGACY2016}), so as an approximate approach we did not include a background in our fits.
This means our study conservatively overestimates the ability to model source spectra (which was only done for sources with at least 30 counts).
See Appendix~\ref{app:background} for further discussion of the background and minimum detection threshold.

First, any spectrum with fewer than three counts (the minimum in CCLS*) was set to zero counts.
Thus, three counts is considered a Chandra ``detection.''\footnote{We chose to set the threshold in counts instead of flux since that grounds our study in detector space.}
For the detection study below 30 counts, we had an excess of spectra with few ($<10$) counts.
To correct this, we included a function to randomly reduce the number of sources at the same rate that Chandra would not detect because of background.\footnote{The CCLS notes that there is significant incompleteness below 20 counts \citep{civanoCHANDRACOSMOSLEGACY2016}.}
With this approach, the detection rate for our simulated sample was 1532/2280 (67.2\%).
Fig.~\ref{fig:counts_hist}(a) shows the resulting count distribution for the 1532 detected simulated spectra compared with that of the CCLS* sample.

Our counts distribution matches well for sources with more than $\sim200$ counts.
These are bright sources with high signal-to-noise ratios, for which background is negligible.
Compared with CCLS*, we are missing sources below $\sim100$ counts.
This is likely because our template models are drawn from local \bat-selected AGN, in which obscured and CT sources are more likely to be detected because of their higher energy coverage relative to deep surveys like CCLS (see Fig.\ref{fig:nh_comparison} in Appendix~\ref{app:nh_comparison}).
In Fig.~\ref{fig:counts_hist}(b), it is shown that obscured and CT simulations are skewed to lower counts relative to all other simulations, which is a well known effect \citep[e.g.,][]{brandtNatureSoftXRay2000}.
Fig.~\ref{fig:counts_hist}(c) further shows that heavily obscured sources comprise a significant portion of undetected sources.
There is a spike at $\log\nh=20$ (shown separately in blue) due to the application of the selection function below 30 counts, and the fact that entirely unobscured sources make up a significant portion of our simulations.
Note that all models with $\log\nh\leq20$ are labeled as $\log\nh=20$ in the BASS survey since Galactic \nh\ dominates below that point, and therefore the intrinsic column density cannot be reliably constrained.

Table \ref{tab:counts} summarizes the number of simulations detected.
Fig.~\ref{fig:heatmaps} presents the percentages of spectra with $\geq$30 and $\geq$70 counts, in bins of luminosity and obscuration column density.
{\em In particular, only 5/192 (2.60\%) of CT simulated spectra were detected with at least 30 counts}, the minimum threshold for spectral modeling that we adopted from M16.
This means that only five of the fitted spectra were simulated with CT models because of the inherently low counts in the simulation. 
Note that the Poisson fluctuations introduced by \texttt{fakeit} also affect the counts distribution, so the detection fraction of obscured sources can be considered an upper bound, since the non-detections due to noise would only make the fraction lower. \begin{deluxetable}{ccccc} \label{tab:counts}
    \tablecaption{Number of simulations detected, and those with at least 30 and 70 counts.  Percentages of the total in each row are shown in parentheses.}
    \tablehead{\colhead{} & \colhead{Total} & \colhead{Detected} & \colhead{30+} & \colhead{70+} } 
\startdata
All sims & 2280 & 1532 & 1179 & 751 \\
 &  & (67.2\%) & (51.7\%) & (32.9\%) \\
 \hline
Obscured & 1056 & 493 & 316 & 171 \\
($\log\nh\geq22$) &  & (46.7\%) & (29.9\%) & (16.2\%) \\
\hline
CT & 192 & 17 & 5 & 2 \\
($\log\nh\geq24$) &  & (8.9\%) & (2.6\%) & (1.0\%) \\
\enddata
\vspace{-25pt}
\end{deluxetable}

\subsection{Fitting procedure} \label{subsec:fitting_procedure}

Next, we investigate how well Chandra can recover the parameters of the sources that we detect.
To do so, we follow the CCLS fitting procedure outlined in Sec.~4.1 of M16 for all CCLS sources with at least 30 counts. 
They adopted a four-step fitting procedure, which we summarize here.
In each step, the quality of the fit was measured with the $C$ statistic \citep[$C$;][]{cashParameterEstimationAstronomy1979},\footnote{$C = 2\sum_i \left( m_i - n_i + n_i\ln\left(\frac{n_i}{m_i}\right) \right)$, where $n_i$ ($m_i$) is the the observed (predicted) counts in bin $i$.} which is particularly useful for data sets with low counts per bin, where Gaussian assumptions break down.
Since we did not include background spectra in the simulations, we can use the $C$ statistic, which assumes only Poisson noise.
For each subsequent fit, the criterion for improvement over the previous fit was $\Delta C \equiv C_\mathrm{old} - C_\mathrm{new}>2.71$.
We used the \texttt{XSPEC} \texttt{error} command to calculate 90\% confidence intervals.
All spectra were grouped to at least three counts per bin for fitting.
In what follows, $N_\mathrm{H,gal}$ was set to $2.5\times10^{20}\ \mathrm{cm}^{-2}$, and all redshifts were fixed to the values used in the simulation model. 
\begin{enumerate}
    \item If the spectrum had fewer than 70 counts, we fit to an absorbed power law \texttt{`phabs*(zphabs*zpow)'} with $\Gamma$ fixed to 1.9. Only \texttt{zphabs.nH} (\nh) and \texttt{zpowerlw.norm} were free to vary.
    \item For all sources with at least 30 counts, we then fit the spectrum to the same absorbed power law model as in step 1, but with \texttt{zpowerlw.PhoIndex} ($\Gamma$) also free to vary.
    \item For all sources with at least 30 counts, we then added an additional unabsorbed power law with $\Gamma_2=\Gamma_1$: \texttt{`phabs*(zphabs*zpow + const*zpow)'}.  This represents intrinsic emission that escapes through sight lines of column density lower than that of the dense circumnuclear obscurer and is subsequently scattered into the direction of the detector.  The relative normalization of the secondary power law (\texttt{constant.factor}) is a consequence of both the covering factor of the torus and the viewing angle \citep{mckaigRaytracingSimulationsSoft2023}, and as noted in previous studies \citep[e.g., M16, R17,][]{pecaCosmicEvolutionAGN2023}, it should not exceed a few percent of the primary component (see discussion in Appendix~\ref{app:2pl_discussion}).  Consistent with M16, \texttt{constant.factor} was constrained to be $<0.15$ (private correspondence).\footnote{We also set a lower limit of 0.001 \citep{guptaBATAGNSpectroscopic2021}, in order to not over-index on data points at the highest energies of the Chandra band pass, which would have the most significant background. Any fits that hit the upper or lower limits were discarded.}
    \item For all sources with at least 30 counts, we added a Gaussian component to whatever the best-fit model was at this stage to represent the iron K$\alpha$ line. \texttt{zgauss.LineE} was fixed to 6.4 and \texttt{zgauss.Sigma} was fixed to 0.1.
\end{enumerate}

A summary of how many simulated spectra settled on each fit is presented in Table \ref{tab:fit_summary}.
The goodness-of-fit statistics are discussed in Appendix~\ref{app:fitting_statistics} and the full catalog of the simulation fits is presented in Table \ref{tab:fits}.
Since the BASS catalog reports all best-fit \nh\ values $\leq10^{20}\ \mathrm{cm}^{-2}$ as $10^{20}\ \mathrm{cm}^{-2}$, after fitting, we imposed the same lower limit on all of our best fit \nh\ values and their lower limits for easy comparison with the BASS models.
These fits were considered to be consistent with a column density of zero.
Among the 1179 best fits, 906 had \nh\ values consistent with zero with 90\% confidence, among which 117 had completely unconstrained \nh\ (all possible values fell within the 90\% confidence intervals).
Meanwhile, among simulations with at least 30 counts, 529 were simulated with $\log N_\mathrm{H,sim}\leq20$. \begin{deluxetable}{cccccc} \label{tab:fit_summary}
    \tablecaption{Summary of fitting results.  Fit numbers correspond to steps 1--4 described in Sec.~\ref{subsec:fitting_procedure}.   Percentages of the total in each row are shown in parentheses.  Entries with ``---'' indicate that the column is not applicable by definition of the fit. Entries with ``by def'' underneath them are always true by definition of the fit.}
    \tablehead{\colhead{Fit} & \colhead{Total} & \colhead{$<70$ cts} & \colhead{$\Gamma$ free} & \colhead{2PL} & \colhead{Fe line} } 
\startdata
1 & 355 & 355 & --- & --- & --- \\
 &  & (by def) &  & &  \\
 \hline
2 & 711 & 50 & 711 & --- & --- \\
 &  & (7.0\%) & (by def) & &  \\
\hline
3 & 57 & 17 & 49 & 57 & --- \\
 &  & (29.8\%) & (86.0\%) & (by def) &  \\
\hline
4 & 56 & 6 & 52 & 3 & 56 \\
 &  & (10.7\%) & (92.9\%) & (5.4\%) & (by def) \\
\hline
\hline
ALL & 1179 & 411 & 812 & 60 & 56 \\
FITTED &  & (34.9\%) & (68.9\%) & (5.1\%) & (4.8\%) \\
\enddata
\vspace{-25pt}
\end{deluxetable}

\begin{figure*}
    \gridline{
        \fig{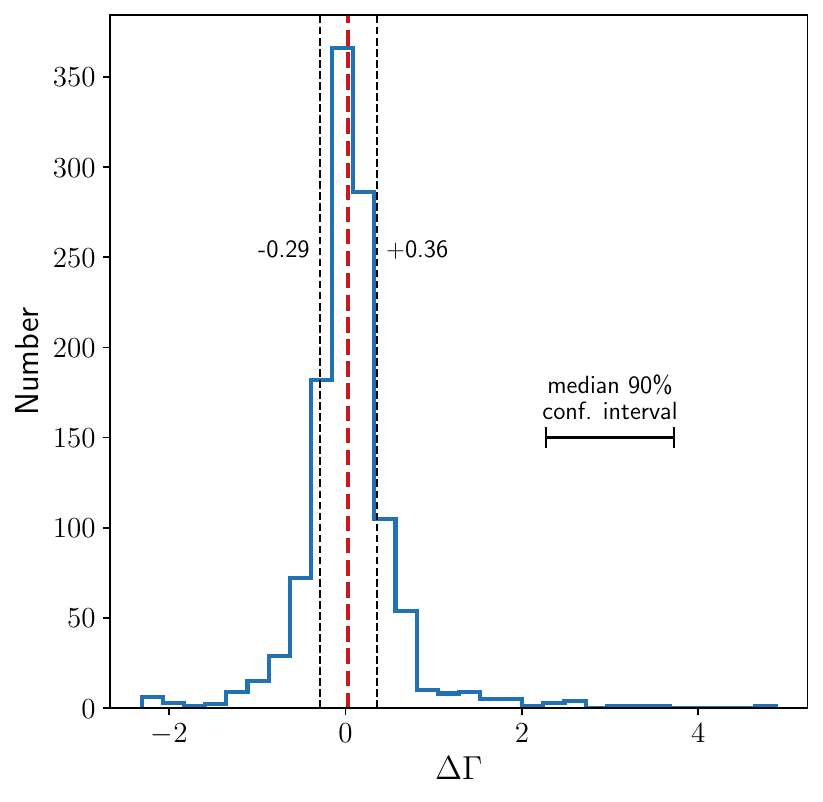}{0.40\textwidth}{(a)}
        \fig{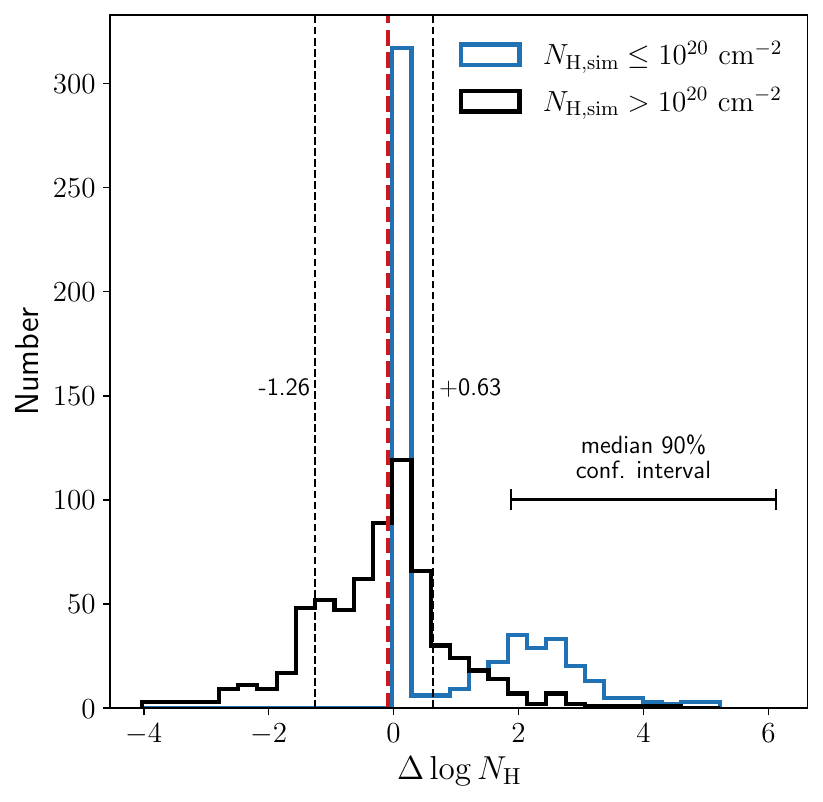}{0.40\textwidth}{(b)}
        }
    \caption{\textbf{Error histograms.} For each of $\Gamma$ and $\log\nh$, we calculate the error of the measured value relative to the simulated value.  \textbf{(a) Accuracy of measured $\Gamma$.}  The vertical red dashed line indicates the median $\Delta\Gamma$ at $+0.03$. 68\% of the spectra lie between the black dashed lines at $-0.29$ and $+0.36$.  The error bar indicates the median size of the 90\% confidence interval computed by \texttt{XSPEC}, which is 0.73. \textbf{(b) Accuracy of best-fit \nh\ (log scale).} For simulations with some obscuration (black), the distribution is peaked around zero with a median of $-0.10$ dex (indicated by the dashed red line) and 68\% of spectra lying between the dashed black lines at $-1.26$ dex and $+0.63$ dex. In contrast, there is a tail of significantly overestimated spectra for simulations with an intrinsic \nh\ consistent with zero (blue). The error bar indicates the median size of the 90\% confidence interval computed by \texttt{XSPEC} (among those that were finite), which is 2.12 dex.\label{fig:delta_param_dist}}
\end{figure*}
\begin{figure*}
    \gridline{
        \fig{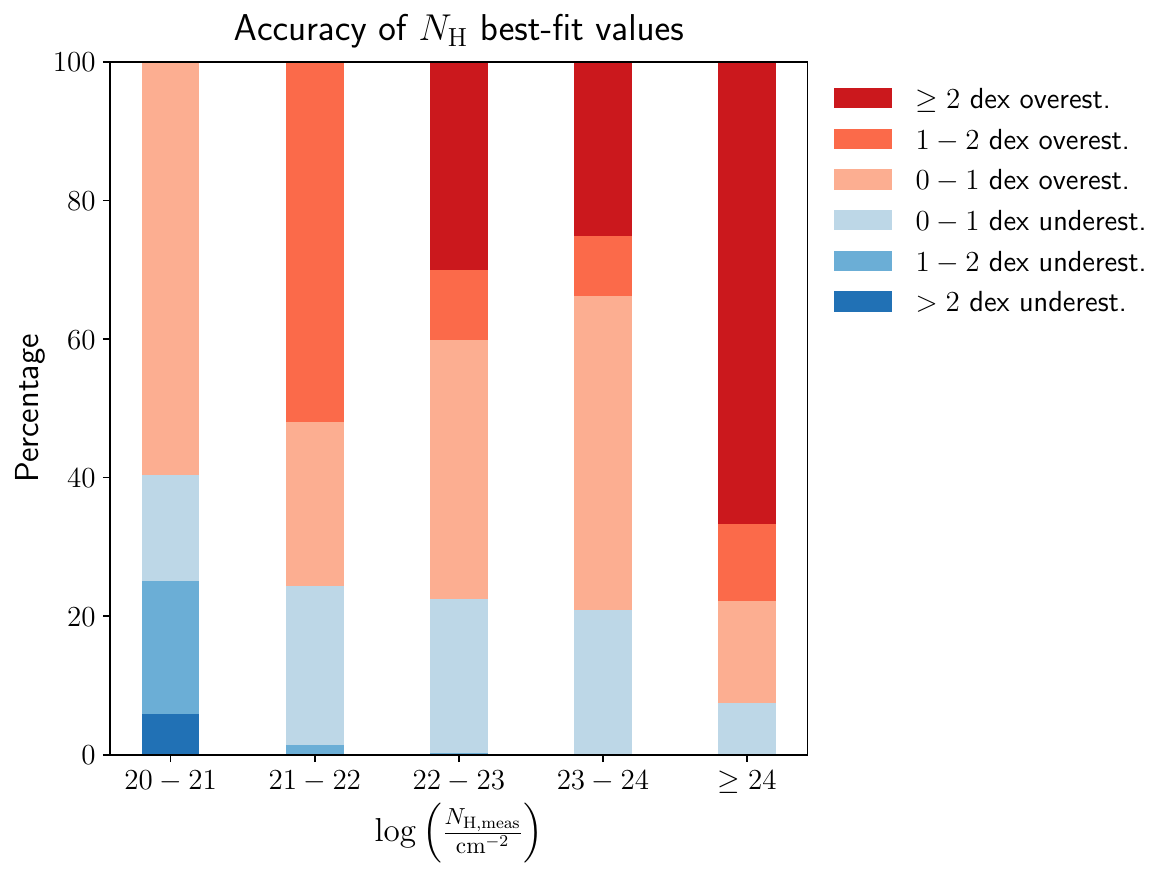}{0.52\textwidth}{(a)}
        \fig{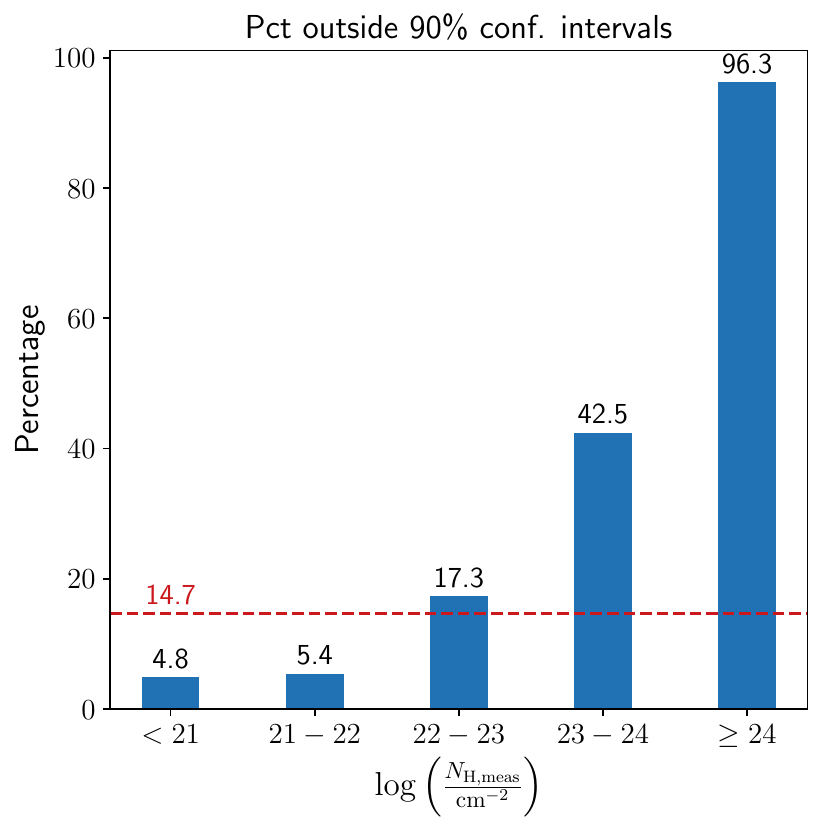}{0.38\textwidth}{(b)}
        }
    \caption{\textbf{CT overestimation.} We investigate the accuracy of the best-fit \nh\ estimations and the 90\% confidence intervals. \textbf{(a) Accuracy of best-fit \nh\ measurements.} In bins of best-fit measured \nh, we show the proportion of fits that are accurate to different dex values. Among the 27 spectra that infer CT space ($\log N_\mathrm{H,meas}\geq24$), 21 (77.8\%) are overestimated by $>$ 1 dex. \textbf{(b) The proportion of \nh\ fits that are inaccurate within 90\% confidence intervals.} The percentage of fitted spectra in each \nh\ bin whose \nh\ was incorrectly determined with 90\% confidence. The red dashed line indicates the percentage for the entire fitted dataset.\label{fig:nh_errors}}
\end{figure*}

\subsection{Parameter recovery: photon index} \label{subsec:parameter_recovery_gamma}

To quantify how well parameters were recovered, we focused on the photon index $\Gamma$ and the absorption column density \nh.
For $\Gamma$, we compute the error $\Delta\Gamma = \Gamma_\mathrm{meas} - \Gamma_\mathrm{sim}$.
The histograms of the results are shown in Fig.~\ref{fig:delta_param_dist}(a).
$\Delta\Gamma$ is peaked around zero, with a median value of +0.03, and 68\% of sources within the interval $[-0.29, 0.36]$.
With a Spearman's correlation coefficient of ($r_\mathrm{S}(1177)=0.076, p=0.009$), a mildly significant correlation was found between $|\Delta\Gamma|$ and $z_\mathrm{sim}$, indicating that the photon index is less accurately measured for sources at higher redshift.
There was a significant negative correlation between $|\Delta\Gamma|$ and net counts ($r_\mathrm{S}(1177)=-0.240, p<10^{-16}$),\footnote{We used the Spearman's correlation coefficient ($r_\mathrm{S}$), which is based on ranks of the data rather than the actual values, because the absolute error, net counts, and log\nh\ are not normally distributed. We used Pearson's coefficient ($r_\mathrm{P}$) when measuring the relationship between $\Delta\Gamma$ and $\Gamma_\mathrm{sim}$ (in any case, the Spearman's coefficient also indicates a correlation between those two variables).} which is expected, since higher quality data should have less error in the measured slope.
There was a negative correlation between $\Delta\Gamma$ and $\Gamma_\mathrm{sim}$, ($r_\mathrm{P}(1177)=-0.249, p<10^{-5}$), indicating that softer (harder) slopes tend to be overestimated (underestimated).
Additionally, a mild correlation was found between (signed) $\Delta\Gamma$ and $\log N_\mathrm{H,sim}$ ($r_\mathrm{S}(1177)=0.0659, p=0.024$).
Among the fits with $\Gamma$ as a free parameter, 69.21\% (562/812) were accurate within the 90\% confidence intervals calculated by \texttt{XSPEC}.
Among those that were outside of the 90\% confidence intervals, 40.00\% (100/250) were underestimated and 60.00\% (150/250) were overestimated.

\subsection{Parameter recovery: column density} \label{subsec:parameter_recovery_nh}

Similarly, we computed $\Delta\log\nh = \log N_\mathrm{H,meas} - \log N_\mathrm{H,sim}$.
Note that this error is in logarithmic space.
The results are shown in Fig.~\ref{fig:delta_param_dist}(b).
There was a positive correlation between $\Delta\Gamma$ and $\Delta\log\nh$ ($r_\mathrm{S}=0.358, p<10^{-4}$), indicating that underestimated $\Gamma$ values tend to be paired with underestimated obscuration values.
Furthermore, among the 250 fits (with $\Gamma$ free to vary) for which $\Gamma_\mathrm{sim}$ did not lie within the 90\% confidence intervals of $\Gamma_\mathrm{meas}$, 67 of them (26.8\%) measured $N_\mathrm{H}$ outside of the 90\% confidence intervals.
Meanwhile, among the 562 fits that measured $\Gamma$ correctly within the 90\% confidence intervals, 36 of them (6.4\%) measured $N_\mathrm{H}$ outside the 90\% confidence intervals.
In all but two cases, when $\Gamma$ was below the 90\% confidence interval (i.e., the slope was estimated to be too soft), $\nh$ values were also underestimated. 
These correlations are the result of a degeneracy in power-law fits to AGN X-ray spectra, which is further discussed in Sec.~\ref{subsec:overestimated_nh}.

The error in the best-fit $\log\nh$ shows a clear excess of overestimated sources.
The overestimation of \NH\ in X-ray surveys is a well-known issue \citep[e.g., Appendix A of][]{akylasXMMNewtonChandraMeasurements2006} and is taken into account in several X-ray population studies \citep[e.g.,][]{simmondsXZDerivingRedshifts2018, pecaXRayRedshiftsObscured2021, pecaCosmicEvolutionAGN2023}, and in this study we quantify the effect for a large sample for the first time. 

Furthermore, we know that only five CT simulations were detected with at least 30 counts---only 0.42\% of the 1179 total that were fitted.
Meanwhile, 472 spectra (40.03\% of fitted sources) had best-fit $\log N_\mathrm{H,meas}\geq 22$, and 993 (84.22\%) were obscured within 90\% confidence intervals.
Among those, 27 (2.29\%) had best-fit $\log N_\mathrm{H,meas}\geq 24$, and 158 (13.40\%) were CT candidates within 90\% confidence intervals.
Thus, many of these column densities were overestimated.

Interestingly, the error in the best-fit \nh\ is centered about approximately zero (median of $-0.10$ dex, while the median 90\% confidence interval is 2.12 dex wide) for all spectra with $\log N_\mathrm{H,sim}>20$ (although the distribution is broad, with 68\% of fits in the interval $[-1.26, 0.63]$).
The excess of significantly overestimated column densities arises predominantly from simulations of completely unobscured models ($\log N_\mathrm{H,sim}\leq20$, consistent with zero).
A spike in zero-error \nh\ values can also be seen for the $\log N_\mathrm{H,sim}\leq20$ sample.
This is due to the lower limit imposed at $\log\nh=20$.

Fig.~\ref{fig:nh_errors}(a) shows the distribution of errors in best-fit \nh.
While most spectra with $\log N_\mathrm{H,meas}\leq 21$ are within one dex of the simulated value, the proportion of significantly overestimated increases in each subsequent bin.
Among the spectra measured to be CT, 66.6\% (18/27) are actually unobscured sources ($\log\nh\leq22$) whose column density is overestimated by at  least two dex, suggesting that X-ray measurements of  low signal to noise AGN with fitted values indicating  high obscuration are unreliable.

Conversely, among the five CT spectra detected with at least 30 counts, two were measured to be CT, one was measured to be obscured, and two were measured to have \nh\ consistent with zero with upper limits on $\log\nh$ less than 24.

If instead of the best-fit \nh, we look at the 90\% confidence intervals, we find that the \nh\ for nearly all highly obscured and CT sources are inaccurate.
While the measured \nh\ is accurate for 95.0\% of simulations of  unobscured AGN ($\log\nh<22$), only 70.8\% of those measured as obscured, and 3.7\% of those estimated to be CT, are accurate within the error bars.
See Fig.~\ref{fig:nh_errors}(b).
No similar patterns were found among the net counts or $z_\mathrm{sim}$ distributions. 

\subsection{False CT sources} \label{subsec:2pl_results}
\begin{figure}
    \centering
    \includegraphics[width=0.48\textwidth]{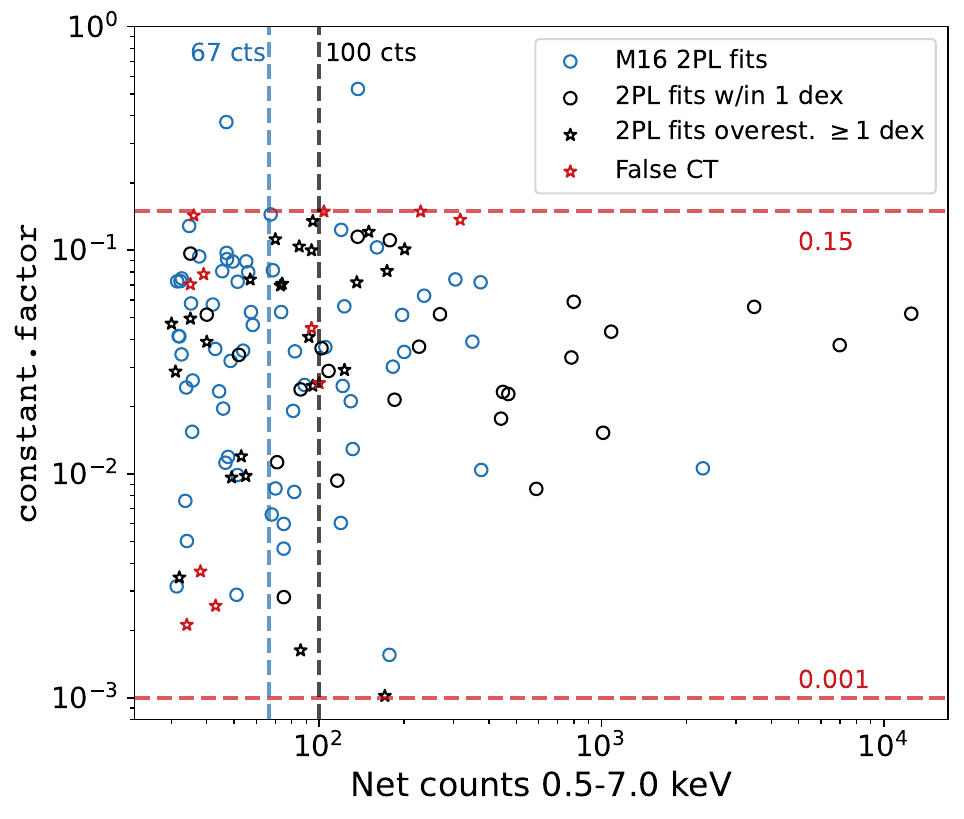}
    \caption{\textbf{Properties of 2PL fits.}  For each of our 60 2PL fits, we show the relative normalization of the secondary power law versus the net counts in the 0.5--7.0 keV band.  2PL fits that measured \nh\ accurately to within 1 dex are shown in black circles, while those that were inaccurate by at least 1 dex are shown in black stars. Red stars represent the false CT fits, which had best-fit $\log\nh\geq24$ and $N_\mathrm{H,sim}\leq20$, and all of which settled on a 2PL fit. The 67 2PL fits of M16 are shown as blue circles for comparison.  The red dashed lines indicate the upper and lower limits on the secondary normalization that we set during our fitting procedure, and the vertical dashed lines indicate the median counts of the full simulation sample (black; 100 cts) and the full sample of M16 (blue; 67 cts). \label{fig:2pl_norms_comp}}
\end{figure}

To investigate the phenomenon of overestimated column densities, we restrict our attention to the 11 spectra that had $\log N_\mathrm{H,sim}\leq 20$ and $\log N_\mathrm{H,meas}\geq 24$ (the ``false CT sources'').
We note that none of the false CT sources had 90\% confidence intervals that extended below $\log\nh=23.5$.
Moreover, the false CT sources were not necessarily those with few counts: 4/11 had counts above 100, the median of the 1179 fitted simulations.

All of these fits settled on a double power law (2PL) fit (step 3 described in Sec.~\ref{subsec:fitting_procedure}).
4/11 had a fixed photon index and none included an iron K$\alpha$ line.
Fig.~\ref{fig:2pl_norms_comp} shows their secondary normalizations versus net counts, compared with the other 2PL fits in our simulated sample as well as those of M16.
It is notable that almost all of our fits that settled on a secondary normalization very near the upper limit turned out to be false CT source
Among the 145 fits that had \nh\ overestimated by at least 2 dex, 21.4\% fit to a 2PL, compared with 2.5\% for the rest of the fits.
Conversely, 51.6\% of the 60 2PL fits overestimated \nh\ by at least 2 dex, compared with only 10.1\% of all other fits.
Fig.~\ref{fig:2pl_comparison} in Appendix~\ref{app:2pl_discussion} shows a histogram of $\Delta\log\nh$ for these fits.

In general, for single power law fits, there is degeneracy between $\Gamma$ and \nh\, since an unobscured softer power law can mimic the behavior of an obscured source and a harder continuum (perhaps with a reflected component).
This explains the correlation reported in Sec.~\ref{subsec:parameter_recovery_nh}, wherein overestimates of $\Gamma$ were correlated with overestimates in \nh.
The photon index distribution of the false CT sources skews high relative to the simulation sample.
To assess this, the $\Gamma_\mathrm{meas}$ of false CT sources were compared to the full simulation sample using a modified Z-score test\footnote{$Z_\mathrm{mod}(X_i) = (X_i - \mathrm{median}(X))/\mathrm{MAD}(X)$, with $\mathrm{MAD}(X) = \mathrm{median}|X_i - \mathrm{median}(X)|$.
We used the 95\textsuperscript{th} percentile of the modified Z-score values as our test statistic.} with bootstrapped significance ($10^5$ samples).
The false CT subset exhibited significantly higher $\Gamma_\mathrm{meas}$ values ($p=0.0007$).
This suggests that the necessity of a secondary power law is due to the sharper drop-off of the primary component.

We refit the 11 false CT spectra using the same procedure outlined in Sec.~\ref{subsec:fitting_procedure}, but without step 3, so we did not allow a 2PL fit.
After doing so, 9/11 of the spectra were fitted to correct \nh\ values within 90\% confidence intervals.
Note that the $\Gamma_\mathrm{meas}$ shifted to lower values for these fits; using the same method as above, we found no significant difference between the $\Gamma_\mathrm{meas}$ values and those of the entire sample ($p=0.05$), consistent with the fact that a 2PL with higher $\Gamma_\mathrm{meas}$ values is degenerate with a single, softer power law.

The 2/11 refits that still overestimated \nh\ were BASS ID 136 at $z_\mathrm{sim}=1.86$ and BASS ID 376 at $z_\mathrm{sim}=3.74$, each of which had fewer than 40 counts.
Upon inspection, both seem to be a case of the \nh-$\Gamma$ degeneracy discussed above for single power law fits.
The former fit to values of $\Gamma=1.9$ (fixed) and $\log\nh\in[21.76,23.27]$, while the BASS simulation model had $\Gamma=1.72$ and $\log\nh=20$.
The latter fit to values of $\Gamma=5.20$ and $\log\nh\in[23.68,24.52]$, while the BASS simulation model had $\Gamma=1.97$ and $\log\nh=20$.

Examples of the false CT spectra and their refits are presented and discussed in detail in Appendix~\ref{app:false_ct_examples}.

\subsection{Refitting with a torus model} \label{subsec:torus_refit}

As pointed out in \cite{lanzuisiChandraCOSMOSLegacy2018}, phenomenological models are not optimized to correctly identify CT AGN since they do not take into account Compton scattering of photons nor the geometry of the obscuring torus.
This is in addition to the \nh-$\Gamma$ degeneracy mentioned above.
These considerations become especially important for $\log\nh\geq23$ \citep[e.g.,][]{yaqoobXRayTransmissionCold1997}.
Therefore, the phenomenological modeling of M16 is followed by a study of 67 of the sources with $\Gamma_\mathrm{meas}\leq1.4$ and $\log N_\mathrm{H,meas}\geq 23$, presented in \cite{lanzuisiChandraCOSMOSLegacy2018}.
These CT candidates were fit to the physically motivated \texttt{MYTorus} model \citep{murphyXraySpectralModel2009}, and using a novel Monte-Carlo parameter estimation technique, 41.9 of the candidates were measured to be CT.

We used the \texttt{MYTorus} model implemented in \cite{lanzuisiChandraCOSMOSLegacy2018} to refit the 11 false CT sources.
For simplicity, we use the \texttt{XSPEC error} command to calculate 90\% confidence intervals on \nh\, instead of the more sophisticated approach used in that study.
A detailed description of the model and its rationale can be found in \cite{lanzuisiChandraCOSMOSLegacy2018}, which we briefly summarize here.
The model included all three \texttt{MYTorus} components (the zeroth order continuum, the reprocessed continuum, and the fluorescent emission lines) as well as a secondary power law restricted to be $<5\%$ of the normalization of the primary continuum (i.e., \texttt{phabs * (const * zpow + torusScat + torusLines + zpow * torusZeroCont}).
The inclination angle was fixed to $75^\circ$, the photon index was fixed to 1.8, the redshift was fixed to the value used for the simulated spectrum, and all \nh\ parameters were tied to each other.
This left three free parameters in the fit: \nh, the normalization of the primary power law, and the relative normalization of the secondary power law.

Zero of the 11 previously false CT sources were measured to be CT within 90\% confidence intervals, and 10/11 were correctly measured to have \nh\ consistent with zero.
Best-fit values can still be unreliable, as 6/11 spectra still had $\log N_\mathrm{meas}>20$, and one had $\log N_\mathrm{meas}>23$.
This is expected since the physical torus models assume some degree of obscuration at the adopted fixed inclination angle.
These results emphasize both the importance of refitting CT candidates to a physical torus model (e.g., \texttt{MYTorus}; \texttt{BNtorus}; \texttt{borus}, \citealt{balokovicNewSpectralModel2018, balokovicNewToolsSelfconsistent2019}; and others), as well as using robust error estimators, such as Monte Carlo or Bayesian methods, to constrain \nh.

\subsection{Parameter recovery: Fe line} \label{subsec:fe_line}

163 of the 380 selected BASS models included a Gaussian component to represent the narrow Fe K$\alpha$ line and thus, 978 simulations included a Gaussian component.
Among the simulated spectra with at least 30 counts, 371 included a Gaussian component, 23 of which were recovered.
The remaining 33 fits that included a Gaussian component were not simulated with one, suggesting that those spectra were over-fit to spurious features.


\section{Discussion} \label{sec:disc}

\subsection{Obscuration bias} \label{subsec:obscuration_bias}
\begin{figure*}
    \gridline{
        \fig{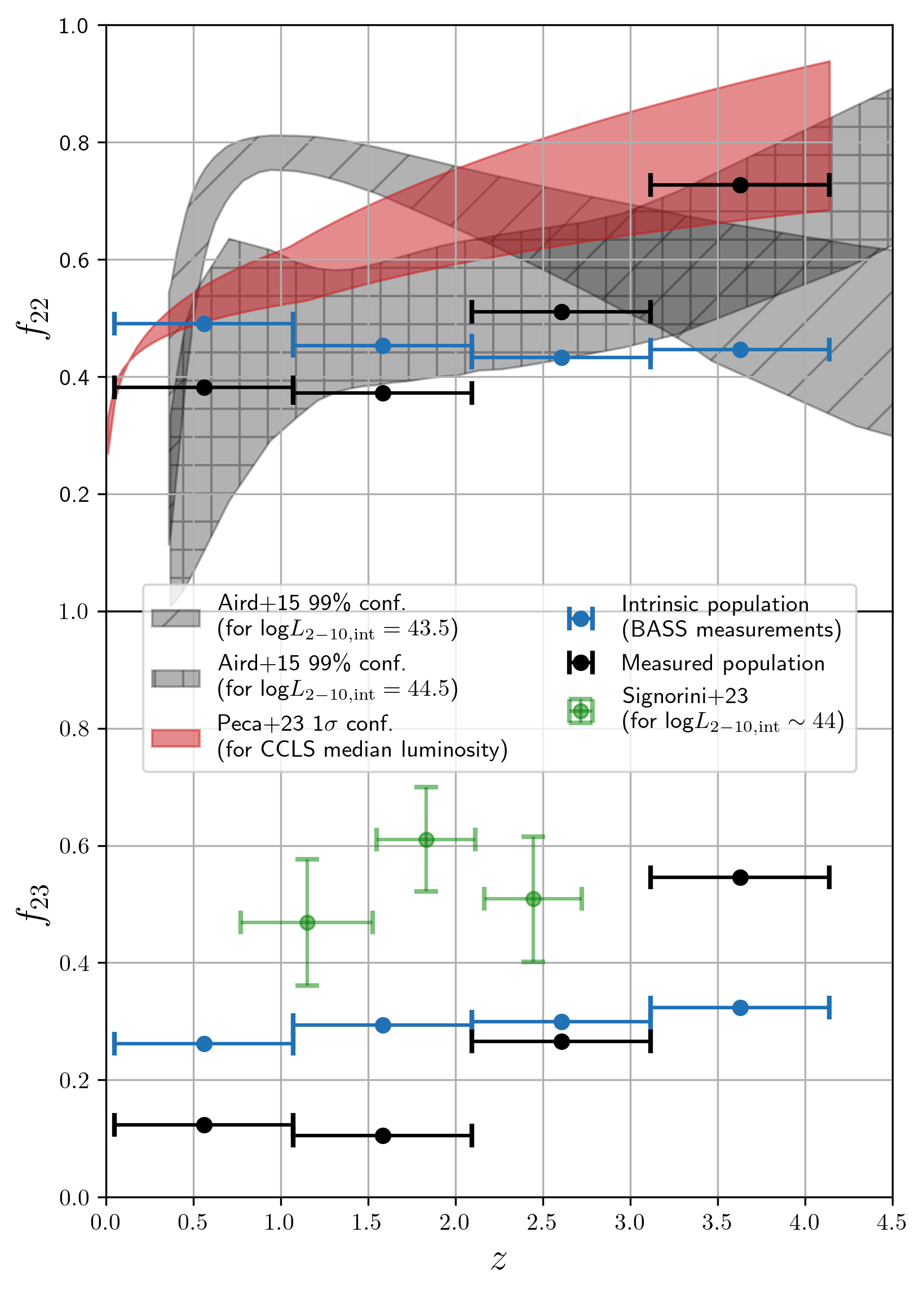}{0.49\textwidth}{(a)}
        \fig{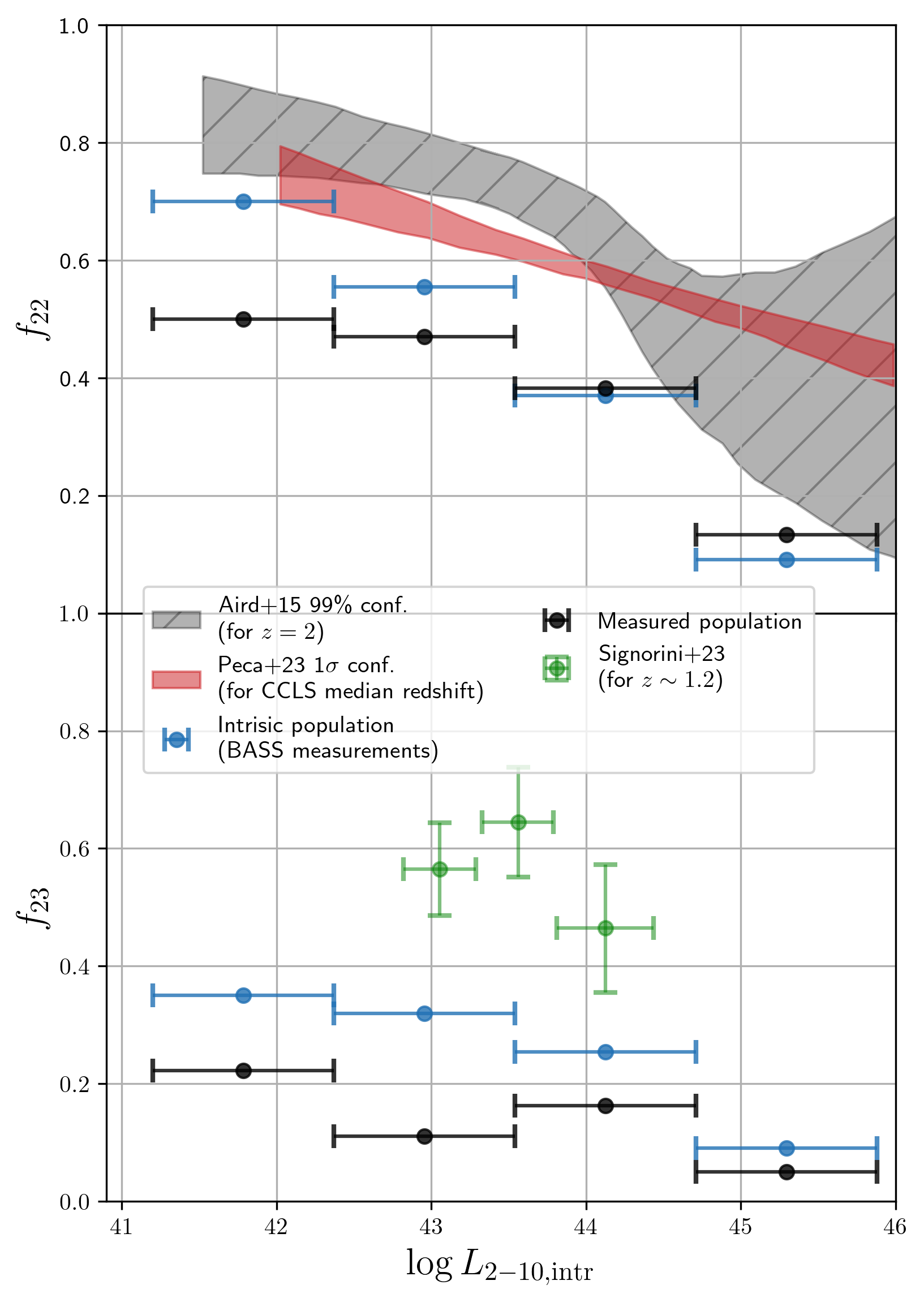}{0.49\textwidth}{(b)}
        }
\caption{\textbf{Implications for the evolution of the obscured fraction.} These plots show the implication of our study for the cosmic evolution of obscured AGN. $f_{22}$ and $f_{23}$ indicate the fraction of obscured $\log\nh\geq22$ and $\log\nh\geq23$ AGN in each bin, respectively. The actual \nh\ of the BAT AGN luminosity analogs is shown in blue and the measured population when observed in COSMOS using the fitting of Sec.~\ref{subsec:fitting_procedure} is shown in black. These are overlaid on top of the evolution models from \cite{airdEvolutionXrayLuminosity2015} and \cite{pecaCosmicEvolutionAGN2023} for comparison. Note that those studies only model $f_{22}$ up to $\log\nh = 24$. For the $f_{23}$ plots, we include the measurements of \cite{signoriniXrayPropertiesObscured2023} for comparison. \textbf{(a) Redshift evolution.}  The obscured fraction of the luminosity-matched analogs decreases with redshift, but measurement biases show an increasing trend, consistent with other X-ray studies.  \textbf{(b) Luminosity function.} The obscured fraction of the actual population exhibits a sharp decline with intrinsic 2--10 keV luminosity, while the measured population shows a gentler decline in line with observed samples.  Both of these results show the significant biases in measured populations. \label{fig:obsc_evolution}}
\end{figure*}

Our results suggest that Chandra is inefficient at identifying obscured and CT AGN in the redshift range of 0.5--3.0.
Only 493/1056 (17/192) simulated obscured (CT) sources were detected, and among those with at least 30 counts, 315/316 (3/5) were detected to be obscured (CT) within 90\% error bars, out of 993 (150) total simulated spectra that were measured to be obscured (CT) with 90\% confidence.
Fig.~\ref{fig:heatmaps} can be interpreted as a coarse-grained selection function to recover the intrinsic populations of AGN at different luminosities and column densities, based on the detected numbers (assuming the \nh\ and intrinsic luminosities are themselves accurately measured).
This suggests that many sources currently classified as obscured and CT, especially those at low luminosity, probably have overestimated column densities.
One potential cause of this overestimation is the misuse of the 2PL model, discussed below in Sec.~\ref{subsec:overestimated_nh} and Appendix~\ref{app:2pl_discussion}. 
While the \nh\ bias towards higher column densities has been well demonstrated on single sources \citep[e.g.,][]{marchesiComptonthickAGNsNuSTAR2018, tanimotoNuSTARObservations522022}, our study represents the first time it is quantified in a large ($N>100$) study of an entire AGN population.

The results in Fig.~\ref{fig:nh_errors} highlight the danger of using best-fit values to characterize the survey data.
Even the $22\leq\log\nh<23$ bin, which correctly measures the \nh\ to 90\% confidence for 82.7\% of sources, has best-fit values that are inaccurate by over 1 dex for nearly 40\% of sources.
The current approach of using best-fit values for a distribution that is highly biased, either due to the low count rates or limitations in fitting models, results in many derived values that are significantly outside their predicted errors.
Due to its systematic nature, this important bias must be taken into account while deriving population statistics and when selecting sources based on best-fit properties.

As shown in Fig.~\ref{fig:heatmaps}, a large fraction of AGN with $\log L_\mathrm{2-10, int}<44.5$ and $\log\nh>23$, the sources that produce most of the observed 2--10  keV X-ray background \citep{buchnerOBSCURATIONDEPENDENTEVOLUTIONACTIVE2015, anannaAccretionHistoryAGNs2019, pecaCosmicEvolutionAGN2023}, are not detected by Chandra in the CCLS.
Thus inferences about \nh\ evolution based on this data set and similarly short observations are model dependent and potentially biased, having to rely on the much deeper, but much smaller solid angle, fields of the CDF-N and CDF-S. 

Our methodology of placing the local AGN population on a patch of sky into the high-redshift universe does not account for any potential change to the intrinsic population of AGN over cosmic time.
Many X-ray surveys report an increase in the fraction of obscured AGN, and a decrease in the overall AGN number density, with redshift \citep[e.g.,][]{francaHELLAS2XMMSurveyVII2005, treisterEvolutionObscurationActive2006, cecaCosmologicalPropertiesAGN2008, hasingerAbsorptionPropertiesEvolution2008, uedaStandardPopulationSynthesis2014, airdEvolutionXrayLuminosity2015, lanzuisiChandraCOSMOSLegacy2018, anannaAccretionHistoryAGNs2019, pecaCosmicEvolutionAGN2023, mattheeLittleRedDots2024}.
In Fig.~\ref{fig:obsc_evolution}, we show how the measured obscured fraction compares with the simulated population, in bins of both redshift and intrinsic 2--10 keV luminosity.
This demonstrates the biases that can occur in X-ray modeling of mildly obscured populations of AGN.
The X-ray fitting techniques tend to overestimate the obscured fraction at higher redshifts and at higher luminosities.
This leads to a systematic overestimate of the increase in obscuration with redshift and of the obscured fraction at higher luminosities. 

Surprisingly, the BAT AGN luminosity analogs redshifted into COSMOS are very similar to the high redshift population when folded in with the measurement biases despite being at $z=0$.
This suggests that the trend of increasing obscuration with redshift may be much less significant than previously believed.
Since the increase in \nh\ with redshift has been corroborated in other wavelengths, including by ALMA millimeter wavelength studies of host galaxy contributions \cite[e.g.,][]{gilliSupermassiveBlackHoles2022}, it is unlikely that the increase is in doubt, but our study suggests that it is less significant than X-ray surveys suggest due to modeling biases.

Note that the selection effect we are reporting is a result of using the Chandra response files, and not due to any particular spectral fitting.
While we have used the CCLS exposure times and matched the selection function below 30 counts to mimic the effect of CCLS background, any flux-limited Chandra survey is expected to exhibit a similar bias.
This ultimately stems from low signal-to-noise and limited band pass.

\subsection{Overestimated column densities} \label{subsec:overestimated_nh}

We found that many of the simulated sources measured to be obscured and CT had overestimated column densities.
This problem was most acute for completely unobscured sources (those with $\log\nh\leq20$), and is at least partially attributable to 2PL fits, as shown for the ``false CT'' spectra in Sec.~\ref{subsec:2pl_results}.
Upon inspection of these spectra, we find that (as is generally known) the unabsorbed secondary power law, even at a low normalization fraction, effectively compensates for the extreme absorption of the primary power law.
If the intrinsic energy at which the obscuring material becomes optically thin is beyond the Chandra energy range, then the fit can settle on virtually any high \nh\ to accommodate just a few data points and then use the secondary power law to recover the unobscured continuum.
Effectively, for an unabsorbed source, the \nh\ of the primary component is degenerate with the normalization of the secondary component \citep[as noted by][]{treisterSpaceDensityComptonThick2009}.
Note that this has the added effect of potentially inflating estimates of X-ray luminosity.

Furthermore, because of the reflection and reprocessing of X-ray photons that shine from the corona onto broad-line region, torus, and accretion disk, spectra often exhibit a high energy ``Compton reflection hump'' peaking at a rest frame energy of 20--40 keV \citep{reynoldsComptonReflectionIron1999}.
For false CT spectra with $z\gtrsim2$, this may explain the excess flux at the hard end of the Chandra sensitivity band for high redshift sources, which the 2PL fit misinterprets as a hump where the primary component dominates over the secondary.
\cite{pecaXRayRedshiftsObscured2021} point out that for low count data in this regime, the effective area also plays a role, since the effective area of Chandra decreases dramatically below 2 keV, particularly in the later years of the mission (see Fig.~\ref{fig:eff_area}), and at $z\gtrsim2.5$ the 7.1 keV Fe absorption edge falls at $E\lesssim2$ keV.
Thus, even while the fit passes the $\Delta C$ criterion for a better-fit statistic, the model lacks a correct physical meaning.
Increasing the $\Delta C$ criterion from 2.71 to 6.0 reduces the number of false CT sources from 11 to 4, and increasing the criterion to 9.0 further reduces the number of false CT sources to one.
Future work studying the addition of a \texttt{pexrav} component to model the Compton reflection hump for obscured sources in fitting pipelines like the one adopted in this paper from M16 \citep[e.g.,][]{liuXRaySpectralAnalyses2017, pecaCosmicEvolutionAGN2023} may reveal additional improvements.

To faithfully reproduce M16, we set upper and lower limits on the relative normalization of the secondary power law.
As has been noted in previous studies \citep[e.g.,][]{marchesiChandraCOSMOSLegacySurvey2016,ricciBATAGNSpectroscopic2017, pecaCosmicEvolutionAGN2023}, the secondary component of the 2PL should not exceed a few percent of the primary component, and higher normalizations can lead to pathological fits.

It has been previously suggested by \cite{akylasXMMNewtonChandraMeasurements2006} that AGN at high redshift are likely to have their \nh\ overestimated by Chandra and XMM-Newton.
This is either due to inaccurate photometric modeling, or noisy data in the low-energy bins, which for high redshift (i.e., low count) data, can be misinterpreted as significant obscuration in single power law fits.
Our study sheds light onto another caveat of \nh\ estimations in high redshift data when a 2PL model is used.

A detailed discussion of how 2PL fits can confuse unobscured spectra for heavily obscured ones, and consequences of allowing extreme values for the relative normalization can be found in Appendix~\ref{app:2pl_discussion}.
We present examples with visualized spectra in Appendix~\ref{app:false_ct_examples}.

While we followed the particular procedure of M16 here, various permutations of 2PL fits are widely used in spectral modeling of X-ray surveys \citep[e.g.,][]{brightmanComptonThickActive2014, buchnerOBSCURATIONDEPENDENTEVOLUTIONACTIVE2015, liuXraySpectralProperties2016, liuXRaySpectralAnalyses2017, pecaCosmicEvolutionAGN2023}.
Each survey differs in the details, but most have a main model that is a single or double power law being applied to low count data in Chandra's limited band pass, and the fitting bias we report is expected to apply.
Our work does not address the potential effects of AGN variability \citep{czernyObscurationModelVariability2000}, which can also significantly affect AGN obscuration measurements in surveys that stack multiple pointings \citep{liuXRaySpectralAnalyses2017}.
Instead, we quantify the accuracy of a particular family of fitting procedures, assuming some very well-known intrinsic population based on local measurements.

\subsection{Implications for the AXIS mission} \label{subsec:axis}

Our investigation highlights the magnitude of AGN obscuration bias in X-ray surveys at $E<10$ keV for $z\lesssim5$, and the need for a next-generation X-ray mission to constrain this parameter space.
The Advanced X-ray Imagine Satellite \citep[AXIS;][]{reynoldsOverviewAdvancedXray2023} is a NASA Probe mission concept that will achieve an order-of-magnitude improvement over Chandra's effective area in the 0.2--10 keV band pass (see Appendix \ref{app:eff_area}), an almost two orders of magnitude improvement in high angular resolution survey grasp (i.e., field of view at $<$2\arcsec $\times$ effective area), and a CCD energy resolution of $\sim150$ eV at 6 keV.
Its low background due to a low earth orbit , its large collecting area, and high spatial resolution will allow AXIS to significantly contribute to our understanding of AGN populations at high redshifts.

To test the effectiveness of a mission like AXIS in detecting obscured AGN, we re-simulated our data set with the AXIS \texttt{.rmf} and field-of-view-averaged \texttt{.arf},\footnote{version from July 2023, \url{https://blog.umd.edu/axis/}} using an exposure time of 375 ks, which corresponds to the proposed AXIS ``medium deep'' field \citep{marchesiMockCatalogsExtragalactic2020,reynoldsOverviewAdvancedXray2023, pecaXrayRedshiftsObscured2024}.
Spectra were loaded into \texttt{XSPEC} and channels with energies 0.5--10.0 keV, where the AXIS detectors effective area will be greatest, were used.
We detected at least three counts in 2223/2280 of the simulated sources (97.5\%, compared to 67.2\% with Chandra) and at least 30 counts in 98/192 (51.0\%, compared to 2.6\% with Chandra) of those that were CT.

\begin{figure*}
    \gridline{\fbox{\fig{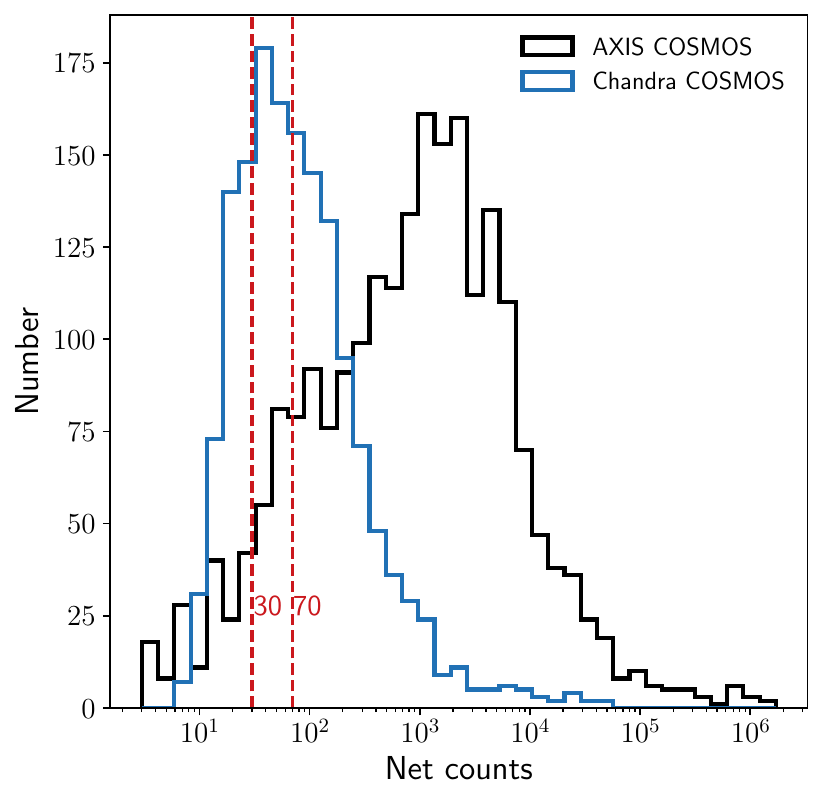}{0.49\textwidth}{(a)}}
        \fig{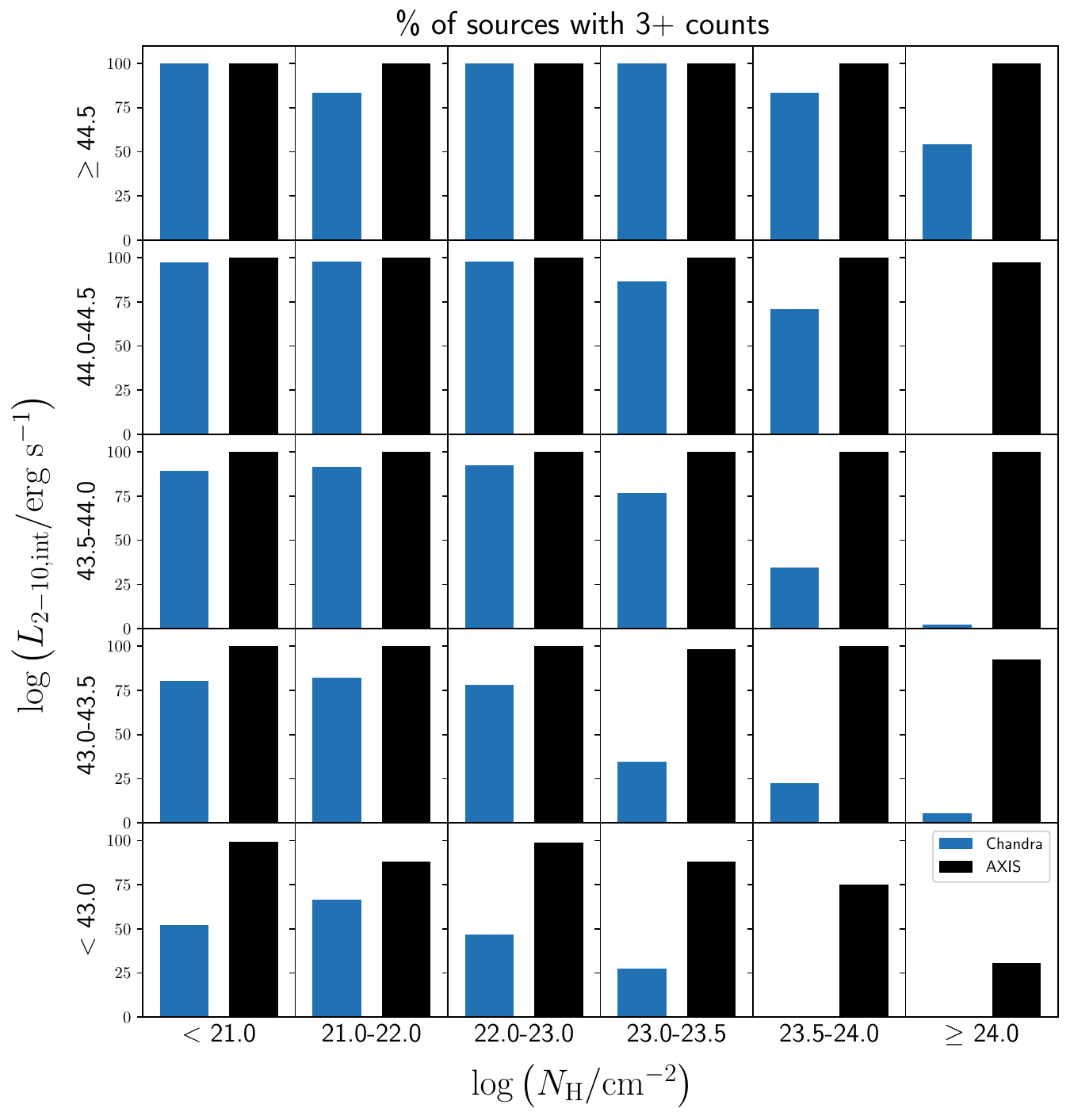}{0.49\textwidth}{(b)}}
    \gridline{\fig{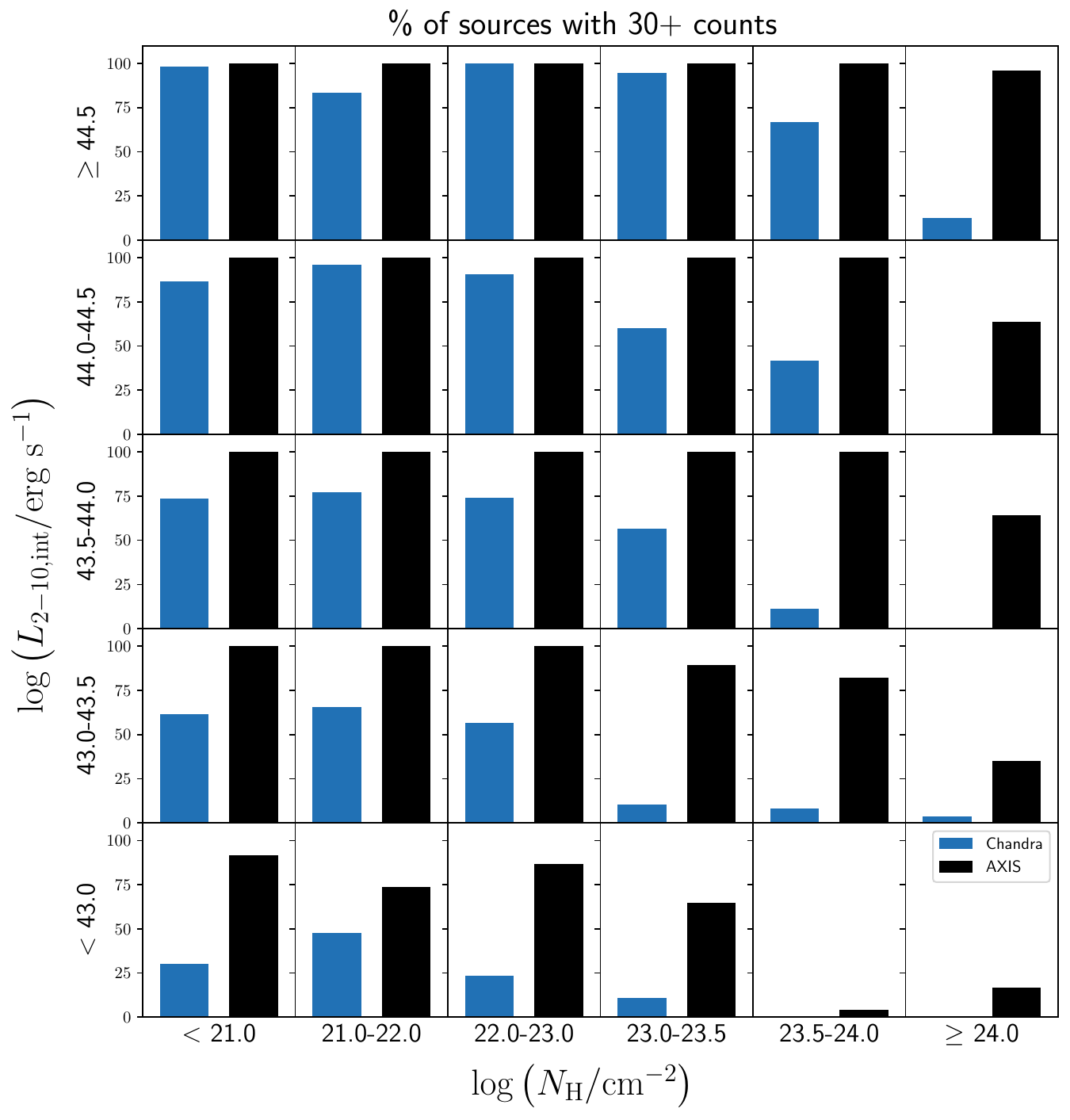}{0.49\textwidth}{(c)}
        \fig{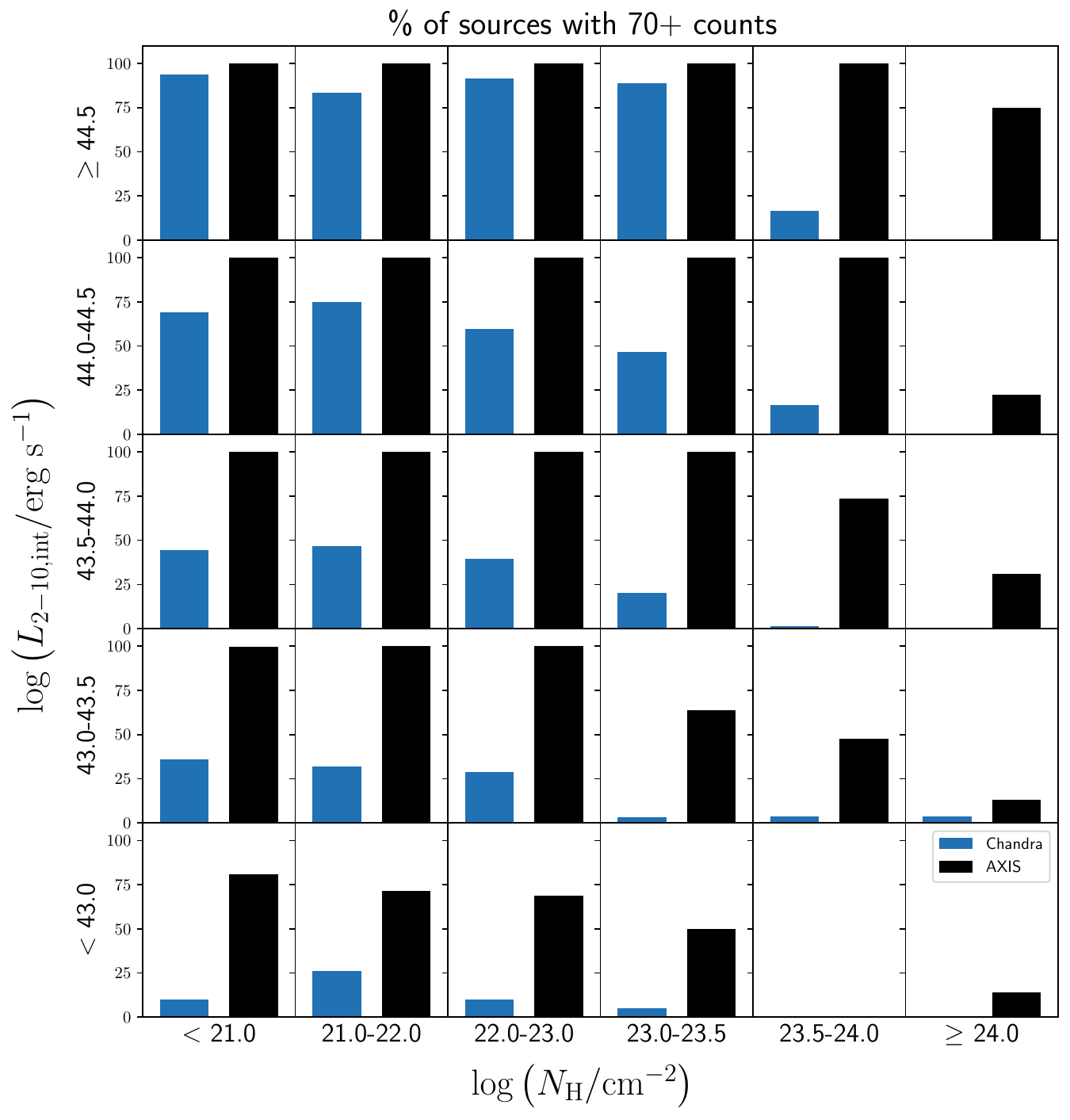}{0.49\textwidth}{(d)}}
\caption{\textbf{AXIS detections of the simulated dataset} organized in bins of intrinsic X-ray luminosity (vertical) and hydrogen column density (horizontal). Contrast with Fig.~\ref{fig:heatmaps}. \textbf{(a) Counts distribution.} Distribution of AXIS net counts at 0.5--10.0 keV (black), compared with that of our Chandra simulations at 0.5--7.0 keV (blue). \textbf{(b)--(c) Percentage of AXIS simulations with 3+, 30+, and 70+ counts.} The percentage of AXIS simulations in each bin is shown in black, compared with the Chandra percentage in blue.\label{fig:axis_heatmaps}}
\end{figure*}

Fig.~\ref{fig:axis_heatmaps} compares the AXIS results with our CCLS simulation results.
Note that since the simulated data set was generated to match the CCLS luminosity distribution, Fig.~\ref{fig:axis_heatmaps}(a) does not include the many lower luminosity sources that would be detected in such a study.
Panels (b)-(d) show that in each bin of \nh\ and luminosity, AXIS detects a larger fraction of sources than does Chandra.
This is most notable for obscured and CT sources.

\section{Conclusions} \label{sec:conc}

In this study, we set out to answer two questions in deep-field AGN surveys: (1) how effective is Chandra at detecting obscured AGN? and (2) how accurately do we measure the intrinsic spectral properties of the AGN?
We employed a novel method of leveraging models of well-understood local AGN from the BASS all-sky survey, and using them as low redshift templates to simulate the higher median redshift CCLS dataset.
We carefully filtered the BASS models to accurately match properties of the CCLS sample, allowing us to compare measured values against the known simulated values.  This simulation method can easily be used to study other missions and surveys.

Our main findings are:
\begin{itemize}
    \item Only \textbf{67.2\%} of all simulated AGN sources were detected (Fig.~\ref{fig:heatmaps}), with fewer than half of obscured and just \textbf{8.9\%} of Compton-thick (CT) sources identified.  Even fewer obscured and CT sources had sufficient counts for rudimentary spectral modeling, highlighting biases in detecting absorbed AGN in deep surveys.  While these results were obtained using the CCLS survey depth, similar results should be expected for any flux-limited Chandra survey, since they are attributable to relatively low count data and limited band pass.
    \item Among spectra recovered with enough counts for spectral modeling with the methodology of M16, column density errors are small but exhibit significant scatter (\(\Delta \log N_\mathrm{H} = -0.10^{+0.73}_{-1.16}\))  (Fig. \ref{fig:delta_param_dist}b).   Unobscured sources (\(\log N_\mathrm{H,sim} \leq 20\)) often have column densities significantly overestimated, with 11 "false CT sources" misclassified due to limitations in 2PL models.  Single power-law models generally estimate column densities at least as reliably as 2PL models in low-count data. Torus models offer improved constraints on column densities for CT candidates and help identify inaccurate 2PL fits.
    \item Therefore at CCLS survey depths, conclusions about the redshift distribution and fraction of obscured sources may be unreliable depending on the level of bias from non-detections and biased spectral modeling (Fig.~\ref{fig:obsc_evolution}).  Next-generation missions like AXIS, however, would significantly improve overall AGN detection rates from \textbf{67.2\% to 97.5\%} and CT source detection from \textbf{2.6\% to 51.0\%} (Fig.~\ref{fig:axis_heatmaps}).

\end{itemize}

While the results of our fitting procedure can likely be generalized to studies with similar model setups as M16, caution should be exercised before comparing with more complex models which are encumbered with much stronger assumptions and more complex systematics.
Note however, that such models may over-fit Chandra data and are not always justified by the Chandra data quality.
  
These considerations highlight the limitations of narrow-band low-count X-ray data and oversimplified spectral models suggesting that additional multi-wavelength analyses are critical.
For luminous systems that are AGN-dominated in the IR, IR colors and ratios of X-ray to IR flux may give rough estimates of line-of-sight obscuration, though studies of nearby AGN have found a very large scatter \citep{pfeifleBASSXXIIINew2022}.
Another observation from low redshift AGN is that nearly all AGN with a broad \Hbeta\ emission line have low X-ray column densities (94\%, $\log\nh<22$, \citealt{kossBATAGNSpectroscopic2017}) though the misclassified population may increase with redshift.
Further studies and simulations are needed to find the best way to use broadband X-ray data with SED model fits and priors on model fitting to best estimate X-ray column density.


\begin{acknowledgments}
    The authors are grateful to: Giorgio Lanzuisi and Stefano Marchesi for their assistance in understanding the CCLS study and for sharing the CCLS spectra; Laura Blecha, Turgay Caglar, Lea Marcotulli, Edmund Hodges-Kluck, and Meredith Powell for their valuable insights and feedback; the anonymous referee for
    constructive feedback.
    We acknowledge support from SAO Chandra archival grant (AR9-20015X) and NASA through ADAP award 80NSSC22K1126 and 80NSSC19K0749 (M.K.); 
    the Israel Science Foundation through grant No. 1849/19 (B.T.); 
    the European Research Council (ERC) under the European Union's Horizon 2020 research and innovation program through grant agreement No. 950533 (B.T.);
    FONDECYT Regular 1230345 (C.R), 1241005 (F.E.B., E.T.) and 1200495 (F.E.B, E.T.); ANID grants CATA-Basal FB210003 (C.R., F.E.B., E.T.); Millennium Science Initiative Program  – ICN12\_009 (F.E.B.); Fondecyt Iniciacion grant 11190831 (C.R.); YCAA Prize Postdoctoral Fellowship and NASA grant 80NSSC22K0793 (M.B.); the China-Chile joint research fund (C.R.).
    The work of D.S.~was carried out at the Jet Propulsion Laboratory, California Institute of Technology, under a contract with NASA.
\end{acknowledgments}
    
\software{
    Astropy     \citep{astropycollaborationAstropyCommunityPython2013, astropycollaborationAstropyProjectBuilding2018}, 
    XSPEC (v12.11.1; \cite{arnaudXSPECFirstTen1996})
}

\appendix

\section{Instrument effective area} \label{app:eff_area}
\begin{figure*}
    \includegraphics[width=0.98\textwidth]{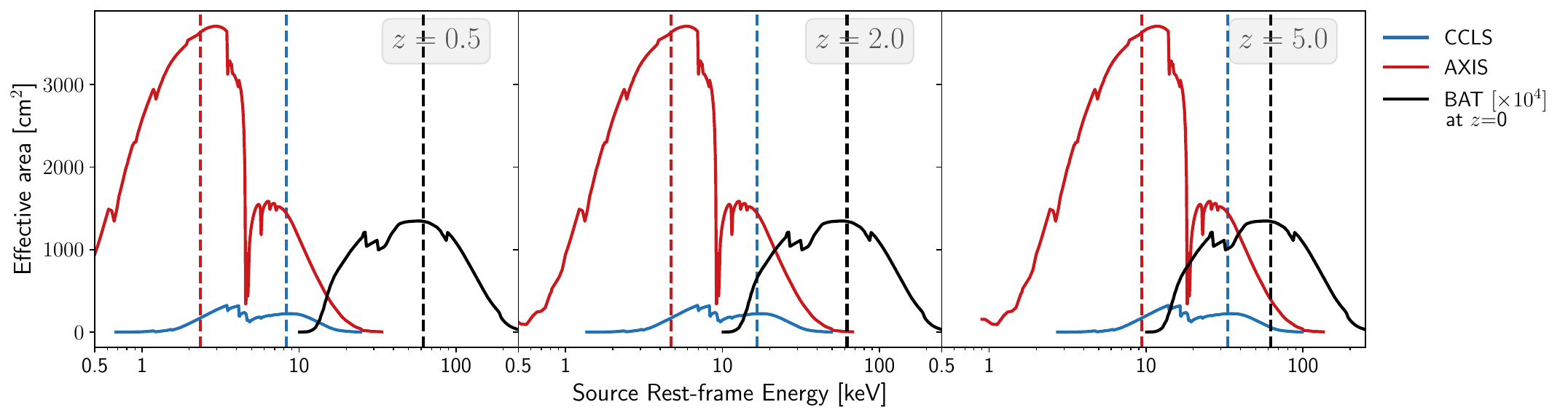}
    \caption{\textbf{Effective area of Chandra ACIS-I and AXIS as a function of redshift.} The effective area of Chandra ACIS-I and AXIS at three different redshifts, compared to that of \bat\ at redshift zero. The \bat\ curve is scaled by $10^4$ for easier visualization. The dashed lines indicate the average energy of the instrument response, weighted by the effective area. The AXIS curves were generated using the field-of-view averaged \texttt{.arf}, and the ACIS curves were generated using a representative \texttt{.arf} file for a CCLS target (\texttt{lid\_151}) during Chandra cycle 14. \label{fig:eff_area}}
\end{figure*}
In principle, soft X-ray telescopes should be as effective at detecting obscured AGN as NuSTAR or \bat\ for deep surveys, since the rest frame energy of the source increases at higher redshifts.
However, to reach energies of 20--30 keV at which heavily obscured sources peak, observations need to be at $z\geq5$, while CCLS and CDF-S mainly observe sources at $z<3$.
In Fig.~\ref{fig:eff_area}, we illustrate how the effective area of Chandra at different redshifts compares to that of \bat\ at $z=0$.
The dashed lines indicate the energies with peak effective area.
Note that at $z=2$, the median redshift of CCLS, Chandra is still below 20 keV, while at $z=5$, it begins to overlap with \bat.
Also note that the effective area of \bat\ is more evenly distributed across its band pass, while Chandra is preferentially weighted towards detector frame energies below 2 keV. 
The consequence of this soft weighting and the fact that a typical $\Gamma$ = 1.4 power law obscured source will have a factor of 25 more counts at 0.5 keV than at 5 keV means that a flux-limited sample will be highly biased to the softest unobscured sources.

We also include the effective area of AXIS for comparison.
Note its order-of-magnitude improvement over Chandra over the entire band pass.

While we focus on effective area in this work, it is important to note that sensitivity, or the minimum detectable flux above background, depends on several factors.
These include not only the effective area, but also the background level and the energy range over which the detector operates.
For extragalactic surveys targeting faint sources, sensitivity can often be a more critical consideration.


\section{BASS Models}\label{app:bass_models}
\begin{figure}
    \includegraphics[width=0.48\textwidth]{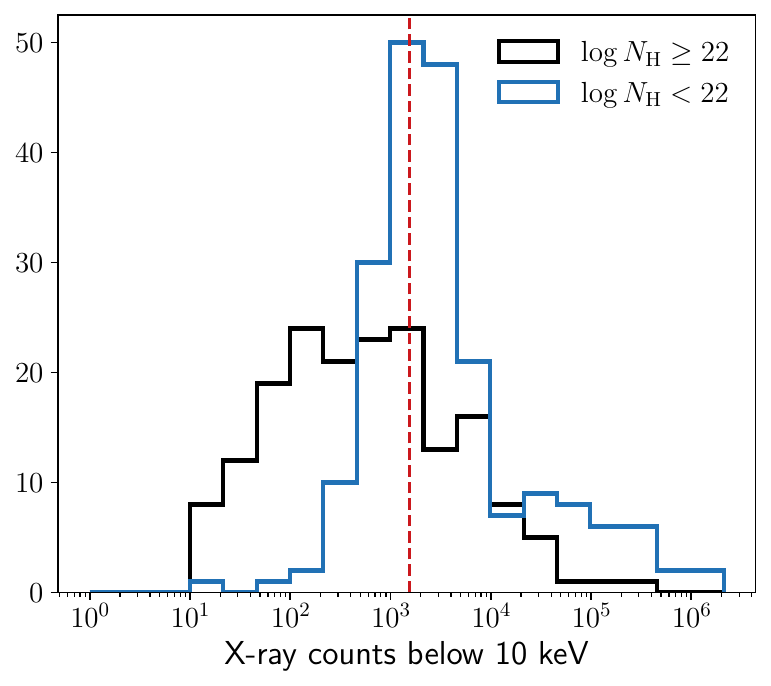}
    \caption{\textbf{Counts distribution of BASS models used.} Photons below 10 keV from which the BASS models were built are from combined ASCA, Chandra,  Suzaku, Swift/XRT, and XMM-Newton data. The blue (black) histogram represents the spectra used to build the unobscured (obscured) BASS models for our simulations. A total of 380 models were used. The dashed red line indicates the median of 1545.5 counts. \label{fig:bass_counts}}
\end{figure}
We briefly summarize the BASS X-ray models that were used as templates to generate our simulation set, as outlined in Sec. \ref{subsec:source_selection}.
We refer the reader to R17 Sec. 4 for full details.

The BASS X-ray models were built from broadband spectra using \bat\ data in the ultra-hard X-ray band, and combined ASCA, Chandra,  Suzaku, Swift/XRT, and XMM-Newton data below 10 keV.
More details can be found in R17.
The counts distribution of the spectra used to build the 380 AGN template models used in our simulations is shown in Fig.~\ref{fig:bass_counts}.
The median number of counts is 1545.5.
Note that both obscured (black) and unobscured (blue) models are built from high-fidelity spectra.

There are four categories of BASS X-ray spectral models: unobscured (A), obscured (B), blazars (C), and other (D).
Since our aim is to simulate the COSMOS field, we only consider A and B models, which have eight and nine variants, respectively.
All variants of these models include the following \texttt{XSPEC} components: \texttt{tbabs} (to model Galactic absorption), \texttt{constant} (cross-calibration constant to normalize between different instruments), \texttt{zphabs}, \texttt{cabs}, and \texttt{pexrav}.
Variants of model A include at least one of the following \texttt{XSPEC} components: \texttt{zxipcf}, \texttt{zpcfabs}, \texttt{bbody}, and \texttt{apec}.
Variants of model B include at least one of the following \texttt{XSPEC} components: \texttt{zxipcf}, \texttt{zpcfabs}, \texttt{cutoffpl}, \texttt{apec}, and \texttt{constant*cutoffpl}.
In A models, the reflection parameter of \texttt{pexrav} was set to be non-negative in order to take into the account the primary X-ray emission and reprocessed radiation at the same time.
In B models, the reflection parameter of \texttt{pexrav} was set to be negative, since in these cases, the reflection component is disconnected from the primary X-ray emission and assumed to be unobscured.

A \texttt{zgauss} component representing the narrow Fe K$\alpha$ line was added to all spectra for which it could be constrained.
297/698 of the BASS template models, and 163/380 of the models selected for simulations, had a Gaussian component to represent the narrow Fe K$\alpha$ line.

The typical errors in $\log\nh$ and $\Gamma$ for these models is shown in Table~\ref{tab:bass_errors}. \begin{deluxetable}{CcCC} \label{tab:bass_errors}
    \tablecaption{Typical errors in the 380 selected BASS models. Each entry shows median value of the width of the 90\% confidence interval, and the values within which the medians of 68\% of models lie.  Note that the width of the $\log\nh$ confidence interval is in logarithmic space.  Also note that since all sources in the BASS catalog with $\log\nh\leq20$ are considered to be consistent with zero, many of the unobscured sources have zero width in the $\log\nh$ measurement.}
    \tablehead{ & Total & \colhead{$\log\nh$ width} & \colhead{$\Gamma$ width} } 
\startdata
\log\nh<22 & 204 & 0.000^{+0.360}_{-0.000} & 0.240^{+0.210}_{-0.140} \\
 \hline
22\leq\log\nh<23 & 68 & 0.170^{+0.140}_{-0.100} & 0.395^{+0.535}_{-0.165} \\
\hline
23\leq\log\nh<24 & 76 & 0.200^{+0.140}_{-0.100} & 0.590^{+0.480}_{-0.240} \\
\hline
\log\nh\geq24 & 32 & 0.345^{+0.595}_{-0.205} & 0.450^{+0.220}_{-0.250} \\
\hline
\hline
\mathrm{ALL} & 380 & 0.130^{+0.230}_{-0.130} & 0.340^{+0.390}_{-0.210} \\
\enddata
\vspace{-25pt}
\end{deluxetable}


\section{Simulated vs. CCLS* column density distribution}\label{app:nh_comparison}
\begin{figure}
    \includegraphics[width=0.48\textwidth]{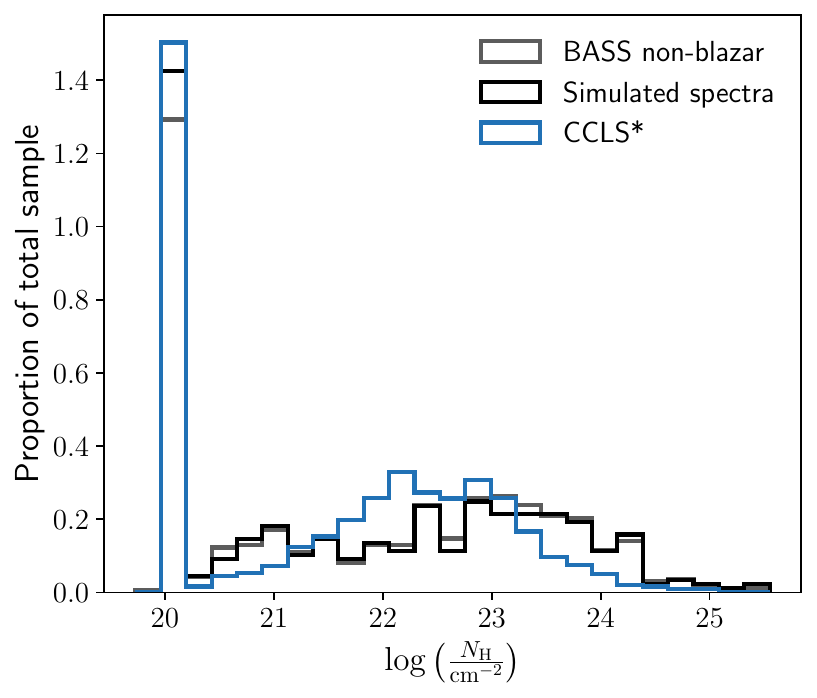}
    \caption{\textbf{\nh\ distributions of the simulated sample and CCLS.} The \nh\ distribution of the BASS models selected for simulations (black) is compared with that of the CCLS (blue). The latter \nh\ distribution was obtained as described in the text. We also include the entire sample of BASS non-blazar AGN (gray). The bin at $\log\nh=20$ includes all sources with $\log\nh\leq20$. \label{fig:nh_comparison}}
\end{figure}

We did not attempt to match the CCLS* \nh\ distribution in our simulated data set.
While this has the potential drawback of not being able to correct the sample size to account for sources missed due to obscuration in the CCLS* sample, it allows us to remain agnostic as to the obscuration bias that we are attempting to measure.
The emphasis of this paper is that, assuming some intrinsic \nh\ distribution, which we draw from the local BASS population, the recovered obscured fraction is different and quantifiable as a function of \nh\ and luminosity.
A comparison of the simulated and CCLS* \nh\ distributions is shown in Fig.~\ref{fig:nh_comparison}.
The CCLS histogram in the figure was obtained from M16, after filtering out all Galactic sources and those with undetermined redshift.
We replaced the \nh\ of all sources in that catalog that were refit in \cite{lanzuisiChandraCOSMOSLegacy2018} (the ``CT candidates'') with the updated values.
These 1855 sources are a subset of the 2291 that comprise CCLS*.
A minimum value of $\log\nh=20$ was imposed for direct comparison between the samples.
We also include the full catalog of BASS non-blazar AGN and note that the 380 selected models are drawn evenly from the \nh\ distribution of entire BASS catalog.


\section{Poisson noise and background}
\label{app:background}

In this appendix, we discuss the potential selection effects introduced by the \texttt{fakeit} procedure, as well as the effect of background photons on our results.

\subsection{\texttt{fakeit} procedure}

To test the selection effects endemic to the \texttt{fakeit} procedure itself, we re-simulated all 2280 spectra, but as simple absorbed power laws \texttt{`phabs*zphabs*zpow'}.
The column density, photon index, and intrinsic luminosity were set by the BASS model, while Galactic absorption, norms, redshift, and exposure times were set as described in Sec. \ref{subsec:source_selection}.
We ignored any Gaussian component.
Even though this is not an ideal model for heavily obscured AGN, the point of this test is to see how the Poisson noise introduced by \texttt{fakeit} affects the ability to recover parameters and not necessarily to accurately model AGN.

1215 of these simulated spectra had at least 30 counts, and we fit these using only the first two steps of the fitting procedure from Sec.~\ref{subsec:fitting_procedure}.
Since the same model was used in both the simulations and the fits, any error in recovering parameters is due to low counts and the introduction of Poisson noise alone.

\nh\ was accurately fit to 90\% confidence for 1098/1215 (90.4\%) of all fits, in 812/874 (92.9\%) of all unobscured simulations, 286/341 (83.9\%) of all obscured simulations, and 7/7 (100\%) of all CT simulations.
The accuracy of \nh\ measurements seems to correlate more with counts than with obscuration: 215/253 (85.0\%) of simulations with fewer than 50 counts, 310/351 (88.3\%) of simulations with 50--99 counts, 153/169 (90.5\%) of simulations with 100--149 counts, and 420/442 (95.0\%) of simulations with at least 150 counts were measured accurately to 90\% confidence.
Note that in the study presented in this paper, \nh\ was accurate to 90\% confidence in 85.3\% of fits, and accuracy did not correlate with counts.

\subsection{Background photons}

Background photons are not included in the simulations, and therefore we do not subtract the background of the simulated spectra or do background spectral modeling.
When determining the detection fraction, we compute the net counts in each simulated spectrum and compare it with the predetermined threshold of three counts.
We expect many spurious sources with at least three counts, and the additional cut that we perform in Sec.~\ref{subsec:detection_cts} for sources with fewer than 30 counts can be interpreted as a comparison with a galaxy catalog or some other cross-reference validation.

For the Chandra ACIS-I detector in cycle 14, the background rate in the 0.5--7.0 keV band for a 2\arcsec\ detection radius\footnote{According to Fig. 2 of \cite{civanoCHANDRACOSMOSLEGACY2016}, most of the sources lie in an aperture of 2\arcsec--4\arcsec.  2\arcsec thus represents the edge of the bin.} with aperture correction is $\sim9\times10^6\ \mathrm{cts}/\mathrm{s} = 1.4$ cts per aperture on average in 160 ks.
Based on the confidence limits in Table 1 of \cite{gehrelsConfidenceLimitsSmall1986}, a 3$\sigma$ detection with 1.4 background counts requires $\sim12$ net counts.
Note that even zero background photons would require seven counts for a detection, so the three count detection criterion results in a conservative overestimate for our detection fraction.
Moreover, according to Fig. 9 of \cite{civanoCHANDRACOSMOSLEGACY2016}, $>99\%$ of the sample has $\gtrsim6$ counts in the 0.5--7.0 keV band and $\sim90\%$ have $\gtrsim10$ counts. 
We compensate for this likely overestimate by matching the CCLS counts function for detections of fewer than 30 counts.

\cite{civanoCHANDRACOSMOSLEGACY2016} use the CIAO \texttt{WAVDETECT} tool \citep{fruscioneCIAOChandraData2006} to determine detected sources, and accept a false-positive detection probability that corresponded to about 10 in every 150 sources ($\sim7\%$).
This corresponds to a detection threshold of about 8 counts for an average of 1.4 background counts per 2\arcsec\ aperture.

We do not expect our spectral fitting results to be significantly affected by the lack of background.
Using Poisson statistics,\footnote{$P(X=k) = e^{-\lambda} \lambda^k/k!$ for a mean and variance of $\lambda$.} if we again assume a mean of 1.4 counts per aperture, then out of 2280 sources (the number of simulations we generated) we expect (562, 787, 551, 257, 90, 25, 6, 1) systems with (0, 1, 2, 3, 4, 5, 6, 7, 8) background counts in their apertures, or a 5\% chance of a source having at least five background counts.
Since we only modeled spectra with at least 30 counts, and our conclusions about fitting results apply just as well to spectra with thousands of counts, we do not expect the lack of background to significantly bias our spectral fitting.


\section{Fitting statistics}
\label{app:fitting_statistics}
\begin{figure}
    \includegraphics[width=0.48\textwidth]{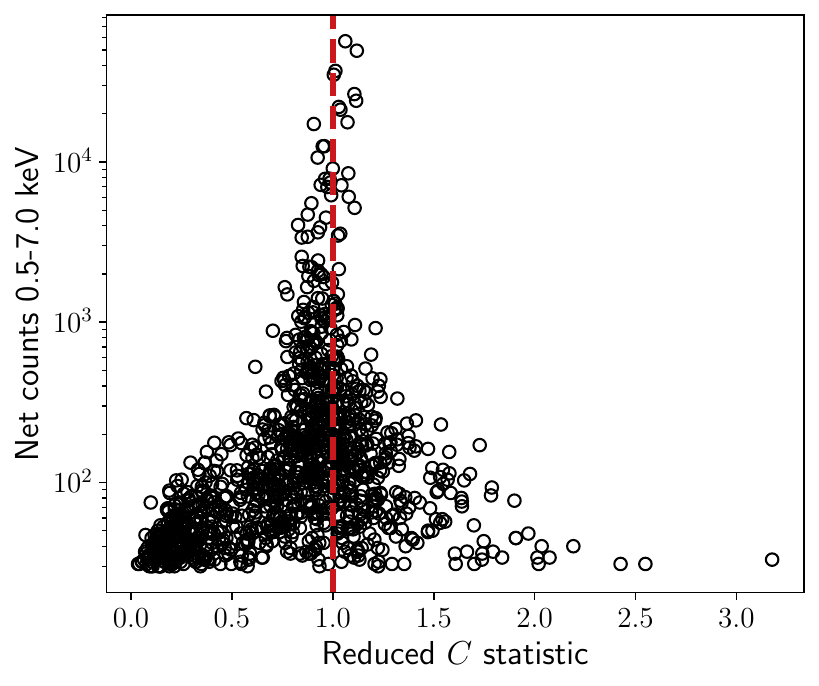}
    \caption{\textbf{Fitting statistics for the simulated data set.} The reduced $C$ statistic is shown along with the net counts of each simulated spectrum.  The red dashed line is drawn at 1. The excess of under-determined fits ($C<1$) for simulations with fewer than 70 counts may be due to the lack of background in our study. \label{fig:cstat}}
\end{figure}

The fit statistics for the 1179 fitted simulations are shown in Fig.~\ref{fig:cstat}.
The reduced $C$ statistic, defined as $C$ divided by degrees of freedom, is a goodness of fit metric.
The reduced value $C\sim1$ implies a good fit.
Reduced $C<1$ ($>1$) implies an under-determined (over-determined) fit.
The median reduced $C$ is 0.87 for sources with at least 30 counts and 0.93 for those with at least 70 counts.
The standard deviations are 0.39 and 0.26, respectively.
An excess of under-determined fits is largely due to spectra with $<70$ counts.
This may be because we did not simulate or model a background spectrum.
This distribution did not change significantly with the more stringent criteria ($\Delta C$ of 6.0 and 9.0).


\section{Detailed discussion of double power law model}
\label{app:2pl_discussion}
\begin{figure}
    \includegraphics[width=0.48\textwidth]{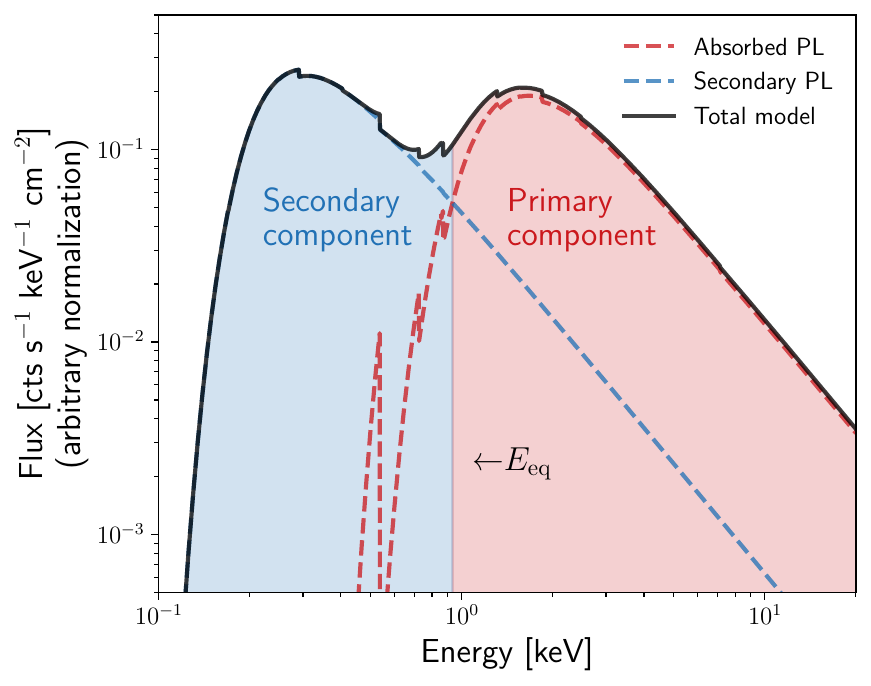}
    \caption{\textbf{Double power law model.} A typical 2PL model with an obscured primary component (red) and an unobscured secondary component with an identical slope (blue). Here the normalization of the secondary component is chosen to be 5\% of the primary normalization. The main feature of this model is the ``hump'' at which the primary component begins to dominate over the secondary component. $E_\mathrm{eq}$ is the energy at which the two power laws are equal, and above which the primary component dominates.
    \label{fig:2pl_model}}
\end{figure}
Here we discuss the effectiveness of a 2PL for capturing AGN obscuration in X-ray spectra.
\subsection{Why a secondary power law?}
The primary power law ($KE^{-\Gamma}$) continuum is observed in the CCLS redshift range at 0.5--20 keV, and since finer features of the spectrum cannot be recovered in low signal-to-noise data, this is what is modeled for basic fits.
When there is a high column density ($\log\nh\gtrsim{20}$), the power law is multiplied by factors of $e^{-\nh\sigma(E)}$, where $\sigma(E)$ is the photoelectric cross section that depends on both energy and species, to account for absorption \citep{morrisonInterstellarPhotoelectricAbsorption1983}.
However, even in obscured cases, there are often photons at energies below the photoelectric cutoff due to photons leaking through the obscuring material and other factors \citep[the physical cause of this soft component is discussed in many papers, including in, e.g., Secs. 4.1.3 and 5.4.2 of R17 and in][for BASS AGN]{guptaBATAGNSpectroscopic2021}.
A single power law can then fit these soft photons simultaneously by underestimating \nh\ and lowering the power law index $\Gamma$, giving a relatively flat model with an inaccurate physical interpretation.
To remedy this while still minimizing the number of fit parameters, a secondary, unobscured power-law component can be added.
Since the soft photons are believed to be sourced from the same coronal emission, the index of the secondary power law is tied to that of the primary, and the normalization can either be free to vary, or even fixed to some fraction of the primary normalization (effectively assuming some coupled covering and scattering fraction for the obscuring material) to maintain the same number of free parameters \citep[e.g.,][]{uedaSuzakuObservationsActive2007}.
This fraction is typically on the order of a few percent \citep[e.g. M16, R17,][]{bianchiNatureSoftXray2007, guptaBATAGNSpectroscopic2021, mckaigRaytracingSimulationsSoft2023, pecaCosmicEvolutionAGN2023}.
The 2PL thus provides a model that is more physically accurate than a single power law while still accommodating low-count data because it has few free parameters when it is statistically required.

\subsection{Column density degeneracy}

The main feature in an absorbed spectrum modeled by a 2PL is the ``hump'' where the primary obscured power law dominates over the unobscured secondary power law.
The location of the hump is determined by the \nh\ of the model.
For low column densities, the hump coincides almost exclusively with the peak of the secondary component; as \nh\ increases, the hump separates from the peak of the secondary component and moves toward higher energies.
We define the energy at which the two power laws are equal, and after which the primary component dominates, as $E_\mathrm{eq}$, as shown in Fig.~\ref{fig:2pl_model}

As a result, for a given detector band pass, each redshift determines a range of \nh\ values that can or cannot be captured by a 2PL (assuming high fidelity data throughout the detector band pass).
If $E_\mathrm{eq}$ is at or above the high end of the sensitivity band, too few photons from the primary component are detected and \nh\ is unconstrained from above.
If $E_\mathrm{eq}$ is at or below the low-energy sensitivity cutoff, then the secondary component is not sufficiently represented, then \nh\ is unconstrained from below.
Thus there is a degeneracy between a single power law and 2PL model if $E_\mathrm{eq}$ is outside of the detector band pass, since in these cases, the spectrum will appear as a single, unabsorbed power law.
This can be interpreted as either the primary component of a relatively unobscured spectrum or as the soft leakage from a highly obscured source.
The degeneracy is further exacerbated by the Compton reflection hump that is often observed in X-ray spectra---a slight excess at the hard end of an unobscured spectrum can be interpreted as the hump of the primary component in a highly obscured spectrum at high redshift.
These degeneracies are of less concern in broadband spectra, which cover a wider range of $E_\mathrm{eq}$ in the detector sensitivity band.
\begin{figure*}
    \includegraphics[width=0.99\textwidth]{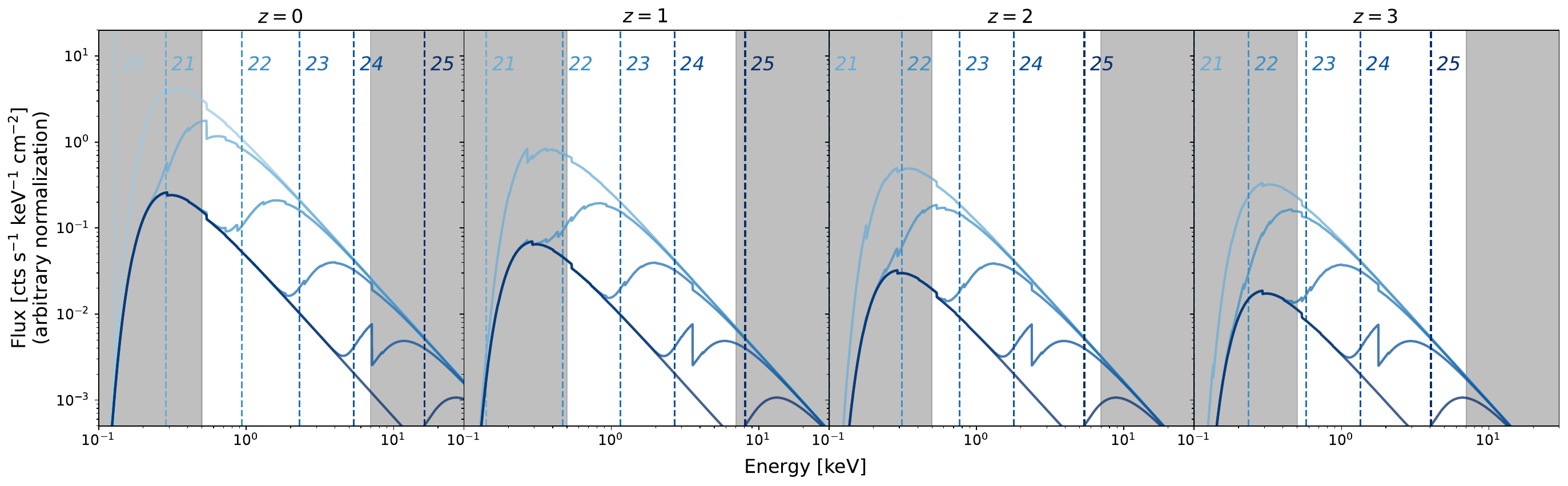}
    \caption{\textbf{Double power laws at varying redshift and \nh.}  We overlay a 2PL model onto the Chandra energy band, at four redshifts. The vertical dashed lines indicate $E_\mathrm{eq}$, the energy at which the primary power law equals the secondary power law, and beyond which the primary component dominates. The shaded regions are beyond the Chandra energy band. For each redshift, the \nh\ values for which $E_\mathrm{eq}$ is well within the white region are the \nh\ values that we expect to be well-recovered by a 2PL model. Other \nh\ values are subject to degeneracy between a heavily obscured model with a significant secondary component and an unobscured single power law.
 \label{fig:2pl_multipanel}}
 \end{figure*}
 \begin{figure}
    \includegraphics[width=0.48\textwidth]{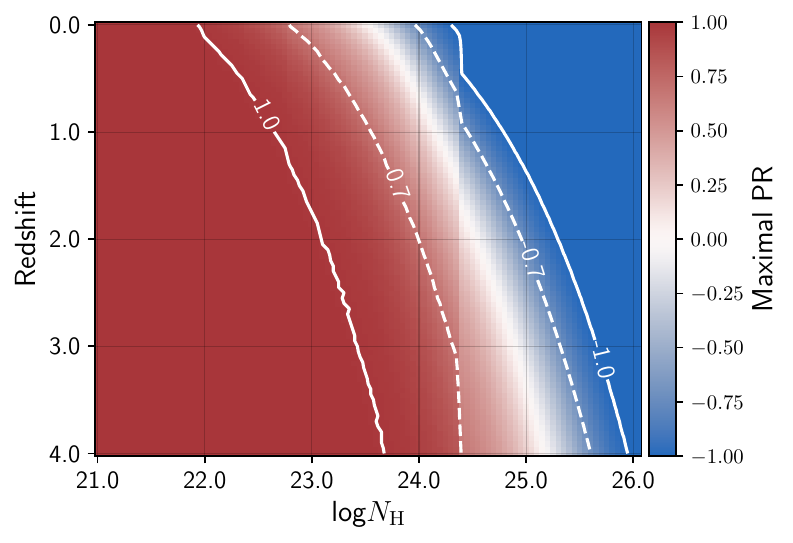}
    \caption{\textbf{Primary ratio as a function of redshift and \nh.} We compute the maximal primary ratio, as defined in the text, for CCLS simulations using a 2PL model with a 5\% secondary normalization at a range of redshifts and \nh. We used the CCLS response files and a high exposure time of $10^{12}$ s to minimize Poisson noise and accurately measure the ratio.  The area between the white lines is where the 2PL is expected to accurately recover \nh\ at a given redshift, assuming high fidelity data.  Outside of the solid white lines (PR$=\pm1$), only a single power law is detectable, which can be interpreted as either the secondary component of a highly obscured source or the primary component of an unobscured source. The discontinuity at $\log\nh\sim24.3$ is due to the absorption edge in the detector response that can be seen in the $\log\nh=24$ models of Fig.~\ref{fig:2pl_multipanel}.
    \label{fig:2pl_heatmap}}
\end{figure}
\begin{figure}
    \includegraphics[width=0.48\textwidth]{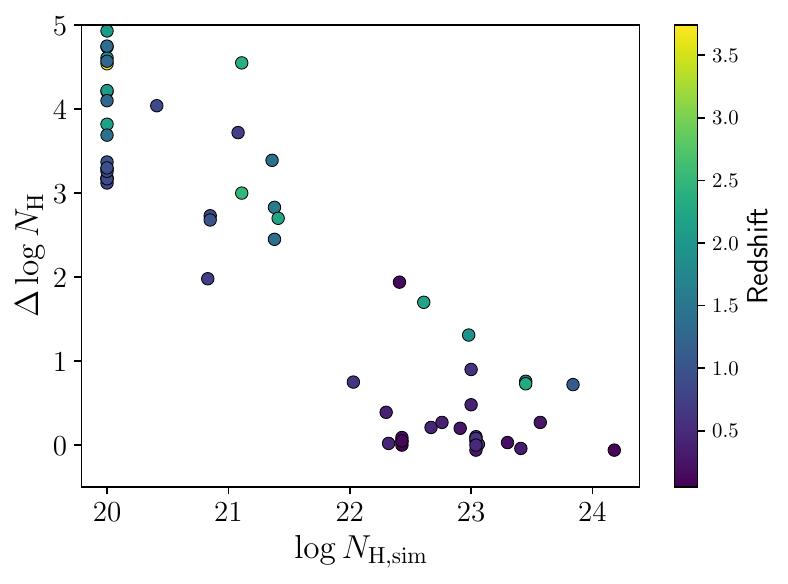}
    \caption{\textbf{\nh\ accuracy of the double power law fits as a function of $N_\mathrm{H,sim}$ and $z$.} Among the 60 simulated spectra that were fit to a 2PL, those with $z<1$ begin to accurately measure \nh\ for column densities of $\log\nh\gtrsim22$, while the higher redshift simulations ($z\gtrsim2$) are measured accurately within 1 dex for $\log\nh\gtrsim23$.  This is consistent with the predictions of the heatmap in Fig.~\ref{fig:2pl_heatmap}.  Note that the ability of the 2PL to accurately estimate \nh\ depends on the actual PR, while the heatmaps predict the maximal PR for a given model. This explains the low redshift fits at higher $\Delta\log\nh$.
    \label{fig:2pl_nh}}
\end{figure}
 
We illustrate this phenomenon in Fig.~\ref{fig:2pl_multipanel}.
The vertical dashed lines indicate $E_\mathrm{eq}$ for different values of $\log\nh$, and the shaded regions are beyond the 0.5--7.0 keV band pass of Chandra.
For each redshift panel, we expect spectra with \nh\ values whose $E_\mathrm{eq}$ is within the white region to be well-recovered by a 2PL.
Note that $E_\mathrm{eq}$ is a function of the secondary power law normalization; here we have assumed a value of 5\%, consistent with the upper limit of the prescriptions in M16 and R17.\footnote{This model only accounts for photoelectric absorption in the primary component. Note that at $\nh\gtrsim10^{23}$, Compton scattering suppresses it significantly further. This effect is taken into account by the \texttt{XSPEC cabs} model. However, our goal here is to evaluate the merits and faults of the 2PL as used in studies like M16.}

To quantify this limitation of the 2PL fit, we compute a ``primary ratio'' (PR), which we define as \begin{equation}
    \mathrm{PR} = \frac{n_\mathrm{p} - n_\mathrm{s}}{n_\mathrm{p} + n_\mathrm{s}},
\end{equation} where $n_\mathrm{p}$ is the number of primary component photons ($E>E_\mathrm{eq}$) and $n_\mathrm{s}$ is the number of secondary component photons ($E<E_\mathrm{eq}$).
PR close to +1 (-1) indicates that nearly all detected photons are from the primary (secondary) component.
For a given model, we can calculate the ``maximal PR,'' the theoretical PR assuming high quality data simulated from the model, and then given data, we can also calculate the actual PR once $E_\mathrm{eq}$ is computed from the model.
The actual PR is what indicates the validity of the fit (see example in Sec.~\ref{app:subsec:example2}). 
We expect that the closer a PR is to zero, the better-suited a 2PL model is to fitting the spectrum.
PR is a function of the detector response, the detector band pass, the secondary normalization, redshift, and \nh.
We simulate a 2PL spectrum with a 5\% secondary normalization using the CCLS response files, and compute the maximal PR for a range of redshifts and \nh.
The results are shown in Fig.~\ref{fig:2pl_heatmap}.

We test this understanding by looking at our simulated data set.
Among those simulations that settled on a 2PL, we look at $\Delta\log\nh$ as a function of $N_\mathrm{H,sim}$ and redshift.
According to Fig.~\ref{fig:2pl_heatmap}, simulations for $z\sim0$ should only be accurate for $\log N_\mathrm{H,sim}\gtrsim22$, with the threshold increasing at higher redshifts, assuming the highest-possible data quality.
Indeed, this expected trend can be observed in Fig.~\ref{fig:2pl_nh}.
This prediction is also consistent with the example spectrum presented in Sec.~\ref{app:subsec:example1}: the simulated and measured \nh\ values are on opposite sides of the solid white lines of Fig.~\ref{fig:2pl_heatmap} for the redshift of the simulation.

\begin{figure}
    \includegraphics[width=0.48\textwidth]{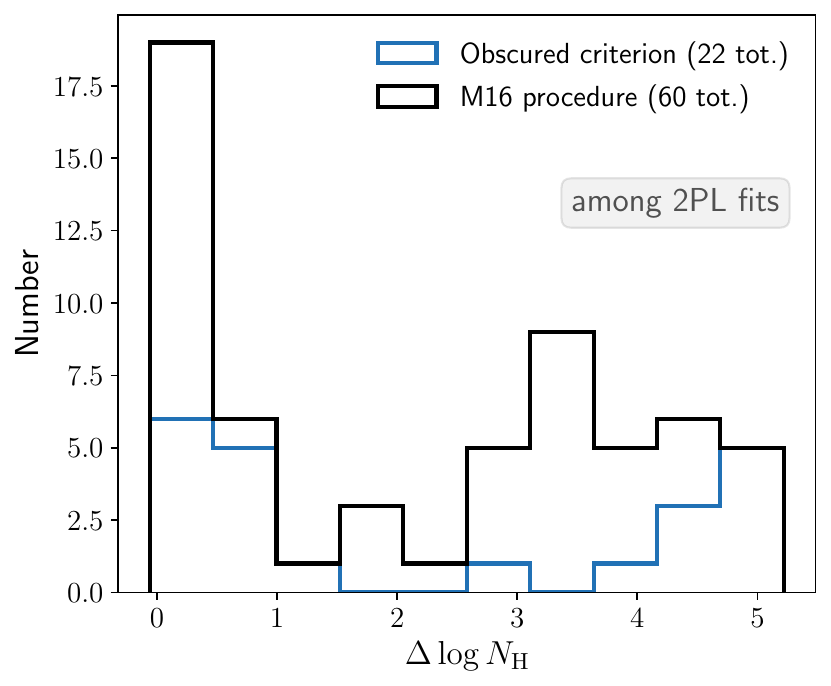}
    \caption{\textbf{\nh\ accuracy of the double power law fits using the obscuration criterion.} A comparison of the $\Delta\nh$ histograms for the 2PL fits in the original procedure (black) and when only attempting the 2PL model for sources with the additional criterion that the single power law estimates $\log N_\mathrm{H,meas}\geq23$ (blue). While the obscured criterion yields fewer fits with moderately overestimated \nh, the criterion does not suppress severely overestimated \nh\ sources.
    \label{fig:2pl_comparison}}
\end{figure}
One potential way to break this degeneracy is to only try the 2PL fit for spectra that are measured to be obscured by a single power law fit.
We refit all spectra according the M16 procedure outlined in Sec.~\ref{subsec:fitting_procedure}, but with the additional check that step 3 (the 2PL fit) was only performed if $\log N_\mathrm{H,meas}\geq23$ to 90\% confidence after that power law fit.
22/60 2PL fits, and 9/11 false CT, remained.

Fig.~\ref{fig:2pl_comparison} compares the \nh\ accuracy of 2PL fits with and without this additional criterion.
While the criterion succeeds in suppressing moderately overestimated \nh\ fits, it still results in many sources with $\Delta\log\nh\gtrsim 4$, all of which are completely unobscured.
The fits with $1\lesssim\Delta\log\nh\lesssim 4$ turn out to be primarily sources with $20<\log N_\mathrm{H,sim}<23$.
If the criterion is made even stricter such that step 3 is only performed if the best-fit $\log N_\mathrm{H,meas}\geq23$, then, by construction, the 2PL only overestimates \nh\ for those sources which the single power law also overestimates, but we find that there is no improvement in accuracy.

After studying different scenarios with the 2PL, including setting a lower threshold for the secondary normalization and stricter $\Delta C$ criteria, we find that with the limited band pass of Chandra, a single power law always estimates \nh\ at least as well as a 2PL.

\subsection{Normalization of the secondary power law}

As noted in previous studies \citep[e.g.,][]{marchesiChandraCOSMOSLegacySurvey2016,ricciBATAGNSpectroscopic2017, pecaCosmicEvolutionAGN2023}, the secondary component of the 2PL should not exceed a few percent of the primary component.
The methodology developed for this study presents an opportunity to explore this additional important limitation of the 2PL.
When we repeated our fitting procedure from Sec. \ref{subsec:fitting_procedure}, but without any constraints on \texttt{`constant.factor'} in the 2PL, we found 243/1179 2PL fits, compared with 60/1179 when we imposed upper and lower limits.
We found 53 false CT sources (see Sec.~\ref{subsec:2pl_results}), compared with 11 in our study, and again, each of them settled on 2PL fits. 
Among them, 17 had relative normalization $>15\%$ and six had a relative normalization $<0.001\%$.

In the case of a high secondary normalization, the depression between the primary and secondary power laws is shallow, and the secondary power law can be used to fit a few noisy data points to an unabsorbed spectrum, and the model no longer physically represents an obscured AGN X-ray continuum.
Each of the four fits with secondary normalizations greater than 50\% were relatively high signal-to-noise spectra ($>150$ counts---these can be fit to more sophisticated models).
An example of a nonphysical 2PL fit with a high secondary normalization is shown below in Sec.~\ref{app:subsec:example3}.

In the case of a very low secondary normalization, there are two factors that can make the model nonphysical.
The first is that arbitrarily low relative normalizations can couple with arbitrarily high normalizations, and arbitrarily high \nh\, for the primary power law, resulting in the secondary component filling the entire band pass.
The second is that, for relatively low count data, when primary normalizations that are physical, extremely low flux corresponds to low signal-to-noise, and is not physically significant.
An example of a nonphysical 2PL fit with a low secondary normalization is shown below in Sec.~\ref{app:subsec:example4}.
\begin{figure}
    \fig{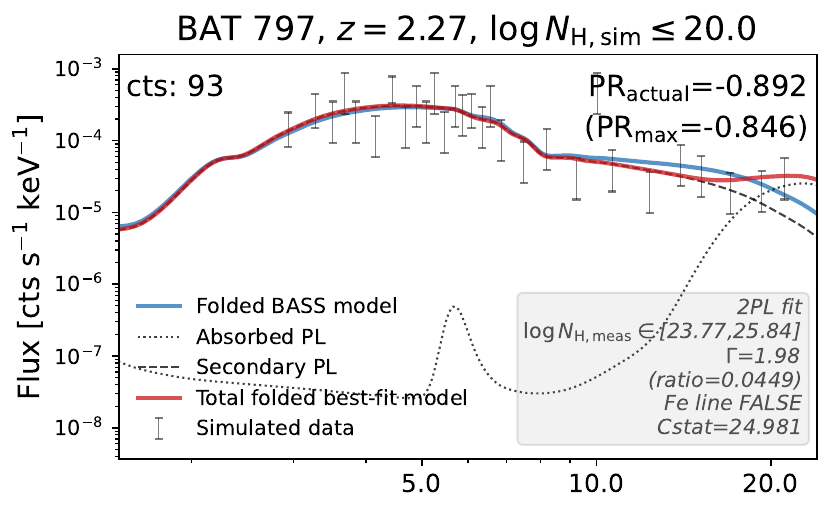}{0.49\textwidth}{}
    \vspace{-40pt}
    
    \fig{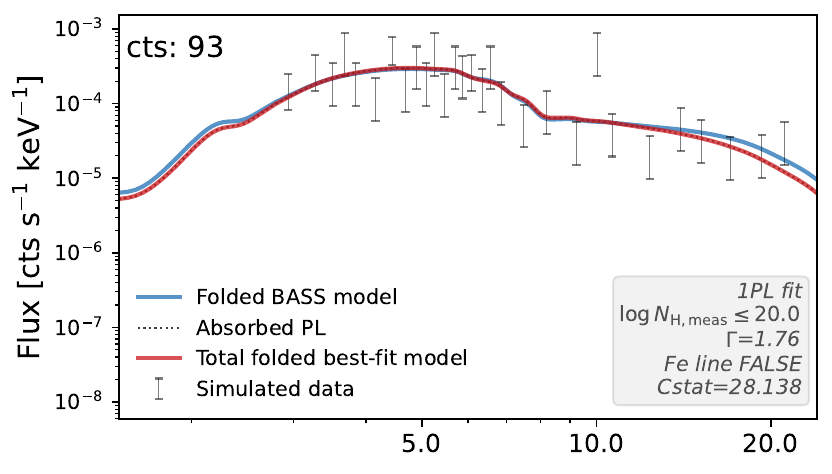}{0.49\textwidth}{}
    \vspace{-30pt}
    
    \fig{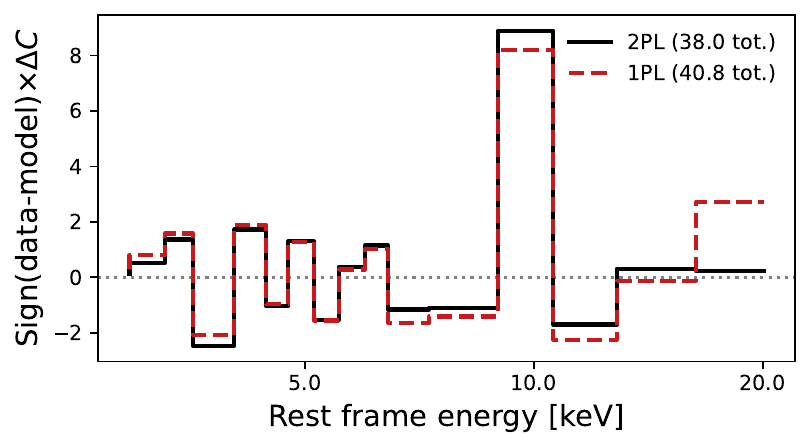}{0.49\textwidth}{}
    \vspace{-25pt}
    \caption{\textbf{False CT example 1: high redshift with a bump.} See Appendix~\ref{app:false_ct_examples} for an explanation of the panels and Sec.~\ref{app:subsec:example1} for a discussion of this example. \label{fig:fakect_example1}}
\end{figure}
\begin{figure}
    \fig{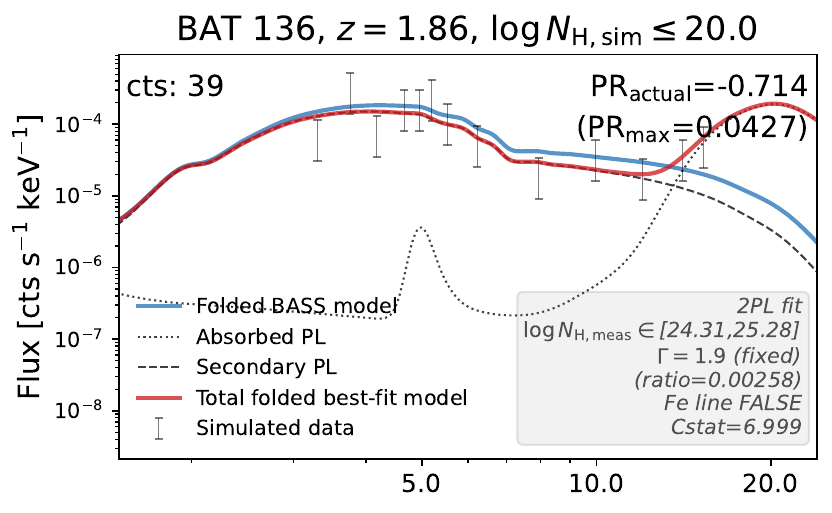}{0.49\textwidth}{}
    \vspace{-40pt}
    
    \fig{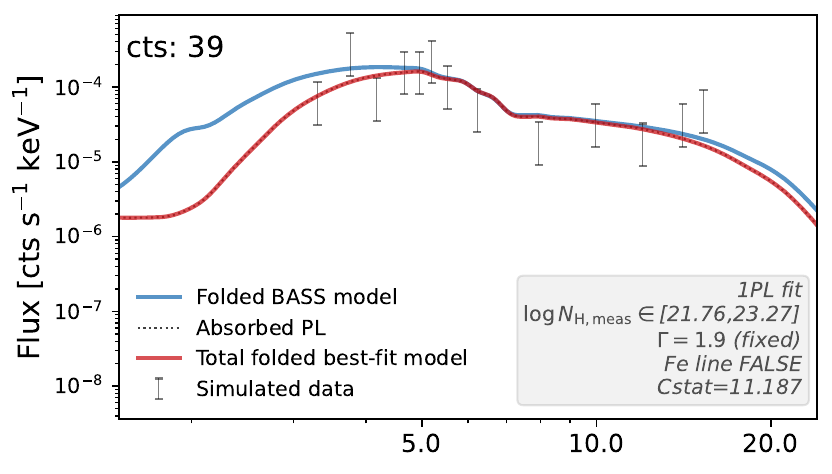}{0.49\textwidth}{}
    \vspace{-30pt}
    
    \fig{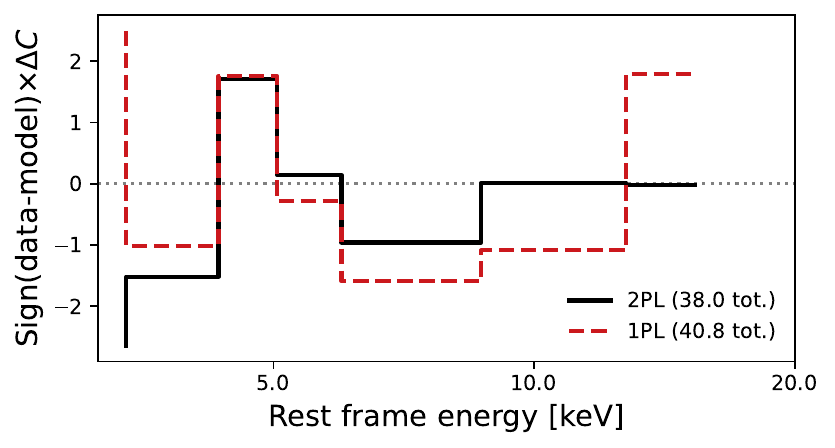}{0.49\textwidth}{}
    \vspace{-25pt}
    \caption{\textbf{False CT example 2: out of band pass, few photons.} See Appendix~\ref{app:false_ct_examples} for an explanation of the panels and Sec.~\ref{app:subsec:example2} for a discussion of this example. \label{fig:fakect_example2}}
\end{figure}
\begin{figure}
    \fig{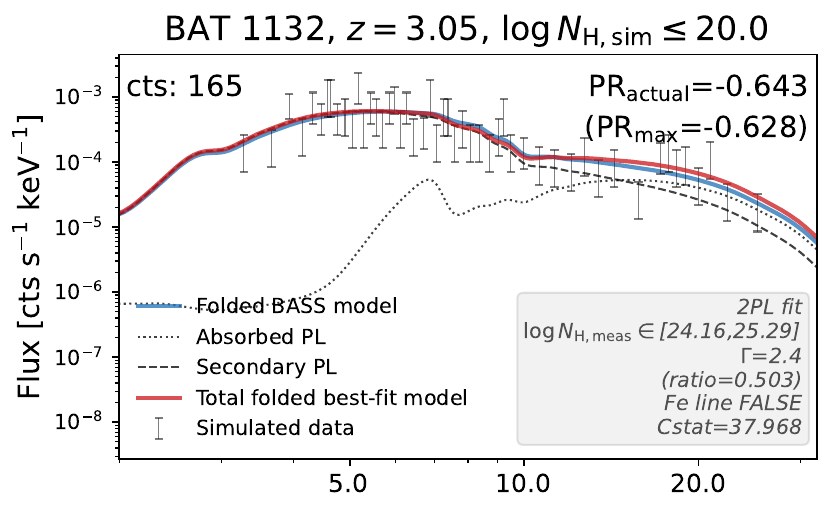}{0.49\textwidth}{}
    \vspace{-40pt}
    
    \fig{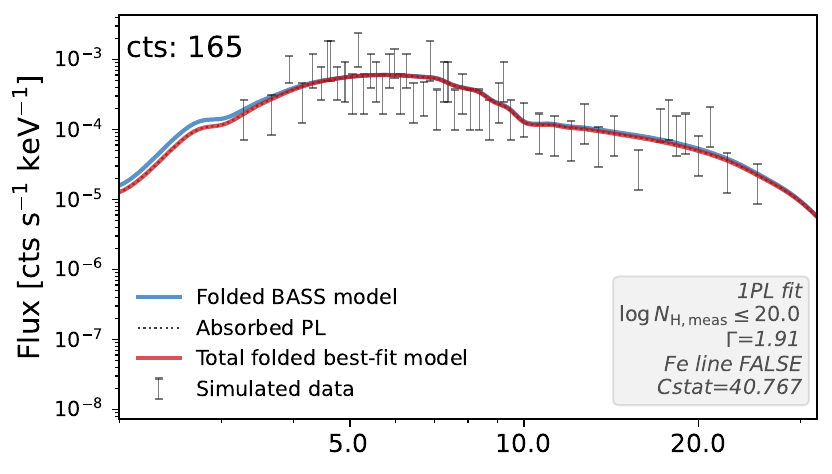}{0.49\textwidth}{}
    \vspace{-30pt}
    
    \fig{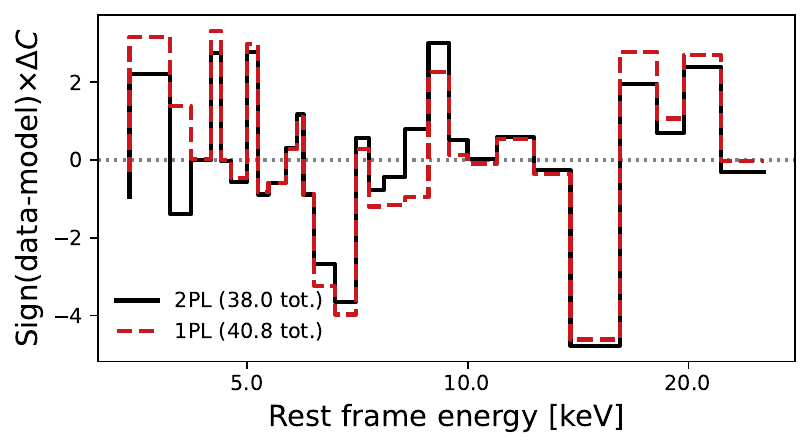}{0.49\textwidth}{}
    \vspace{-25pt}
    \caption{\textbf{False CT example 3: high secondary normalization.} See Appendix~\ref{app:false_ct_examples} for an explanation of the panels and Sec.~\ref{app:subsec:example3} for a discussion of this example. \label{fig:fakect_example3}}
\end{figure}
\begin{figure}
    \fig{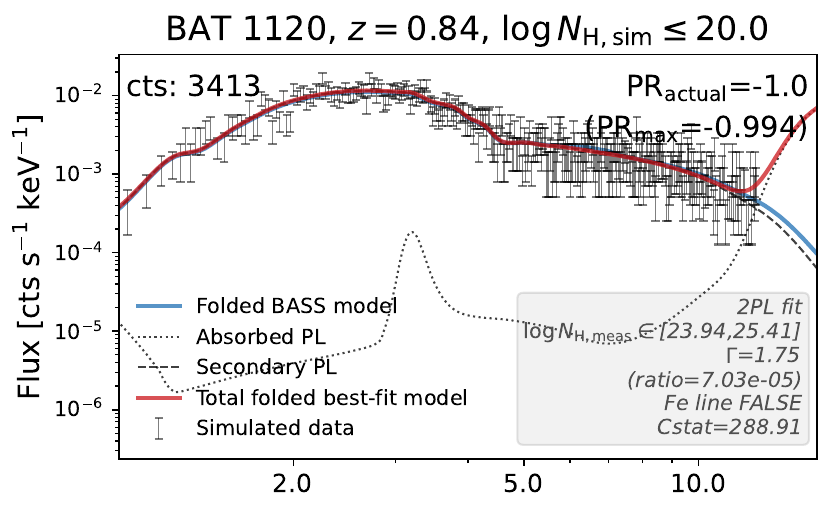}{0.49\textwidth}{}
    \vspace{-40pt}
    
    \fig{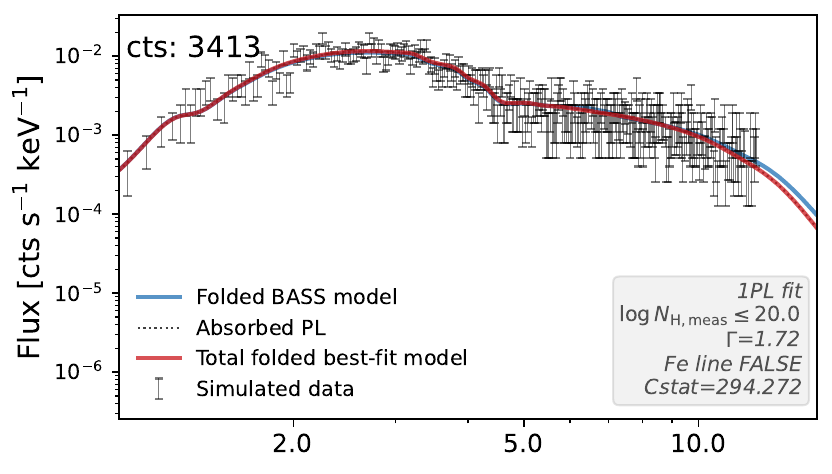}{0.49\textwidth}{}
    \vspace{-30pt}
    
    \fig{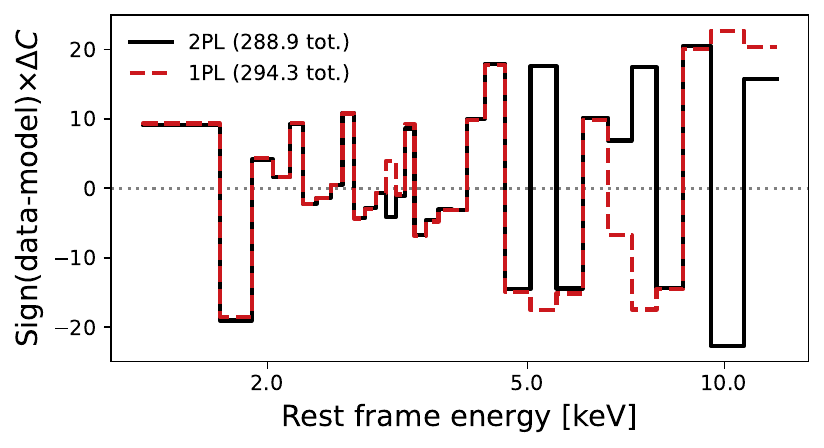}{0.49\textwidth}{}
    \vspace{-25pt}
    \caption{\textbf{False CT example 4: low secondary normalization.} See Appendix~\ref{app:false_ct_examples} for an explanation of the panels and Sec.~\ref{app:subsec:example4} for a discussion of this example. \label{fig:fakect_example4}}
\end{figure}

\subsection{Prescriptions}

To ensure that 2PL models accurately estimate the \nh\ of AGN spectra, it is important to consider the source redshift, the detector band pass, and the slope of the continuum, all of which determine the severity of the degeneracy between unobscured and heavily obscured spectra.
This is quantified by the PR metric, which determines the regions of parameter space within which we expect the 2PL model to accurately measure \nh.
For fixed $\Gamma$, and secondary normalization, the Chandra band pass allows for a range of \nh\ approximately 2 dex wide at each redshift for the which the 2PL is appropriate, as illustrated in Fig.~\ref{fig:2pl_heatmap}.
Outside of that range there is degeneracy between very low and very high column densities.
Furthermore, while a given 2PL may have a suitable maximal PR, the actual PR based on the data may be poorer.

A 2PL model should also be avoided for spectra with high signal-to-noise spectra, where more degrees of freedom can be employed.
Often, careful visual inspection, testing with different $\Delta C$ criteria, and testing with different limits on the secondary normalization can reveal nonphysical fits.


\section{Illustrative examples of false CT sources}
\label{app:false_ct_examples}

Here we provide a description of Figs.~\ref{fig:fakect_example1}--\ref{fig:fakect_example4}, each of which show illustrative examples of false CT sources.
In the top panels, the simulated spectrum is shown as error bars showing each bin's Poisson error of flux.
All spectra are binned to a minimum of three counts per bin. 
The blue solid curves show the BASS template from which the spectrum was simulated, the dotted and dashed lines show the best-fit model components, and the red solid curve shows the total fit model.
The best-fit parameters are shown in the gray text boxes.
$\log N_\mathrm{H,meas}$ shows the 90\% confidence interval for the measured intrinsic obscuration in the case that the spectrum was measured to have $\log N_\mathrm{H,meas}>20$.
Note that $\log\nh\leq 20$ is consistent with zero.
\begin{description}
    \item[Top] The 2PL fit settled on by the fitting algorithm described in Sec.~\ref{subsec:fitting_procedure}.  This fit overestimates the \nh\ of the completely unobscured simulation, since the secondary power law can compensate for the extreme absorption of the primary component. Both the actual and maximal primary ratios (PR, see Appendix~\ref{app:2pl_discussion}) are shown on the top-right of the panel. Note that $|\mathrm{PR}_\mathrm{actual}|$ close to unity indicates that a 2PL model cannot, in principle, recover \nh\ in these cases. Further note that while a model may in principle have a maximal PR close to zero, the data quality ultimately determines the actual ratio of primary to secondary photons.
    \item[Middle] The fit when a secondary power-law component is not allowed, which more accurately measures \nh\ despite a higher overall $C$ statistic.
    \item[Bottom] Comparison of the contribution to the fitting statistic in each energy bin. In each case, the 2PL (black) has an overall lower $C$ statistic than the single power law (red dashed). These panels are adaptively binned for easier visual inspection.
\end{description}

\subsection{Example 1: high redshift with a bump} \label{app:subsec:example1}

Our first false CT example is shown in Fig.~\ref{fig:fakect_example1}.
In this example, there is a slight bump at the high-energy end that the 2PL captures.
By comparing the data with the BASS model (blue), it is seen that this bump is due to Poisson noise.
The residual plot in the bottom panels shows that the high data point at the end contributes most to the more favorable $C$ statistic of the 2PL over the single power law.
Note that a similar effect may occur in AGN spectra with a Compton reflection hump.
This example illustrates the degeneracy between $\log\nh\lesssim23$ and $\log\nh\gtrsim25.5$ for $z=2.27$ shown in Fig.~\ref{fig:2pl_heatmap}.
Note that even spectra with $\gtrsim100$ counts are subject to this degeneracy.
The PR of -0.892 quantifies the dominance of secondary photons over primary photons in this 2PL model.
The single power law in the middle panel accurately classifies this source as unobscured, and by visual inspection can be seen to recover the template model with high accuracy, despite the higher overall $C$ statistic.

\subsection{Example 2: out of band pass, few photons} \label{app:subsec:example2}

Our second false CT example is shown in Fig.~\ref{fig:fakect_example2}.
This example illustrates how visual inspection of spectra can reveal obvious mismatches to data that are not captured in fitting statistics, since the model clearly suppresses the primary power law just enough to push it out of the band pass of this data.
This leaves only the unobscured secondary component in the band pass, which is equivalent to fitting to just a single unobscured power law.
Here we see a case where, in principle, the 2PL model should be able to discern the best-fit $N_\mathrm{H,meas}$ (PR$_\mathrm{max}\sim0$), but the paucity of photons in the actual data render the 2PL insufficient ($|\mathrm{PR}_\mathrm{actual}|>0.7$).
Although the best-fit $\log N_\mathrm{H,meas}$ for the single power law is still an overestimate for this simulation, it performs significantly better.
It is easy to see the cause of this overestimate: two of the first three bins fall below the model curve due to random noise.
At high redshift, any small dip in the spectrum represents a very high obscuration because it's a higher energy bin intrinsically.
Note that this example would still be incorrectly fit for a more stringent $\Delta C$ criterion.

\subsection{Example 3: high secondary normalization} \label{app:subsec:example3}

The false CT example shown in Fig.~\ref{fig:fakect_example4} illustrates how a high secondary normalization results in a shallow depression between the two power laws.
This allows the model to fit an \nh\ that would not typically be degenerate at that redshift by a 2PL with an appropriately low secondary normalization and effectively turns the secondary power law into a fudge factor that can slightly improve a fitting statistic.
Indeed, the PR is not quite as high in this example, but the obscuration is still overestimated.
Notice that the most improvement to the fit statistic is in the first couple of bins, where the secondary power law boosts the model up slightly to better match the BASS template model.

\subsection{Example 4: low secondary normalization} \label{app:subsec:example4}

In the false CT example shown in Fig.~\ref{fig:fakect_example3}, we can see the degeneracy between $\log\nh\lesssim22.5$ and $\log\nh\gtrsim24.5$ for $z=0.84$ shown in Fig.~\ref{fig:2pl_heatmap}.
Note that by allowing an extremely low secondary normalization, even very high-fidelity spectra are subject to this degeneracy.
This example also illustrates how visual inspection of spectra can reveal obvious mismatches to data that are not captured in fitting statistics, since the model clearly suppresses the primary power law just enough to accommodate the band pass of this data.
The secondary power law thus takes up the entire data, indicated by $\mathrm{PR}_\mathrm{actual}=-1$.
This is achieved by a combination of an unreasonably high primary normalization ($2.135\ \mathrm{cts}\ \mathrm{keV}^{-1}\ \mathrm{s}^{-1}\ \mathrm{cm}^{-1}$ at 1 keV, compared with typical values $\lesssim1\times10^{-4}\ \mathrm{cts}\ \mathrm{keV}^{-1}\ \mathrm{s}^{-1}\ \mathrm{cm}^{-1}$ at 1 keV) and an extremely high obscuration of $\log\nh=25.21$.
Note that this example would still be incorrectly fit for a more stringent $\Delta C$ criterion.
In any case, high-fidelity spectra like this can generally be fit with more sophisticated models.


\section{Data tables} \label{app:data_tables}

Our simulated data set was constructed from 380 BASS AGN template models.\footnote{The detailed catalog and data can be found at \url{https://www.bass-survey.com/}} 
The selection of these models is described in Sec.~\ref{subsec:source_selection}.
Table~\ref{tab:templates_table} lists the BAT catalog ID of each model, as well as the six redshifts and exposure times selected for each to generate our simulated CCLS catalog.

Simulation properties and the results of the fitting routine described in Sec.~\ref{subsec:fitting_procedure} are tabulated in Table \ref{tab:fits}.
These are the naive fit parameters before any refitting described in the results and discussion sections.

\begin{deluxetable*}{CDDDDDDDDDDDDDD}
    \tablecaption{\textbf{Template models for the simulated data set.} Each BAT ID model was selected based on its intrinsic 2--10 keV luminosity, and simulated at six different redshifts and exposure times. 
    Redshifts and exposure times were chosen as described in Sec.~\ref{subsec:source_selection}.\label{tab:templates_table}}
        \tablewidth{0pt}
            \tabletypesize{\scriptsize}
        \tablehead{
            \colhead{batID} & \multicolumn2c{$\log L_\mathrm{2-10,int}$} & \multicolumn2c{log\nh} & \multicolumn2c{z1} & \multicolumn2c{z2} & \multicolumn2c{z3} & \multicolumn2c{z4} & \multicolumn2c{z5} & \multicolumn2c{z6} & \multicolumn2c{exp1} & \multicolumn2c{exp2} & \multicolumn2c{exp3} & \multicolumn2c{exp4} & \multicolumn2c{exp5} & \multicolumn2c{exp6}
            \\
            & \multicolumn2c{(erg s$^{-1}$)} & \multicolumn2c{(cm$^{-2}$)} & & & & & & & & & & & & & \multicolumn2c{(ks)} & \multicolumn2c{(ks)} & \multicolumn2c{(ks)} & \multicolumn2c{(ks)} & \multicolumn2c{(ks)} & \multicolumn2c{(ks)}
        }
            \decimals
        \startdata
                1 & 43.42 & 22.19 & 0.63 & 0.73 & 1.01 & 0.76 & 1.09 & 1.56 & 169.92  & 71.79 & 157.76 & 172.92 & 170.11 & 162.63 \\
                2 & 43.61 & 20.00 & 1.02 & 2.56 & 1.63 & 0.84 & 1.94 & 2.50  & 27.40 & 162.74 & 171.17 & 268.40 & 121.21 & 180.44 \\
                4 & 43.04 & 22.86 & 1.00 & 1.20 & 1.16 & 0.95 & 1.24 & 0.82 & 167.36 & 161.05  & 36.13 & 152.89 & 151.93 & 154.02 \\
                5 & 44.01 & 22.61 & 2.88 & 2.98 & 1.51 & 1.35 & 2.19 & 3.09 & 163.24 & 171.64 & 148.57 & 169.99 & 165.49 & 115.70 \\
                6 & 43.23 & 20.48 & 0.53 & 0.97 & 1.71 & 1.48 & 0.62 & 0.67 & 161.41 & 170.91 & 167.86 & 107.66 & 170.76 & 170.14 \\
                7 & 43.57 & 23.56 & 2.12 & 2.20 & 1.42 & 1.87 & 2.01 & 0.92 &  48.79 & 134.86 & 175.13 & 207.95 & 165.18 & 163.84 \\
                8 & 44.17 & 21.04 & 1.59 & 1.10 & 2.78 & 1.98 & 1.21 & 1.83 & 178.15 & 158.09 & 120.07  & 86.11 & 163.80 & 204.62 \\
                10 & 44.40 & 21.98 & 4.13 & 2.04 & 1.39 & 1.21 & 1.73 & 3.07  & 40.43 & 113.54 & 156.24  & 77.90 & 152.88 & 175.29 \\
                14 & 43.68 & 20.00 & 1.86 & 0.73 & 0.72 & 1.02 & 2.33 & 1.78 & 161.63 & 160.84 & 161.07 & 176.04 & 178.22  & 44.74 \\
                16 & 44.39 & 20.00 & 1.69 & 2.71 & 1.34 & 2.31 & 0.94 & 2.29 & 121.83 & 175.50  & 80.27 & 157.78 & 173.77 & 118.27 \\
        \enddata
    \tablecomments{Table \ref{tab:templates_table} is published in its entirety in the machine-readable format. A portion is shown here for guidance regarding its form and content.}
\end{deluxetable*}
    
\startlongtable
\begin{deluxetable*}{llCl}
	\tablecaption{Contents of simulation catalog \label{tab:fits}}
	\tablewidth{0pt}
        \tabletypesize{\scriptsize}
	\tablehead{
		\colhead{Num} & \colhead{Label} & \colhead{Units} & \colhead{Explanation}
	}
	\startdata
	1 & batID &  & BAT ID of source
        \\
        2 & SwiftID & & Swift-BAT 70-month hard X-ray survey source name
        \\
        3 & specZ\_BASS & & Spectroscopic redshift of BASS template source
        \\
        4 & logL2-10-intr\_sim & \mathrm{erg\ s}^{-1} & Logarithm of intrinsic luminosity in the 2--10 keV range of BASS template source
        \\
        5 & logNH\_sim & \mathrm{cm}^{-2} & Logarithm of column density of the neutral obscuring material of BASS template source
        \\
        6 & Gamma\_sim & & Photon index of the primary X-ray continuum of BASS template source
        \\
        7 & gauss\_sim & & Flag for Gaussian in BASS template source
        \\
        8 & z\_sim & & Redshift of simulated spectrum
        \\
        9 & counts & \mathrm{cts} & 0.5--7.0 keV net counts of simulated spectrum
        \\
        10 & detection & & Detection status of simulated spectrum
        \\
        11\tablenotemark{*} & logNH\_meas & \mathrm{cm}^{-2} & Measured logarithm of column density of the neutral obscuring material of simulated spectrum
        \\
        12 & b\_logNH\_meas & \mathrm{cm}^{-2} & Lower bound, 90\% confidence interval for logNH\_meas
        \\
        13 & B\_logNH\_meas & \mathrm{cm}^{-2} & Upper bound, 90\% confidence interval for logNH\_meas
        \\
        14 & Gamma\_meas & & Measured photon index of the primary X-ray continuum of simulated spectrum
        \\
        15 & f\_Gamma\_meas & & Flag on Gamma\_meas, indicating a fixed value during fitting
        \\
        16 & b\_Gamma\_meas & & Lower bound, 90\% confidence interval for Gamma\_meas
        \\
        17 & B\_Gamma\_meas & & Upper bound, 90\% confidence interval for Gamma\_meas
        \\
        18 & Gamma\_ratio\_meas & & Ratio of Gamma\_2 to Gamma\_1 in 2PL model of simulated spectrum
        \\
        19 & pl\_norm & \mathrm{cts/keV/s/cm}^2 & Measured norm of the PL
        \\
        20 & gauss\_norm & \mathrm{cts/keV/s/cm}^2 & Measured norm of the Fe line Gaussian
        \\
        21 & cStat\_meas & & Best fit statistic of simulated spectrum
        \\
        22 & red\_cStat\_meas & & Best fit statistic / dof of simulated spectrum
        \\
        23 & best\_fit & & Model code for best fit of simulated spectrum
        \\
        24 & error\_flag & & \texttt{XSPEC} error code for confidence intervals on \nh
	\enddata
\tablenotetext{*}{This and all subsequent columns are only for fitted sources, which are those with at least 30 counts}
\tablecomments{Table \ref{tab:fits} is published in its entirety in the machine-readable format.}
\end{deluxetable*}


\bibliography{zotero_export}{}
\bibliographystyle{aasjournal}


\end{document}